\documentclass[11pt, a4paper]{article}

\usepackage{graphicx}
\usepackage{bm}
\usepackage{amsmath, amssymb,theorem,verbatim}
\usepackage[round]{natbib} 
\usepackage{setspace}
\usepackage{multirow} 
\usepackage{hyperref}
\usepackage{bm}
\usepackage{algpseudocode}
\usepackage{textcomp}

\long\def\comment#1{}

\newtheorem{theorem}{Theorem}[section]

\newtheorem{lemma}{Lemma}[section]

\newtheorem{note2}{Note}[section]{\bfseries}{\itshape}

\begin{document}

\author{Paul Fearnhead\thanks{Department of Mathematics and Statistics,
Lancaster University,
Lancaster LA1 4YF, UK. Email: \href{mailto:p.fearnhead@lancaster.ac.uk}{p.fearnhead@lancaster.ac.uk}.} \and Piotr Fryzlewicz\thanks{Department of Statistics, London School of Economics, London WC2A 2AE, UK. Email: \href{mailto:p.fryzlewicz@lse.ac.uk}{p.fryzlewicz@lse.ac.uk}.}}

\title{{\it Change-Point Detection and Data
Segmentation}
Chapter: The Multiple Change-in-Gaussian-Mean Problem}

\oddsidemargin=0.25in
\evensidemargin=0in
\textwidth=6in
\headheight=0pt
\headsep=0pt
\topmargin=0in
\textheight=9in

\maketitle

\begin{abstract}
A manuscript version of the chapter ``The Multiple Change-in-Gaussian-Mean Problem'' from the book ``Change-Point Detection and Data Segmentation'' by Fearnhead and Fryzlewicz, currently in preparation. All R code and data to accompany this chapter and the book are gradually being made available through \url{https://github.com/pfryz/cpdds}.
\end{abstract}


\tableofcontents

\vspace{50pt}

In \cite{fearnhead2022detecting}, we were concerned with the problem of testing for, and detecting, a single change-point
in the mean of a noisy univariate data sequence.
In this chapter, we deal with the task of detecting
potentially multiple change-points.

The main added challenges of the multiple change-point problem in comparison with the AMOC problem studied in \cite{fearnhead2022detecting} are estimating the number of change-points, and computing the estimate of the number and locations of change-points in a computationally efficient way. The increased computational challenge should not surprise, as we now have a multivariate optimisation problem to solve. In addition, it may be more difficult to estimate the change-point locations with the same accuracy as in the AMOC problem, the intuitive reason being that the presence of a change-point may, depending on the method of detection and the landscape of the signal, adversely impact location estimation for its neighbour.
These aspects make the multiple change-point problem much harder than the AMOC setting of \cite{fearnhead2022detecting}.

One way of appreciating the additional difficulty is that in the multiple change-point problem, the space of all possible piecewise-constant signals is so large that it is often unclear
what sorts of tools are the most appropriate to perform detection in a given noisy dataset. For example, suppose we were to localise the problem, in the sense of only considering a subsample of the data, in the hope of detecting at most a single change-point in it, using a tool available to us from the AMOC setting (such as the CUSUM statistic). If we took 
a subsample that was too large, we might end up with a stretch of the data containing multiple change-points, which could make our AMOC detection tool suboptimal. Conversely, if we localised with too small a window, our detection power might suffer and we could end up not having enough data points to detect any change-points, even if present.
This may suggest that it could be beneficial to look for an optimal window size to localise with, but this optimal size may vary from one part of the signal to the next, as the change-points may be spread unevenly across the domain of the signal.

The multiple change-point detection problem can logically be split into two sub-problems: estimating the number, and estimating the locations of the change-points.
Some methods available in the existing literature have been designed to solve these two sub-problems at once, some explicitly separate them, and some can be represented in either of these ways and return the same output with either representation. For any given approach,
having separate implementations of solutions to these two sub-problems may be helpful as this then enables
flexible pairing, across different approaches, of modules for model choice (i.e. estimation of the number of change-points) and location estimation.

This chapter covers a basic setting in which we assume that we have observations of a univariate piecewise-constant signal plus independent, identically distributed, zero-mean Gaussian noise, that is
\begin{equation}
\label{ch4:eq:univ_mult}
X_i = f_i + \varepsilon_i,\quad i = 1, \ldots, n,
\end{equation}
with the following properties.
\begin{enumerate}
\item
The signal vector $\mathbf{f} = (f_1, \ldots, f_n)^T$ is deterministic and piecewise-constant, with $N_0$ ($0 \le N_0 \le n-1$) change-points, denoted $\tau^0_1, \ldots, \tau^0_{N_0}$, for which
$f_{\tau^0_j} \neq f_{\tau^0_j+1}$. For notational completeness, we let $\tau^0_0 = 0$, $\tau^0_{{N_0}+1} = n$.
\item
The noise vector $\bm{\varepsilon} = (\varepsilon_1, \ldots, \varepsilon_n)^T$ is a realisation of the $n$-variate normal distribution with $\mathbb{E}(\bm{\varepsilon}) = \mathbf{0}_n$ and
$\mbox{Var}(\bm{\varepsilon}) = \sigma^2\mathbf{I}_n$, where $\mathbf{0}_n$ is a vector of zeros of length $n$, the matrix $\mathbf{I}_n$ is the identity matrix of dimension $n \times n$, and $\sigma^2$ is a positive constant.
\end{enumerate}

Occasionally, in this chapter we will be making references to works whose results apply to larger classes of distributions for $\varepsilon_i$, which include $N(0, \sigma^2)$ as a special case.

Our main task in this chapter is the estimation of the number $N_0$ of change-points, if unknown, and the estimation of their locations $\tau^0_1, \ldots, \tau^0_{N_0}$; these will
always be assumed unknown.
There are some pedagogical advantages in beginning our exploration of estimation in multiple change-point problems with the simple stochastic model (\ref{ch4:eq:univ_mult}).
Being arguably the simplest model for multiple change-points, it has seen the most development in the literature, so it offers the widest possible
background for a discussion of the different methodological approaches to estimation.

As illustrated in Section 2.5 of \cite{fearnhead2022detecting}, most change-point detection approaches are sensitive to the degree of serial dependence in the noise in the input data, and need
a good estimate of the dependence structure to perform well. Under the assumption of the absence of serial correlation (which we make here), its estimation can obviously be skipped, which enables us
to attribute the performance of each change-point detection method to the quality of the detection algorithm itself rather than to the accuracy with which it estimates the dependence structure of the noise.
The variance parameter $\sigma^2$ in model (\ref{ch4:eq:univ_mult}) has to be assumed unknown in practical applications. How to estimate
$\sigma^2$, where this is needed, will be discussed in the respective sections below; see also Section \ref{ch4:sec:noise_variance}.

Model (\ref{ch4:eq:univ_mult}) is a useful testing ground for change-point detection methodologies in the sense that if a method does not perform well in the setting of (\ref{ch4:eq:univ_mult})
despite its simplicity, it is unlikely to perform well in more complex stochastic settings (e.g. ones involving correlated or non-Gaussian noise, or outliers).
Even the simple model (\ref{ch4:eq:univ_mult}) can easily pose serious challenges for the state of the art in estimation, especially in situations in which
the number ${N_0}$ of change-points is large, the change-points are located close to each other, or the signal-to-noise ratio is small.

Finally, another way in which model (\ref{ch4:eq:univ_mult}) is relevant is that  some problems can be reduced to the setting of (\ref{ch4:eq:univ_mult}) by a data transformation, e.g. by performing a variance-stabilising transformation or local pre-averaging of the input data, or by fitting a statistical model in a more complex setting and working with the residuals.


\section{Impact of Loss Function and Typical Forms of Assumptions and Results}
\label{ch4:sec:1}


Not every estimation method that produces piecewise-constant estimates of $\mathbf{f}$ in model (\ref{ch4:eq:univ_mult}) is useful for the task of consistently estimating the change-points in $\mathbf{f}$. In particular, some piecewise-constant estimation methods 
are consistent for $\mathbf{f}$ in the mean-square sense, and can therefore be seen as ``signal approximators'', rather than change-point detectors.
Examples of such approaches include the fused lasso \citep{tsrz05} and the taut string methods \citep{dk01}. Both of these penalise the least-squares fit to the data with a penalty on the sum of the absolute first differences of the signal (the $L_1$ penalty); we cover $L_1$-penalised methods in more detail in Section \ref{ch4:sec:l1}. Another example is the wavelet thresholding or shrinkage approach using orthogonal Haar wavelets \citep{dj94}; see e.g. \cite{vid} for an accessible introduction to the use of wavelets in statistics.


In this chapter, we will mainly be focusing on methods that 
aim to estimate
$N_0$ and the locations of all change-points as accurately as possible.
Many of the methods
described will come with large-sample consistency results stating that under certain assumptions on the minimum spacing between change-points and
the minimum magnitudes of the corresponding jumps in $\mathbf{f}$, for $n$ large enough and on an event set whose probability tends to one with $n$, 
the number $N_0$ of change-points is estimated exactly (i.e., $\hat{N} = N_0$) and, if $N_0 \ge 1$, the change-point location estimators are guaranteed
to lie within certain maximum distances from the corresponding true locations. As a minimum consistency requirement, the largest of these distances must be of order $o(n)$. 

As argued in Section 1.2 of \cite{fearnhead2022detecting}, in the AMOC setting with i.i.d. Gaussian noise, if the true change-point location is away from 
either edge of the data by a distance of $\asymp n$ (throughout the book, we use $f \asymp g$ to mean $\exists\,\,C,D>0\,\,\, C|g| < |f| < D|g|$), and if the jump size is bounded away from zero as $n$ increases, then the 
error in estimating the change-point location is $O_p(1)$. (Recall that $V_n = O_p(a_n)$ means for all $\delta > 0$, there exist $M, N$ such that for all $n > N$ we have $P(|V_n / a_n| > M) < \delta$.) This best-possible rate carries over to the multiple change-point setting if the distances between any
pair of consecutive change-points are $\asymp n$, and if the minimum jump size is bounded away from zero as $n$ increases. A heuristic argument as to
why this is true is that there are estimation procedures in the multiple change-point setting that attempt to ``isolate'' change-points as they detect them, i.e. to break up the
$N_0$-dimensional estimation problem into a sequence of $N_0$ one-dimensional problems, and for each of these the error rate for location is $O_p(1)$
as argued in Section 1.2 of \cite{fearnhead2022detecting}. Some of these estimation methods will be reviewed later in this chapter.

From the discussion in \cite{fearnhead2022detecting}, it is easy to see that it is impossible to rule out asymptotic underestimation of $N_0$ without some assumptions on the
minimum distance between change-points and the minimum jump size. This phenomenon happens even in the AMOC setting, not to mention in the more
challenging setting of multiple change-points. A heuristic argument is as follows. Consider the non-centrality parameter, $\nu$, in formula (3) of 
\cite{fearnhead2022detecting}
 for the distribution of the test statistic for a change-point under the alternative of a change,
and suppose for simplicity that the location of a possible change-point is known to be $\tau^0_1$. The power of the test, and hence the probability of correctly
detecting the change-point at $\tau^0_1$ if there is one at that location, is $P(\chi_1^2(\nu) > k)$, as described in formula (4) of 
\cite{fearnhead2022detecting}.
We must
have $k \to \infty$ for consistency in case there is no change-point at $\tau^0_1$. However, if, for example, $\nu$ is bounded from above as $n \to \infty$, then
$P(\chi_1^2(\nu) > k) \not\to 1$ as $n\to\infty$ and hence we do not estimate $N_0$ consistently if $N_0 = 1$. When is $\nu$ bounded from above in $n$? From
formula (3) of
\cite{fearnhead2022detecting},
this will happen (assuming a fixed jump size $\Delta$) if either $\tau^0_1$ or $n - \tau^0_1$ does not grow with $n$, or in other words if the segment size to the left or to the right of the change-point does not increase with $n$. This illustrates the importance of minimum-distance assumptions. Similar considerations apply to the minimum jump size, both in the AMOC and in the multiple change-point setting.

Finally, in addition to convergence in probability, we note that some authors study almost-sure convergence properties of change-point location estimators, but we leave a more detailed discussion of this interesting point to one of the later chapters.

\section{Relationship to Model Choice in Linear Modelling: Similarities and Differences}

\begin{figure}[ht]
\centering
\includegraphics[width=0.95\linewidth]{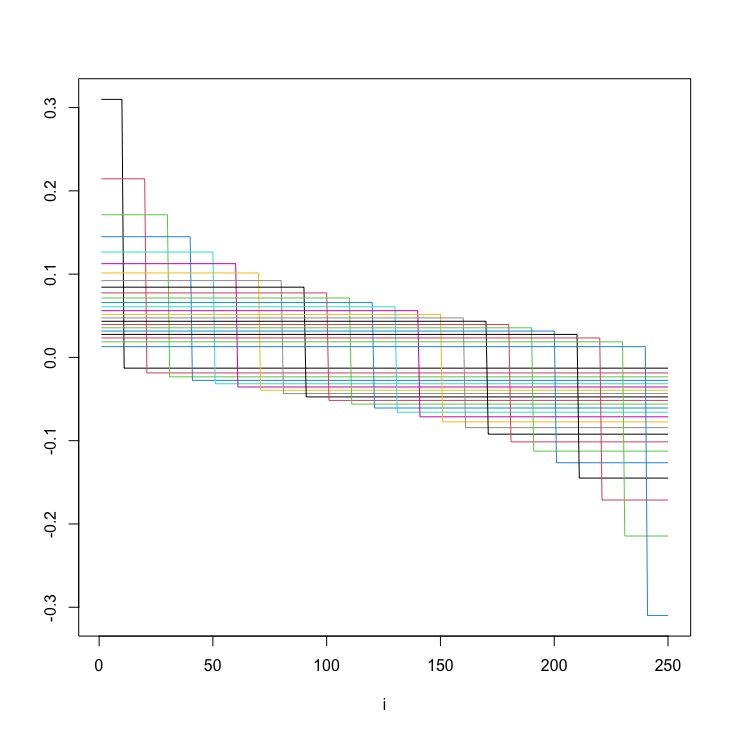}
\caption{Basis vectors $\mathbf{e}^j$ for $n = 250$ and $j = 10, 20, \ldots, 240$.}
\label{fig:ch4-basis-vectors}       
\end{figure}

In this section, we interpret the multiple change-point problem as a problem of model choice in linear regression.
We start by representing the signal vector $\mathbf{f}$ in the Euclidean space $\mathbb{R}^n$ as a linear combination of basis vectors $\mathbf{e}^j$, each of which contains only one
change-point and is normalised in such a way that it sums to zero and square-sums to one. More formally, we define
\begin{equation}
\mathbf{e}^j_i = \left\{  \begin{array}{cc} \sqrt{\frac{n-j}{nj}} & i = 1, \ldots, j, \\ -\sqrt{\frac{j}{n(n-j)}} & i = j+1,\ldots, n, \\
0 & \text{otherwise}, \end{array}  \right.
\end{equation}
for $j = 1, \ldots, n-1$. In addition, define $\mathbf{e}^0$ with
\[
\mathbf{e}^0_i = \sqrt{\frac{1}{n}},\quad i = 1, \ldots, n.
\]
The reader may wonder why we do not define the $\mathbf{e}^j$'s as vectors of the form $(-1, \ldots, -1, 1, \ldots, 1)$ or $(0, \ldots, 0, 1, \ldots, 1)$: this is to ensure that $\sum_{i=1}^n \mathbf{e}^j_i = 0$ and $\sum_{i=1}^n (\mathbf{e}^j_i)^2 = 1$, a common requirement in the linear model selection literature which ensures that the sample correlation between each pair $\mathbf{e}^{j_1}$, $\mathbf{e}^{j_2}$ can simply be computed as $\langle \mathbf{e}^{j_1}, \mathbf{e}^{j_2} \rangle$, their inner product. As an example, a selection of the basis vectors $\mathbf{e}^j$ for $n = 250$ is shown in Figure \ref{fig:ch4-basis-vectors}.

\begin{figure}[ht]
\centering
\includegraphics[width=0.90\linewidth]{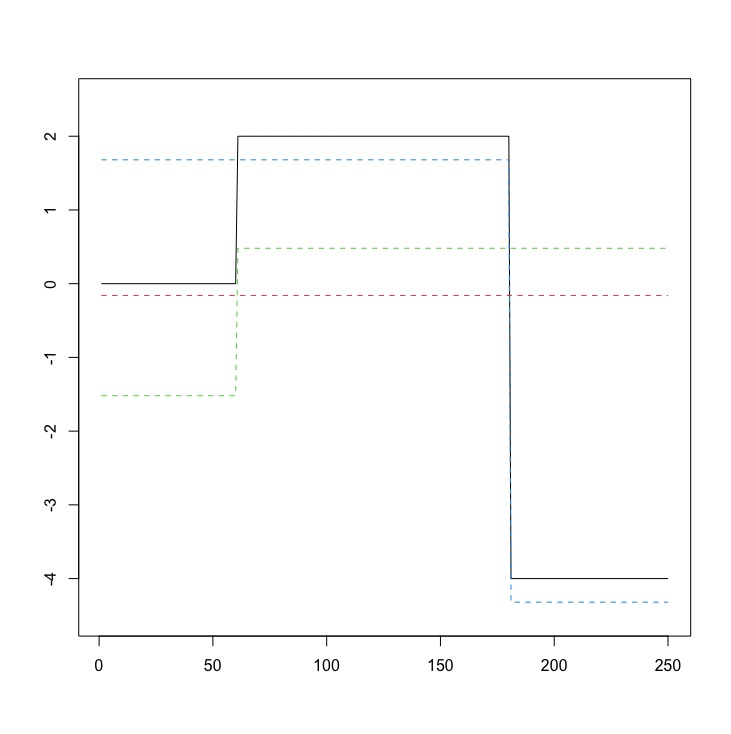}
\caption{Example signal of length 250 with change-points at locations 60, 180 (solid black). Basis vectors times the linear model coefficients that make up the signal: $\beta_0 \mathbf{e}^0$ (dashed red), $\beta_{60} \mathbf{e}^{60}$ (dashed green),
$\beta_{180} \mathbf{e}^{180}$ (dashed blue).}
\label{fig:ch4-basis-vectors-2}       
\end{figure}

With this notation in place, we can represent the signal vector $\mathbf{f}$ as
\begin{equation}
\label{ch4:eq:linmod}
\mathbf{f} = \beta_0 \, \mathbf{e}^0 + \sum_{j \in S_0} \beta_j \, \mathbf{e}^j,
\end{equation}
where $S_0 = \left\{  \tau^0_1, \ldots, \tau^0_{N_0}   \right\}$ is the set of change-point locations. This is obviously equivalent to
\[
\mathbf{f} = \beta_0 \, \mathbf{e}^0 + \sum_{j \in S_0} \beta_j \, \mathbf{e}^j + \sum_{j \in S_0^c} 0 \cdot \mathbf{e}^j,
\]
with $S_0^c = \{1, \ldots, n-1  \} \setminus S_0$, or
\[
\mathbf{f} = \beta_0 \, \mathbf{e}^0 + \sum_{j=1}^{n-1} \beta_j \, \mathbf{e}^j,
\]
where $\beta_j = 0$ for $j \in S_0^c$. An illustration of the representation (\ref{ch4:eq:linmod}) is in Figure \ref{fig:ch4-basis-vectors-2}.

Due to the particular normalisation of the $\mathbf{e}^j$ vectors, the standard setting, in which the jump sizes corresponding to each change-point are fixed in $n$, translates to the condition that $c_1 \le |\beta_j|/\sqrt{n} \le c_2$, $j \in S_0$, for two positive constants $c_1, c_2$.
With this representation, our multiple change-point problem is equivalent to the following: given the linear regression problem
\begin{equation}
\label{ch4:linreg}
\mathbf{X} = \beta_0 \, \mathbf{e}^0 + \sum_{j=1}^{n-1} \beta_j \, \mathbf{e}^j + \bm{\varepsilon},
\end{equation}
identify the set $S_0 = \{j \, : \, \beta_j \neq 0, j > 0 \}$, i.e. the set of change-point locations $\left\{  \tau^0_1, \ldots, \tau^0_{N_0}   \right\}$. Therefore, the multiple change-point detection problem can be interpreted as a sparse variable selection problem, in a ``high-dimensional'' setting in which
the dimensionality of the parameter space equals the number of observations $n$
and the cardinality of the set $S_0$ of the relevant variables equals $N_0$, presumed to be 
much smaller than $n$.

The variable selection problem in sparse high-dimensional linear regression has been studied extensively in the literature, and various methodologies for the recovery
of the set of non-zero variables have been proposed. We note in particular such techniques as the lasso \citep{t96}, the adaptive lasso \citep{z06}, the elastic net \citep{zh05} and the Dantzig selector
\citep{ct07}. We say that a variable selection technique has a sign consistency property if it recovers, with probability tending to one as the number of observations
increases, the set $S_0$, with the correct signs of the constituent $\beta_j$'s.

Any sign consistency result must rely on the design matrix satisfying certain properties. Indeed, if two columns of the design matrix were identical, no method would be able to identify which of the two covariates in question (if either) contributes to the response. This intuition agrees, for example, with \cite{l08}, whose result shows that the lasso has the sign consistency property if the maximum correlation between two different columns
is less than $1/7$, while a similar result for the Dantzig selector requires it to be less than $1/3$.

However, the unusual property of the design matrix in the linear model (\ref{ch4:linreg}) is that it is, in a certain sense, asymptotically collinear. As an example, take two neighbouring
columns $\mathbf{e}^j, \mathbf{e}^{j+1}$ of the design matrix (where $j \ge 1$). The correlation between them is
\begin{eqnarray*}
\frac{\langle  \mathbf{e}^j, \mathbf{e}^{j+1}  \rangle}{\langle  \mathbf{e}^j, \mathbf{e}^{j}  \rangle^{1/2} \langle  \mathbf{e}^{j+1}, \mathbf{e}^{j+1}  \rangle^{1/2}} & = &
\langle  \mathbf{e}^j, \mathbf{e}^{j+1}  \rangle = \frac{1}{n} \left( j \sqrt{\frac{(n-j)(n-j-1)}{j(j+1)}} \right.\\
& - & \left. \sqrt{\frac{j(n-j-1)}{(n-j)(j+1)}}  + (n-j) \sqrt{\frac{j(j+1)}{(n-j)(n-j-1)}} \right).
\end{eqnarray*}
Taking $j = \alpha n$, where $\alpha$ is a constant in $(0, 1)$, the dominant term in the above is 
\[
\frac{1}{n}\left(  \alpha n \sqrt{\frac{(1-\alpha)^2}{\alpha^2}} - \sqrt{\frac{\alpha(1-\alpha)}{(1-\alpha)\alpha}} + (1-\alpha)n\sqrt{\frac{\alpha^2}{(1-\alpha)^2}}   \right) = 1 - \frac{1}{n}.
\]
Therefore, a large portion of the design matrix exhibits asymptotically perfect collinearity between neighbouring columns.

Essentially, this asymptotic collinearity property means that our design matrix $\{  \mathbf{e}^j    \}_{j=0}^{n-1}$ violates the assumptions required for the sign consistency of any variable selection method when applied to problem (\ref{ch4:linreg}). This means that unless $\sigma^2 \to 0$ with $n$, i.e. we are in the (asymptotically) noiseless setting, or if some of the $\beta_j$'s for $j \in S_0$ are larger than $O(\sqrt{n})$, i.e. the corresponding jump sizes increase in $n$, it is impossible for any method to detect the change-point locations exactly. This makes sense as we
know from Section 1.2 of \cite{fearnhead2022detecting} that error-free detection of exact change-point locations is impossible unless the noise variance $\sigma^2$ tends to zero with the sample size $n$ or the relevant jump size increases.

The ``variable selection'' point of view on multiple change-point detection is pursued e.g. in \cite{hll10}, where this viewpoint is enabled by moving from a fused-lasso \citep{tsrz05} formulation of a multiple change-point
detection problem to a standard lasso formulation. We describe this in more detail in Section \ref{ch4:sec:l1}.

\section{Estimators Defined as Optima}
\label{ch4:sec:optima}

One traditional route to solving the multiple change-point detection problem (\ref{ch4:eq:univ_mult}) is to identify it
with the problem of fitting a piecewise-constant function to the data $\mathbf{X} = (X_1, \ldots, X_n)^T$. If the number $N_0$ of change-points is known, in the i.i.d. Gaussian model (\ref{ch4:eq:univ_mult}), such a fit can be carried out via ordinary least squares, which for a Gaussian model is equivalent to maximum likelihood estimation. This results in an $N_0$-dimensional optimisation
problem. Having solved it and obtained the fit, the change-points in the piecewise-constant estimate serve as estimates of the 
change-point locations $\tau^0_1, \ldots, \tau^0_{N_0}$.

This optimisation-based approach to the multiple change-point problem extends to the case of unknown number $N_0$ of
change-points via penalised cost approaches. In these, the estimator is defined as the minimum of a criterion which can be decomposed
as a measure of fit to the data plus a penalty. The role of the penalty is to discourage solutions with particularly many
change-points (the exact fit to the data is given by the solution $\hat{\mathbf{f}} = \mathbf{X}$, which is
of no use as a statistical estimator).

The theoretical consistency analysis is then typically done for this theoretical minimum. Computing the minimum is a different matter
entirely. This section covers the conceptual, theoretical and computational aspects of unpenalised (for $N_0$ known)
and penalised (for $N_0$ unknown) cost approaches to multiple change-point detection.

\subsection{Estimating Change-point Locations When \texorpdfstring{$N_0$}{the Number of Change-points} Is Known}
\label{ch4:sec:n0known}

In this section, we assume that $N_0 > 0$ is known: if not, then we can estimate it consistently as outlined in
the following sections.
For any $\{k_1, \ldots, k_m\} \subseteq \{1, \ldots, n-1\}$,
define
\begin{equation}
\label{ch4:sn}
S_n(k_1, \ldots, k_m) = \sum_{r=1}^{m+1} \sum_{i=k_{(r-1)}+1}^{k_{(r)}} (X_i - \bar{X}_{(k_{(r-1)}+1):k_{(r)}})^2,
\end{equation}
where $k_{(0)} = 0$, $k_{(m+1)} = n$, $0 < k_{(1)} < \ldots k_{(m)} < n$ are the ordered versions of $k_1, \ldots, k_m$, and
$\bar{X}_{a:b} = \frac{1}{b-a+1}\sum_{i=a}^b X_i$, the mean of the data $X_a,\ldots,X_b$.
Consider
\begin{equation}
\label{ch4:eq:taun0known}
(\hat{\tau}_1, \ldots, \hat{\tau}_{N_0}) = \arg\min_{l_1 < \ldots < l_{N_0}} S_n(l_1, \ldots, l_{N_0}),
\end{equation}
with the convention $\hat{\tau}_0 = 0$, $\hat{\tau}_{N_0+1} = n$. Let $\hat{q}_j = \hat{\tau}_j / n$.

The estimators $(\hat{\tau}_1, \ldots, \hat{\tau}_{N_0})$ are the least-squares estimators of the change-point locations. As mentioned above, they are
also the maximum likelihood estimators if the noise $\{\varepsilon_i\}_{i=1}^n$ is i.i.d. standard normal. In this section, we do not discuss
the {\em computation} of $(\hat{\tau}_1, \ldots, \hat{\tau}_{N_0})$; this will be covered in Section \ref{ch4:sec:comp}. On the face of it, the computation
of $(\hat{\tau}_1, \ldots, \hat{\tau}_{N_0})$ may appear to be a difficult problem: the parameter space is discrete and constrained, and the minimisation
is $N_0$-dimensional. Fortunately, as we will discover in Section \ref{ch4:sec:comp}, there are fast algorithms for computing this multivariate minimum.
In this section, we are concerned with the stochastic behaviour of $(\hat{\tau}_1, \ldots, \hat{\tau}_{N_0})$ as a 
theoretically formulated argument-minimum, regardless of how easy or fast it is to compute it.

\cite{ya89} describe the behaviour
of $(\hat{q}_1, \ldots, \hat{q}_{N_0})$ for continuously distributed, i.i.d. noise $\{\varepsilon_i\}_{i=1}^n$ such that $\mathbb{E}(\varepsilon_i) = 0$ and
$\mathbb{E}(\varepsilon_i^6) < \infty$, which includes the i.i.d. standard normal case, the main focus of this chapter. This section summarises
this discussion.
We first quote and discuss a result that states that the estimators $(\hat{q}_1, \ldots, \hat{q}_{N_0})$ are consistent for $(q_1^0, \ldots, q_{N_0}^0)$ (where $q_j^0$ is a rescaled-time version of $\tau_j^0$ and is formally defined in Theorem \ref{ch4:lemconsist} below),
but with no mention of the rates of convergence just yet. We have the following result, adapted from \cite{ya89}.

\begin{theorem}
\label{ch4:lemconsist}
Let the i.i.d. noise $\{\varepsilon_i\}_{i=1}^n$ be such that $\mathbb{E}(\varepsilon_i) = 0$ and $\mathbb{E}(\varepsilon_i^6) < \infty$.
Assume $\tau_j^0 = [n q_j^0]$, where $[\cdot]$ denotes the nearest integer to its argument, and $0 = q_0^0 < q_1^0 < \ldots < q_{N_0}^0 < q_{N_0+1}^0 =1$
are fixed in the rescaled-time interval $[0,1]$ as $n \to \infty$.
Assume also that $|f_{\tau^0_j} - f_{\tau^0_j+1}| > 0$ remains constant with $n$ for each $j = 1, \ldots, {N_0}$. Then
\begin{equation}
\label{ch4:eq:consist}
|\hat{q}_j - q_j^0| = o_p(1), 
\end{equation}
for $1 \le j \le N_{0}$, as $n \to \infty$.
\end{theorem}

We recall that $V_n = o_p(a_n)$ means that for all $\delta > 0$, we have $\lim_{n\to\infty} P(|V_n/a_n| > \delta) = 0$.

In the statement of Theorem \ref{ch4:lemconsist}, the concept of ``rescaled time'' refers to the fact that we are scaling the true change-point locations
$\tau_j^0$ by $n$ to create $q_j^0$, thereby embedding the true domain $\{1, 2, \ldots, n\}$ in $(0, 1]$, and examining the convergence of the change-point
location estimators $\hat{q}_j$ in the new domain $(0, 1]$. This is because apart from trivial situations (such as the variance $\sigma^2$ of the innovations converging to zero
with $n$), it is impossible for $\hat{\tau}_j$ to converge to $\tau_j^0$: we saw in Section 1.2 of \cite{fearnhead2022detecting} that the error of $\hat{\tau}_j$ as a location
estimator is at least $O_p(1)$.

Theorem \ref{ch4:lemconsist} is a bare-bones consistency result for the change-point
location estimators $\hat{q}_j$. Indeed, (\ref{ch4:eq:consist}) can be viewed as a basic sanity check and if it were not to hold, the estimators $\hat{q}_j$
could not be regarded as useful. The key ideas of the proofs of Theorem \ref{ch4:lemconsist} and the following Theorem \ref{ch4:lemrate} are in Section \ref{ch4:sec:n0known:th}, intended for the more theoretically-minded reader.

We now improve on the result of Theorem \ref{ch4:lemconsist} with the following result, also from \cite{ya89}.

\begin{theorem}
\label{ch4:lemrate}
Let the i.i.d. noise $\{\varepsilon_i\}_{i=1}^n$ be such that $\mathbb{E}(\varepsilon_i) = 0$ and $\mathbb{E}(\varepsilon_i^6) < \infty$.
Assume $\tau_j^0 = [n q_j^0]$, where $[\cdot]$ denote the nearest integer to its argument, and $0 = q_0^0 < q_1^0 < \ldots < q_{N_0}^0 < q_{N_0+1}^0 =1$
are fixed in the rescaled-time interval $[0,1]$ as $n \to \infty$.
Assume also that $|f_{\tau^0_j} - f_{\tau^0_j+1}| > 0$ remains constant with $n$ for each $j = 1, \ldots, {N_0}$. Then
\begin{equation}
\label{ch4:eq:consist2}
n|\hat{q}_j - q_j^0| = O_p(1), 
\end{equation}
for $1 \le j \le N_{0}$, as $n \to \infty$.
\end{theorem}

We remark that the result of Theorem \ref{ch4:lemrate} is optimal in the sense that the rate
$n|\hat{q}_j - q_j^0| = O_p(1)$ could not be improved even if the noise
$\{\varepsilon_i\}_{i=1}^n$ were i.i.d. Gaussian, and even if we were in the AMOC setting (as we saw
in Section 1.2 of \cite{fearnhead2022detecting}). Therefore, we can see that Gaussianity is not needed for optimality in this sense and
that the finiteness of the sixth moment suffices.

To better appreciate the result of Theorem \ref{ch4:lemrate}, we consider an example illustrating how the error in estimating the change-point locations changes as we vary $n$.

\begin{figure}[t]
\centering
\begin{minipage}{.5\textwidth}
  \centering
  \includegraphics[width=\linewidth]{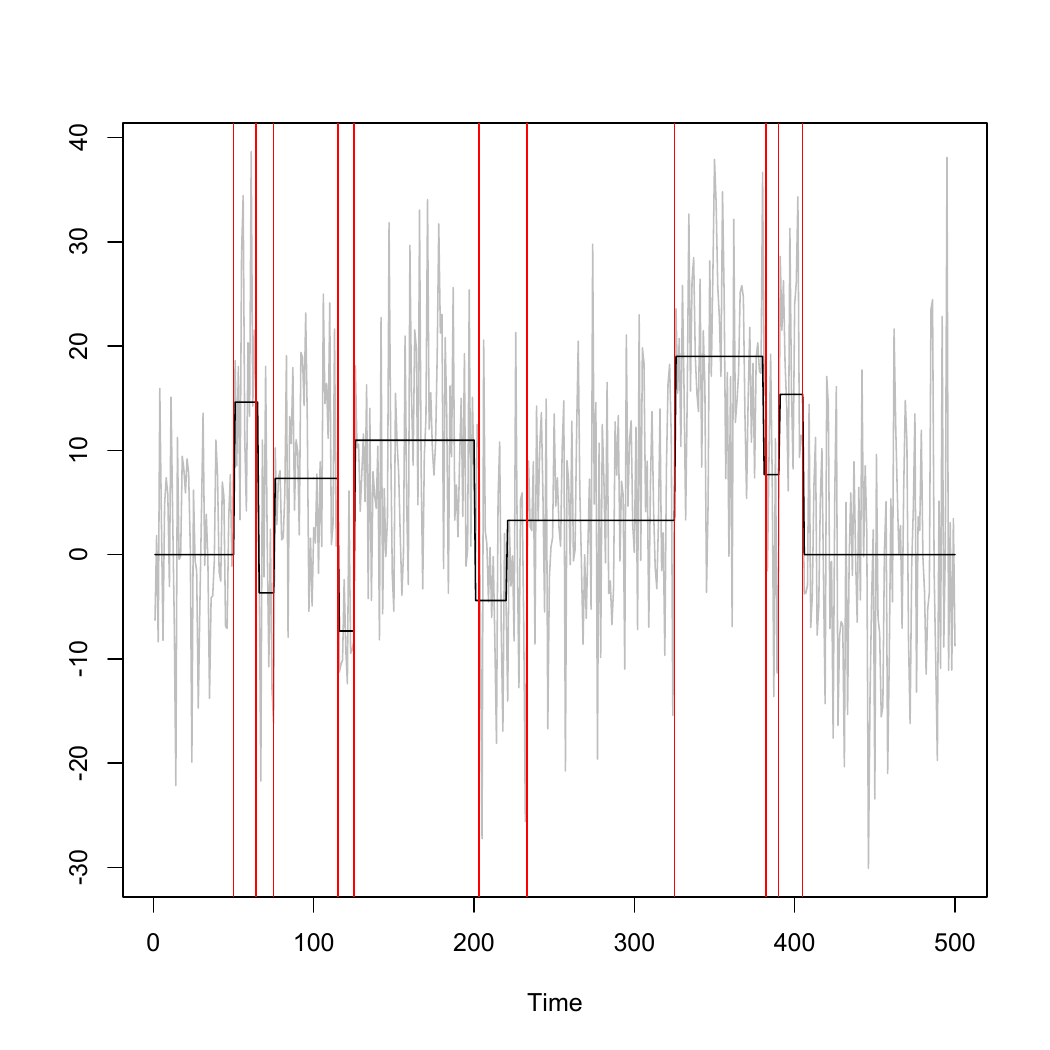}
\end{minipage}%
\begin{minipage}{.5\textwidth}
  \centering
  \includegraphics[width=\linewidth]{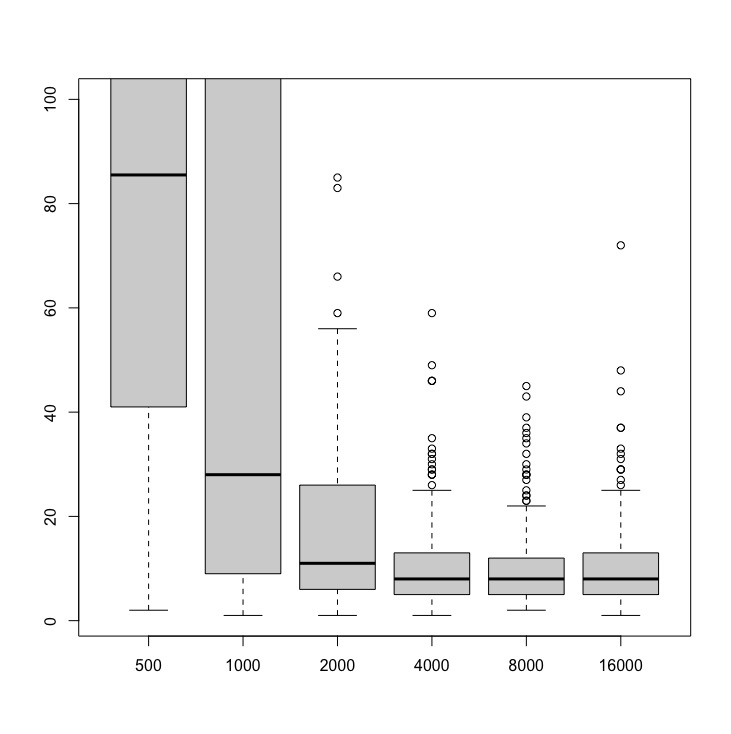}
\end{minipage}\\
\caption{Left: The {\tt blocks} signal for $n=500$ (black), its noisy realisation (grey) and the 11 best change-point estimates via the least-squares criterion (red). Right: box plots of the distribution of $\max_{j=1, \ldots, 11} |\tau_j^0 - \hat{\tau}_j|$ over 250 simulated sample paths, for lengths $n = 500, 1000, 2000, 4000, 8000, 16000$. \label{ch4:fig:blobox}}
\end{figure}

Figure \ref{ch4:fig:blobox} shows the \verb+blocks+ signal from \cite{dj94}, sampled at $n = 500$ points. The signal has 11 change-points. It is contaminated with i.i.d. Gaussian noise with mean zero and variance $\sigma^2 = 100$. We up-sample the signal to lengths $n = 500, 1000, 2000, 4000, 8000, 16000$ and for each length, we estimate the $N_0 = 11$ change-point locations via the least-squares fit in formula (\ref{ch4:eq:taun0known}). We repeat over 250 realisations of the Gaussian noise. As is evident from the right-hand box plots in Figure \ref{ch4:fig:blobox}, the empirical distribution of the quantity
$\max_{j=1, \ldots, 11} |\tau_j^0 - \hat{\tau}_j|$ appears to settle on a limiting distribution. This is consistent with the message of Theorem \ref{ch4:lemrate}, which states that this distance should behave like $n^{-1}$ in rescaled time, and therefore be of a constant order (with respect to $n$) on the original time scale, as $n$ gets larger. The exact limiting distribution of the maximum absolute distance between the estimated change-points and the truth
is a complicated random function of the jump sizes and jump signs, and 
can be obtained from Theorem 1 in \cite{ya89}. We omit this result here. An important observation, though, is that increasing the number of data points between successive change-points increases the power of detecting that there is a change, but does not impact the accuracy of the estimate of the location. The latter is primarily determined by the pattern of the data close to the change-point.

In the above example, we computed the argument minimum in (\ref{ch4:eq:taun0known}) via the Segment Neighbourhood approach \citep{al89}. In R, for a data vector stored in \verb+x+, this can be carried out via the calls
\begin{verbatim}
library(changepoint)
cpts(cpt.mean(x, "None", method = "SegNeigh", Q = 12))
\end{verbatim}
In the above, \verb+Q+ should be set to the number of segments, $N_0+1$. Computation for change-point estimators defined as optima is covered in more detail in Section \ref{ch4:sec:comp}.

\subsection{(*) Estimating Change-point Locations When \texorpdfstring{$N_0$}{the Number of Change-points} Is Known -- Theoretical Details}
\label{ch4:sec:n0known:th}

This section collects some key fact and arguments used in the proofs of Theorems \ref{ch4:lemconsist} and \ref{ch4:lemrate}.

\vspace{10pt}

\noindent {\bf Key ideas of the proof of Theorem \ref{ch4:lemconsist}.} 
The main idea of the proof is simple, and the proof is, in essence, by contradiction. It is possible to get lower bounds on $S_n(l_1, \ldots, l_{N_0})$  in terms of other values of $S_n(\cdots)$ by allowing for more than $N_0$ change-points: as adding change-points to the set $(l_1, \ldots, l_{N_0})$ can only reduce $S_n(\cdots)$.
If (\ref{ch4:eq:consist}) did not hold, then
the value of $S_n(l_1, \ldots, l_{N_0})$ at the estimated change-points
could be lower-bounded
by such an other value of $S_n(\cdots)$, which included as one of its summands the sum of 
squares
\[
\sum_{i=r_s}^{r_e} (X_i - \bar{X}_{r_s:r_e})^2,
\]
where $r_s = [n(q_j^0-\nu)]+1$ and $r_e = [n(q_j^0+\nu)]$, for some $j = 1, \ldots, N_0$ and a (small) fixed constant $\nu \in (0, 1)$.
The fact that the interval $[r_s, r_e]$ included the change-point $q_j^0$ would cause this sum of squares to be ``large'': of order
$\asymp n$. This in turn would cause this lower bound for $S_n(l_1, \ldots, l_{N_0})$ to be substantially larger than
$S_n(\tau^0_1, \ldots, \tau^0_{N_0})$. This demonstrates the impossibility of $S_n(l_1, \ldots, l_{N_0})$ being minimised
at points not satisfying (\ref{ch4:eq:consist}), and therefore leads us to deduce that (\ref{ch4:eq:consist}) must hold.

There is one more aspect of this result that is worth discussing here. Namely, in the reasoning outlined above, 
how can we be sure that the behaviour of the noise $\{\varepsilon_i\}_{i=1}^n$ will not be such that the 
lower bound for $S_n(l_1, \ldots, l_{N_0})$ considered above is spuriously smaller than 
$S_n(\tau^0_1, \ldots, \tau^0_{N_0})$, or at least approaches it, which would cause this reasoning to break down? The answer
lies in the following lemma shown in \cite{ya89}, which is a weaker version of the sharp Gaussian Lemma \ref{ch4:lem:ub} but applying to a wider class of distributions: centred with
a finite $2m$th moment.

\begin{lemma}
\label{ch4:lem:dev2}
Suppose $\epsilon_1, \ldots, \epsilon_n$ are i.i.d. with common mean $0$ and such that $\mathbb{E}(\epsilon_i^{2m}) < \infty$
for some positive integer $m$. Then as $n \to \infty$,
\[
\max_{0\le i < j \le n}  (\epsilon_{i+1} + \ldots + \epsilon_j)^2 / (j-i) = O_p(n^{2/m}).
\]
\end{lemma}

Lemma \ref{ch4:lem:dev2} ensures that with a high probability, in the {\em a contrario} sketch proof outlined above,
the lower bound on $S_n(l_1, \ldots, l_{N_0})$ does not fluctuate ``too much'' and therefore still
remains significantly above $S_n(\tau^0_1, \ldots, \tau^0_{N_0})$.
Lemma \ref{ch4:lem:dev2} is fairly crude and the main tool in its proof is Markov's inequality. What may come as a surprise
is that even the crude Lemma \ref{ch4:lem:dev2}
is {\em sufficient to ensure consistency of the change-point location estimators} (not only for noise with more general distributions than the
Gaussian, but also, as we will see next, with optimal rates).

Looking ahead, the sharp result in Lemma \ref{ch4:lem:ub}, whose proof is fairly complicated, will later improve the bound from Lemma
\ref{ch4:lem:dev2} for the Gaussian distribution. We will see that we will need the sharpness of Lemma \ref{ch4:lem:ub} to correctly estimate the number $N_0$ of
change-points via Schwarz Information Criterion (SIC). The sharpness of Lemma \ref{ch4:lem:ub} needed to estimate $N_0$ well for Gaussian
noise, contrasted with the relative looseness of Lemma \ref{ch4:lem:dev2} sufficient to estimate the change-point locations optimally, and for more general
noise distributions, may signal that estimating the number of change-points is a more difficult problem than estimating their
locations.

\vspace{10pt}

\noindent {\bf Key ideas of the proof of Theorem \ref{ch4:lemrate}.} 
For the result to hold, it is enough to show, for any fixed $\delta > 0$, that there exists $M > 0$ such that $P(n|\hat{q}_j - q_j^0| > M) < \delta$ for $n$ large enough.
There are in essence two sets of results that are used in the proof
of Theorem \ref{ch4:lemrate}:
\begin{enumerate}
\item
The result of Theorem \ref{ch4:lemconsist}.
\item
What \cite{ya89} refer to as the strong law of large numbers,
which ensures that there exist $C > 0$ and integer $M > 0$ such that the event $E_j(n, C, M)$:
\begin{eqnarray}
|\bar{X}_{[n q^0_s]:([n q^0_s]+i)}| & < & C\quad\mbox{for}\quad i = 1, \ldots, [n q^0_{s+1}] - [n q^0_s],\quad s = 0, \ldots, N_0,\nonumber\\
|\bar{X}_{([n q^0_{s+1}]-i):[n q^0_{s+1}]}| & < & C\quad\mbox{for}\quad i = 1, \ldots, [n q^0_{s+1}] - [n q^0_s],\quad s = 0, \ldots, N_0,\nonumber\\
|\bar{X}_{([n q^0_{j}]-i):[n q^0_{j}]} - f_{\tau^0_j}| & < & \delta_1 \quad\mbox{for}\quad i = M, \ldots, [n q^0_{j}] - [n q^0_{j-1}],\nonumber\\
\label{ch4:eq:Ej}
|\bar{X}_{[n q^0_{j}]:([n q^0_{j}]+i)} - f_{\tau^0_{j+1}}| & < & \delta_1 \quad\mbox{for}\quad i = M, \ldots, [n q^0_{j+1}] - [n q^0_{j}],
\end{eqnarray}
where $\delta_1$ is a particular small (fixed, signal-dependent) constant, has a probability larger than $1 - \delta/4$ for $n$ large enough. The conditions (\ref{ch4:eq:Ej}) collectively ensure that the input data, for $n$ sufficiently large, is ``well-behaved'' with an arbitrarily high probability, in a certain uniform sense. More specifically, the first two conditions in (\ref{ch4:eq:Ej}) ensure that the  averages of the data to the left and to the right of each change-point are uniformly bounded by a constant, while the last two conditions ask that the averages of the data to the left and to the right of the $j$th change-point (the change-point singled out in the statement of the theorem) are within a certain specific signal-dependent distance of the corresponding population means, provided that the spans of these averages are at least $M$; the constant $M$ can be arbitrary here, as long as it exists and is fixed (i.e. independent of $n$). Similar conditions ensuring that the input data is well-behaved are commonplace in consistency proofs in the change-point literature. This should not surprise: for example, if the data were not to be well-behaved in the sense of having a cluster of, say, particularly high values, any change-point procedure might be inclined to spuriously detect extra change-points delineating the cluster. Of course, clusters of particularly high values can occur even under Gaussianity, but the probability of such an extreme event is low -- hence the requirement that (\ref{ch4:eq:Ej}) should hold with a probability that is ``high" but less than one.
It would be unrealistic to request that (\ref{ch4:eq:Ej}), or a similar set of conditions, should hold with probability one.
\end{enumerate}

It is insightful to examine, at this point, what exactly is meant by an increasing sample size $n$ in the results of this section.
From the ``rescaled time'' discussion in Section \ref{ch4:sec:n0known}, we can unambiguously deduce that we operate in a stochastic regime in which $\asymp n$
data points are collected between each pair of consecutive change-points. However, one theoretically important question to ask
is whether, as $n$ increases, we ``throw away'' all existing data points between each pair of consecutive change-points and
replace them by new, independent data with length increased by one, or whether we simply add one more new data point to the
existing sequence of observations between each pair of consecutive change-points. Although this is not an issue that is discussed in full in \cite{ya89}, the steps taken in the proof suggests a setting in which as $n$ increases, a new data point gets added to the existing data between each pair of consecutive
change-points.



The main logic of the proof is that from the constraints that describe the set $E_j(n, C, M)$ (formula (\ref{ch4:eq:Ej})), it is possible to show the following.
Let
\[
B(n,\delta_2) = \{ (l_1/n, \cdots, l_{N_0}/n)\quad :\quad 0 < l_1 < \ldots < l_{N_0} < n,\,\, |l_j/n - q_j^0|  < \delta_2   \},
\]
for a suitably chosen, sufficiently small, $\delta_2$,  and
\[
B_j(n, \delta_2, M) = \{  (l_1/n, \cdots, l_{N_0}/n) \in B(n,\delta_2)\quad : \quad l_j - n q_j^0 < -M\}.
\]
By Theorem \ref{ch4:lemconsist}, $(\hat{q}_1, \ldots, \hat{q}_{N_0}) \in B(n,\delta_2)$
with probability larger than $1 - \delta/4$ for large $n$. It is claimed that 
$(\hat{q}_1, \ldots, \hat{q}_{N_0}) \in B_j(n, \delta_2, M)$
with probability less than $\delta/4$ for large $n$. Assuming that this claim holds, we have, for large $n$,
\begin{eqnarray*}
P(n(\hat{q}_j - q^0_j) < -M) & \le & P((\hat{q}_1, \ldots, \hat{q}_{N_0}) \not\in B(n,\delta_2)) \\
& + & P((\hat{q}_1, \ldots, \hat{q}_{N_0}) \in B_j(n, \delta_2, M))\\
& < & \delta/4 + \delta/4 = \delta/2.
\end{eqnarray*}
Similarly, it can be shown that $P(n(\hat{q}_j - q^0_j) > M) < \delta/2$ for large $n$, and so
$P(n|\hat{q}_j - q^0_j| > M) < \delta$ for large $n$. To prove the claim, for every 
$(l_1/n, \cdots, l_{N_0}/n) \in B_j(n, \delta_2, M)$, let 
$(l_1'/n, \cdots, l_{N_0}'/n) \in B(n, \delta_2)$ be such that $l_s' = l_s$ for $s \neq j$ and $l_j' = [n q_j^0]$.

After some fairly elementary algebra (we omit the details here), the constraints in the set $E_j(n, C, M)$ permit us to establish that
\begin{eqnarray*}
\lefteqn{\min_{(l_1/n, \cdots, l_{N_0}/n) \in B_j(n, \delta_2, M)} \{   S_n(l_1, \ldots, l_{N_0}) - S_n(l_1', \ldots, l_{N_0}')  \}}\\
& \ge & \frac{1}{16} |f_{\tau^0_j} - f_{\tau^0_j+1}|^2 \min_{(l_1/n, \cdots, l_{N_0}/n) \in B_j(n, \delta_2, M)} ([n q^0_j] - l_j)\\
& \ge & \frac{1}{16} |f_{\tau^0_j} - f_{\tau^0_j+1}|^2 M > 0,
\end{eqnarray*}
which implies that $(\hat{q}_1, \ldots, \hat{q}_{N_0}) \not\in B_j(n, \delta_2, M)$. Therefore, for large $n$,
\[
P((\hat{q}_1, \ldots, \hat{q}_{N_0}) \in B_j(n, \delta_2, M)) \le 1 - P(E_j(n, C, M)) < \delta/4.
\]
This proves the claim and thus the Theorem.

\subsection{When \texorpdfstring{$N_0$}{the Number of Change-points} Is Unknown: General Principles of Penalised Estimation}
\label{ch4:sec:n0unknown}

In many cases of practical interest, the number $N_0$ of change-points will not be known to the analyst and will need to be estimated from
the data along with the change-point locations. In this section, we discuss general principles behind constructing criteria, penalised for model
complexity, whose minimisation will
yield estimators of both $N_0$ and $(\tau^0_1, \ldots, \tau^0_{N_0})$.

We saw in Section \ref{ch4:sec:n0known} that for a known $N_0$, a valid way to obtain estimators of the change-point locations 
$(\tau^0_1, \ldots, \tau^0_{N_0})$ was to minimise the least-squares criterion $S_n(l_1, \ldots, l_{N_0})$. Some of the criteria
for unknown $N_0$ outlined in the next sections also use $S_n(\cdot)$ as a building block, while some others use the likelihood of the unknown
true model parameters $\theta_0 := (N_0, (\tau^0_1, \ldots, \tau^0_{N_0}), \sigma^2, (f_{\tau^0_1}, \ldots, f_{\tau^0_{N_0}}, f_n))$. Let
a candidate model parameter be denoted by $\theta = (N', (l_1, \ldots, l_{N'}), \tilde{\sigma}^2, (\tilde{f}_{l_1}, \ldots, \tilde{f}_{l_{N'}}, \tilde{f}_n))$.
The likelihood of $\theta$ given the data in model (\ref{ch4:eq:univ_mult}) is
\[
L(\theta | X_1, \ldots, X_n) = (2\pi \tilde{\sigma}^2)^{-n/2} \exp\left\{  -\frac{\sum_{j=1}^{N'+1} \sum_{i=l_{j-1}+1}^{l_j} (X_i - \tilde{f}_{l_j})^2 }{2 \tilde{\sigma}^2}  \right\}.
\]
Taking the logarithm, multiplying by $-1$ and removing any additive terms that do not depend on $\theta$, we obtain the negative log-likelihood $-l(\cdot)$,
defined by
\begin{equation}
\label{ch4:mlik}
-l(\theta | X_1, \ldots, X_n) = \frac{n}{2} \log\,\tilde{\sigma}^2 + \frac{\sum_{j=1}^{N'+1} \sum_{i=l_{j-1}+1}^{l_j} (X_i - \tilde{f}_{l_j})^2 }{2 \tilde{\sigma}^2}.
\end{equation}

Both $S_n(\cdot)$ and $-l(\cdot)$ are measures of fit to the data of a particular candidate model. But while their minimisation can be used
to estimate the change-point locations for a fixed $N'$, an estimate of $N_0$ cannot be obtained by minimising these criteria alone. This is because,
heuristically speaking, the fit of a model to the data improves as its putative number of change-point increases. For example, for the least-squares measure
$S_n(\cdot)$, we generally have (except some trivial cases in which we have equality)
\[
\min_{l_1 < \ldots < l_{N'}} S_n(l_1, \cdots, l_{N'}) > \min_{l_1 < \ldots < l_{N''}} S_n(l_1, \cdots, l_{N''})
\]
for $N' < N''$, as on the right-hand side we are taking the minimum over a larger model class. Therefore, the absolute minimum of $S_n(\cdot)$ is achieved
for $N' = n-1$ and $l_j = j$, or in other words when ``every point is a change-point'', in which case $S_n(l_1, \ldots, l_{N'}) = 0$. Clearly, this is not a sensible
statistical model, as it in no way generates additional insight. A very similar argument can be made for the negative log-likelihood, $-l(\cdot)$.

For this reason, for the analyst wishing to estimate
both $N_0$ and $(\tau^0_1, \ldots, \tau^0_{N_0})$ via minimising a model cost function, a natural and valid approach
is to consider model cost functions of the form
\begin{equation}
\label{ch4:gencrit}
\mbox{fit to data} + \mbox{penalty},
\end{equation}
where the ``fit to data'' part measures the quality of fit of a model candidate (the reader is invited to think of $S_n(\cdot)$ or $-l(\cdot)$ here) and
the ``penalty'' term penalises more complex models, e.g. ones containing more change-points, or containing change-points in particular configurations
which are deemed unlikely (e.g. being located close to one another). In this approach, the preferred model is one that minimises a given criterion of the form (\ref{ch4:gencrit}).

Many readers will be familiar with the penalised approach to model selection 
in other statistical contexts (e.g. model selection in linear regression). In change-point problems, the
most frequently used penalties are proportional to the number of estimated change-points, that is they are of the form $N'\rho(n)$, where $\rho(\cdot)$ is a particular function of the sample size (and possibly $\sigma^2$, if known).
However, more complex penalties exist and we review a selection of the most popular ones later in this chapter.

All methods of this form reviewed in this book will use either $S_n(\cdot)$ or $-l(\cdot)$ as the ``fit to data'' term. We now briefly discuss 
their particular features as well as the similarities and differences between them. If a particular component of
the parameter vector $\theta$ does not appear in the penalty, it can be replaced in $-l(\cdot)$ by its unconstrained maximum likelihood estimator
conditioned on those elements of the parameter vector $\theta$ that do appear in the penalty.
For example, in Schwarz Information Criterion reviewed in detail in Section \ref{ch4:sec:sic} below, the penalty is of the form $N' \log(n)$,
and we therefore make the following replacements in the criterion $-l(\cdot) + N' \log(n)$:
\begin{eqnarray*}
(l_1, \ldots, l_{N'}) & \rightsquigarrow & \arg\min_{l_1 < \ldots < l_{N'}} S_n(l_1, \ldots, l_{N'}),\\
\tilde{\sigma}^2 & \rightsquigarrow & \hat{\sigma}_{N'}^2,\\
\tilde{f}_{l_j} & \rightsquigarrow & \bar{X}_{(l_{j-1}+1):l_j},\\
\end{eqnarray*}
where the quantities on the right-hand sides of these expressions are the MLEs of the quantities on the left under the
assumption that the number of change-points equals $N'$; in particular,
\[
\hat{\sigma}_{N'}^2 = n^{-1} \min_{l_1 < \ldots < l_{N'}} S_n(l_1, \ldots, l_{N'}).
\]
This replacement, and the removal of an additive constant which makes no difference to the minimisation,
reduce the $-l(\cdot)$ function in this particular case to $\frac{n}{2} \log(\hat{\sigma}_{N'}^2)$. This causes Schwarz Information
Criterion to have the form $\frac{n}{2} \log(\hat{\sigma}_{N'}^2) + N' \log(n)$. There is more material on under what conditions Schwarz Information
Criterion leads to consistent model selection in Section \ref{ch4:sec:sic}.

On the face of it, Schwarz Information Criterion is a criterion for the estimation of $N_0$ only, rather than the change-point locations, as these
do not explicitly feature in the expression $\frac{n}{2} \log(\hat{\sigma}_{N'}^2) + N' \log(n)$. However, to obtain this expression, we implicitly
had to estimate the change-point locations for each $N'$ separately by $\arg\min_{l_1 < \ldots < l_{N'}} S_n(l_1, \ldots, l_{N'})$. This specific observation
leads us to make the following more general statement.

\begin{note2}
\label{ch4:notesep}
In penalised problems of the form $-l(\cdot) + \mbox{penalty}$ (or $S_n(\cdot) + \mbox{penalty}$, respectively), if the penalty depends on the number of change-points,
but not on their locations, joint estimation of the number and locations of change-points can be achieved by following the workflow below:
\begin{enumerate}
\item
estimate the change-point locations for each postulated number $N'$ of change-points separately via conditional MLE, by minimising 
$-l(\cdot)$, without a penalty, under the assumption that the number of change-points equals $N'$ (which is equivalent to the minimisation of $S_n(\cdot)$ in the signal + i.i.d. Gaussian noise setting, under the assumption that the number of change-points equals $N'$);
\item
estimate the number $N_0$ of change-points by minimising $-l(\cdot) + \mbox{penalty}$ (or $S_n(\cdot) + penalty$, respectively);
in these expressions, substitute the unknown change-point locations by their MLEs obtained in the previous step.
\end{enumerate}
\end{note2}

This provides a convenient separation of the joint estimation of the number and locations of change-points, for penalties which only depend on the former but not the latter, into unpenalised estimation of the locations for each number first, and then penalised estimation of the number.

The above discussion concerns the case of $\sigma^2 = \mbox{Var}(\varepsilon_i)$ being unknown. It is also useful to study the change-point
problem (\ref{ch4:eq:univ_mult}) in the case of $\sigma^2$ being known, the main reason being that even if it is unknown, it can typically be estimated fairly
accurately, at least when the number $N_0$ of change-points is not very large in relation to the sample size $n$. This can be achieved, for example, by using
the Median Absolute Deviation \citep{h74_4} or Inter-Quartile Range
\citep{rc93_4} estimators suitable for Gaussian data on the input sequence $\{2^{-1/2} (X_{i+1} - X_i) \}_{i=1}^{n-1}$.
Here, the rationale is that most elements of this sequence (more precisely, those for which both $X_i$ and $X_{i+1}$ are taken from the same segment
of constancy) will have mean zero and variance $\sigma^2$, and therefore robust estimators of $\sigma^2$ should be accurate. There is more on the estimation of $\sigma^2$ in Section \ref{ch4:sec:noise_variance}.

Let us now investigate how the minimisation of $-l(\cdot) + \mbox{penalty}$ will simplify in the case of $\sigma^2$ being known. From formula (\ref{ch4:mlik}),
disregarding the (now assumed known) additive term $\frac{n}{2}\log\,\tilde{\sigma}^2 = \frac{n}{2}\log\,\sigma^2$, we can see that the minimisation of 
$-l(\cdot) + \mbox{penalty}$ is equivalent to the minimisation of $S_n(\cdot) + 2 \sigma^2 \mbox{penalty}$. This establishes the equivalence, up to the multiplicative
factor of $2 \sigma^2$, of the likelihood and least-squares approaches in the case of $\sigma^2$ known.

We conclude this section with definitions of a solution path in change-point problems. Note \ref{ch4:notesep} implies
that it is of interest to store sets of change-point estimates $(\hat{\tau}_1^{(N')}, \ldots, \hat{\tau}_{N'}^{(N')})$ for $N' = 1, \ldots, N_U$, where $N_U$
is a certain upper bound. We refer to the list $(\hat{\tau}_1^{(N')}, \ldots, \hat{\tau}_{N'}^{(N')})_{N'=1}^{N_U}$ as the {\em solution path} to a given change-point
problem associated with a particular estimation method. We refer to a solution path as {\em complete} if $N_U$ is the largest possible, i.e. $N_U = n-1$.

We refer to an estimation method as {\em nested} if for each $N' < N''$ we have 
$(\hat{\tau}_1^{(N')}, \ldots, \hat{\tau}_{N'}^{(N')}) \subset (\hat{\tau}_1^{(N'')}, \ldots, \hat{\tau}_{N''}^{(N'')})$.
Estimation methods based on penalised optimisation are typically not nested; heuristically, the reason for this is that estimation (via optimisation) for $N'$ postulated
change-points is conducted separately from estimation for $N''$ postulated change-points, so there is no algorithmic link between the two solution sets and
hence no reason for one to be a subset of the other. However, some other methods introduced later in the chapter will have the nested property. For a
nested estimation method, denote for $N' = 1, \ldots, N_U$,
\[
\hat{\tau}^{(N')}_{j(N')} = (\hat{\tau}_1^{(N')}, \ldots, \hat{\tau}_{N'}^{(N')}) \setminus (\hat{\tau}_1^{(N'-1)}, \ldots, \hat{\tau}_{N'-1}^{(N'-1)}),
\]
the additional estimated change-point when we go from $N'-1$ to $N'$ estimated change-points. The ordered set $\{\hat{\tau}^{(N')}_{j(N')}\}_{N'=0}^{N_U}$ is then referred to as the nested solution path of the nested
estimation method under consideration, and it is said to be complete if $N_U = n-1$.

\subsection{When \texorpdfstring{$N_0$}{the Number of Change-points} Is Unknown: Schwarz Information Criterion}
\label{ch4:sec:sic}

\cite{y88} proposes to estimate the number ${N_0}$ of change-points via Schwarz (a.k.a. the Bayesian) Information Criterion \citep{s78}.
As motivated in Section \ref{ch4:sec:n0unknown},
the
criterion is
\begin{equation}
\label{ch4:sic}
\mbox{SIC}(N') = \frac{n}{2} \log(\hat{\sigma}_{N'}^2) + N' \log(n),
\end{equation}
where $\hat{\sigma}_{N'}^2$ is the Maximum Likelihood Estimator of $\sigma^2$ (see formula (\ref{ch4:sigmahat}) below),
and the estimator is
\[
\hat{N}_{\mathrm{SIC}} = \arg\min_{N'\in\{0, \ldots, N_U\}} \mbox{SIC}(N'),
\]
where $N_U$ is a given fixed upper bound for ${N_0}$.
It is instructive to understand the properties of this estimator and how to demonstrate them.
For each $N' \le N_U$, let $\hat{\tau}_{1, N'}, \ldots,
\hat{\tau}_{N',N'}$ be the values of $\tau_1, \ldots, \tau_{N'}$ that minimise 
$S_n(\tau_1, \ldots, \tau_{N'})$. We have
\begin{equation}
\label{ch4:sigmahat}
\hat{\sigma}_{N'}^2 = \frac{1}{n} S_n(\hat{\tau}_{1, N'}, \ldots, \hat{\tau}_{N',N'}).
\end{equation}
Before proceeding, we remark that if $\hat{N}_{\mathrm{SIC}} = N_0$ (which, under certain assumptions, is the case with probability tending to 1 with $n$, as the rest of this section demonstrates),
the change-point location estimators
$\hat{\tau}_{1, \hat{N}_{\mathrm{SIC}}}, \ldots,
\hat{\tau}_{\hat{N}_{\mathrm{SIC}},\hat{N}_{\mathrm{SIC}}}$ are exactly
the same as the analogous estimators obtained for $N_0$ known, defined in formula (\ref{ch4:eq:taun0known}). This illustrates more concretely
the statement made in Section \ref{ch4:sec:n0unknown} that Schwarz 
Information Criterion is a method for estimating not only $N_0$ but also, implicitly, the change-point locations.
\cite{y88} proves the following result.
\begin{theorem}
\label{ch4:prop:yao1988}
Assume ${N_0} \le N_U$ and $\tau^0_j/n \to q_j \in (0,1)$ for $j = 1, \ldots, {N_0}$ as $n \to \infty$.
Assume also that $|f_{\tau^0_j} - f_{\tau^0_j+1}| > 0$ remains constant with $n$ for each $j = 1, \ldots, {N_0}$. Then
\[
P(\hat{N}_{SIC} = {N_0}) \to 1
\]
as $n \to \infty$.
\end{theorem}
Before outlining the proof of Theorem \ref{ch4:prop:yao1988}, we discuss a few key aspects of its formulation. 
The asymptotic regime assumed in Theorem \ref{ch4:prop:yao1988} is that the change-point locations $\tau^0_j$ are asymptotically
fixed in rescaled time, which means that each sequence $\tau^0_j/n$ has a fixed limit in the interval $(0,1)$.
The significance of this assumption is twofold: firstly, that the spacing between each pair of consecutive true change-points is of asymptotic
order $\asymp n$, which is the largest possible, since the length of the data sample is $n$. Secondly, this trivially implies that the true
number ${N_0}$ of change-points is bounded as a function of $n$.
The assumption of a ``known'' upper bound (here: $N_U$) on the size of the model is common in model selectors that use SIC, as, classically, it is a criterion for model choice from among a finite
set of models \citep{s78}.
Finally, Theorem \ref{ch4:prop:yao1988} requires that the size of each jump remains constant with $n$, rather than being allowed to decrease with
$n$ at a certain speed.

We now digress slightly to discuss the nature and usefulness of asymptotic results in change-point problems, such as that in
Theorem \ref{ch4:prop:yao1988}. Of course, 
asymptotic considerations of this type
have to be taken with a pinch of salt by the applied user.
To start with, in most scenarios involving a posteriori change-point detection, there is only one dataset (of length $n$) available for analysis, so
there is no increasing $n$ regime to speak of in the first place. And even if there were, it would be unrealistic to expect that any change-points remain
fixed in rescaled time -- this would require data sampling between each pair of consecutive change-points each time $n$ increased, and would therefore
preclude typical time series scenarios in which new data were added only at the end of the sample as $n$ increased.

Still, we would argue that asymptotic results of this type can be of some use. Firstly, they offer a basic minimum standard (some readers may like to think of a basic ``sanity check'' here): we would likely be wary of an estimator that was not able to achieve consistency even in this idealised setting, in which results of this type are possible and
achievable. Secondly, consistency results of this type may be useful
for comparing estimators; for example, if estimator A of change-point locations were shown theoretically to be more accurate, or work under less strict assumptions, than estimator B, we might find it easier to
justify our preference for A in practice.

Finally, asymptotic results in change-point problems can provide a practically useful mathematical abstraction of reality. For example, one possible lead
from Theorem \ref{ch4:prop:yao1988} is that SIC may be an accurate model selector in change-point problems in which the data at hand 
exhibit relatively infrequent change-points. Section \ref{ch4:sec:simstu} shows how this observation can manifest itself in practice.



The statement of Theorem \ref{ch4:prop:yao1988}, that $P(\hat{N}_{SIC} = {N_0}) \to 1$ as $n \to \infty$, is typical of results on the accuracy
of model selectors in change-point problems. In fact, as will become apparent from the proof, even more can be said in this and typically in other cases:
the proof will identify a specific subset of $\Omega$, the sample space, on which $\hat{N}_{SIC} = {N_0}$. The subset will be one in which the 
noise $\bm{\varepsilon}$ behaves ``reasonably'' in the sense that a certain suitably rich collection of linear transformations of 
$\bm{\varepsilon}$ uniformly does not exhibit extreme behaviour in a certain precisely defined sense.

\vspace{5pt}

\noindent {\bf Key ideas of the proof of Theorem \ref{ch4:prop:yao1988}.} As with many other proofs of the validity of Schwarz Information
Criterion, the argument here is split into two parts, consisting
in showing why the events (\ref{ch4:hatNless}) and (\ref{ch4:hatNmore}) below are {\em{impossible}}, other than on an event set whose probability tends to zero with $n$:
\begin{eqnarray}
\label{ch4:hatNless}
\hat{N}_{SIC} & < & N_0,\\
\label{ch4:hatNmore}
\hat{N}_{SIC} & > & N_0.
\end{eqnarray}
Before we discuss the above two parts of the proof, we firstly explain how to understand the low-probability event on which 
$\hat{N}_{SIC} \neq {N_0}$. Clearly, if a particular realisation of the noise vector $\bm{\varepsilon}$ is in some sense extreme, 
then we cannot expect most model selection criterions including SIC to identify the correct number of change-points. For example,
imagine that the values of $\varepsilon_1, \ldots, \varepsilon_m$ for a certain $m < n$ randomly happen to be extremely high, much higher than the values
of $\varepsilon_{m+1}, \ldots, \varepsilon_n$. Take $\mathbf{f} = \mathbf{0}_n$. Chances are that many model selection criterions,
SIC included, will mistake this random fluctuation for signal, and will spuriously identify a change-point at location $m$, even though
the true signal $\mathbf{f}$ contains no change-points.

Fortunately, as part of the proof of Theorem \ref{ch4:prop:yao1988}, \cite{y88} shows a key lemma, which is important in two ways:
(a) it shows that the probability of ``extreme'' behaviour of the noise $\bm{\varepsilon}$ (such as that described in the toy example of the previous
paragraph) tends to zero with $n$ uniformly in a certain
strong sense (more on this below), and (b) it is (practically) sufficient for us to be outside this set of extreme noise behaviour to guarantee
$\hat{N}_{SIC} = N_0$ as $n\to\infty$. This key lemma appears as Lemma \ref{ch4:lem:ub} below.

\begin{lemma}
\label{ch4:lem:ub}
Suppose $\epsilon_1, \ldots, \epsilon_n$ are i.i.d. normal with common mean $0$ and variance $\sigma^2$. Then for any $\delta > 0$, as $n \to \infty$,
\[
P\left(\max_{0\le i < j \le n}  (\epsilon_{i+1} + \ldots + \epsilon_j)^2 / (j-i) > 2 (1+\delta) \sigma^2 \log\,n \right) \to 0.
\]
\end{lemma}

This is a remarkable result, and we now explain why. Lemma \ref{ch4:lem:ub} shows that 
outside a set whose probability tends to zero with $n$, all normalised partial sums of an i.i.d. $N(0, \sigma^2)$
sequence $\epsilon_1, \ldots, \epsilon_n$ of the form
\[
z_{i,j} = \frac{\epsilon_{i+1} + \ldots + \epsilon_j}{(j-i)^{1/2}}
\]
are bounded in absolute value by $\sigma \sqrt{2 (1 + \delta) \log\,n}$, for any positive $\delta$ (obviously, the smaller the $\delta$, the higher the probability
that the bound does not hold). Each variable $z_{i,j}$ is again distributed as $N(0, \sigma^2)$, and there are $n^2$ of them. To understand why the
bound of $\sigma \sqrt{2 (1 + \delta) \log\,n}$ may come as a surprise, let us step
back and understand the lowest possible constant $a$ for which
\begin{eqnarray}
\label{ch4:eq:bound}
P(\max_{1 \le i \le n} |\epsilon_i| > \sigma \sqrt{a\,\log\,n}) \to 0.
\end{eqnarray}
Setting $a = 2$ clearly works: using the simple Bonferroni bound and a well-known Mills ratio inequality for the Gaussian distribution, we have
\[
P(\max_{1 \le i \le n} |\epsilon_i| > \sigma \sqrt{2\,\log\,n}) \le \sum_{i=1}^n P( |\epsilon_i| > \sigma \sqrt{2\,\log\,n}) \le \frac{2n}{\sigma \sqrt{2\,\log\,n}} \phi(\sqrt{2\,\log\,n}),
\]
where $\phi()$ is the pdf of the standard normal. This simplifies to $\frac{2}{\sigma \sqrt{2\,\log\,n}}$, which tends to 0 with $n$. Using more advanced
techniques (see e.g. Example 2.29 and Exercise 2.11 in
\cite{w19_4}), it can be shown that no constant $a < 2$
leads to
(\ref{ch4:eq:bound}) being satisfied, and therefore $a = 2$ is the lowest constant
for which (\ref{ch4:eq:bound}) holds.
Now, Lemma \ref{ch4:lem:ub}
tells us that the bound $\sigma \sqrt{2 (1 + \delta) \log\,n}$ for all $n^2$ variables $|z_{i,j}|$
is only very slightly higher than the lowest possible bound
$\sigma \sqrt{2 \log\,n}$ for the $n$ variables $|\epsilon_i|$.
Clearly, if all $z_{i,j}$ were independent, then by the above discussion, the lowest possible bound for
all the $|z_{i,j}|$'s would have to be 
$\sigma \sqrt{2 \log\,(n^2)} = \sigma \sqrt{4 \log\,n} \gg \sigma \sqrt{2 (1 + \delta) \log\,n}$ (bear in mind that any positive $\delta$ is permitted, and we therefore
think of $\delta$ being small). However, the particular dependence structure
in the ensemble of variables $\{z_{i,j}\}_{i,j=1}^n$ causes the bound for this ensemble to be ``only'' 
$\sigma \sqrt{2 (1 + \delta) \log\,n}$.
For completeness, we note that similar bounds for more general sub-Gaussian distributions were independently obtained later by \cite{s95}.

We now outline the two parts of the proof of Theorem \ref{ch4:prop:yao1988}: showing the impossibility of 
(\ref{ch4:hatNless}) and of (\ref{ch4:hatNmore}).

\vspace{10pt}

\noindent {\bf Showing the impossibility of $\hat{N}_{SIC}  <  N_0$.}
If we were to have $\hat{N}_{SIC}  <  N_0$, there would be at least one
true change-point, say $\tau^0_r$ for $1 \le r \le N_0$, with no estimated
change-points in its neighbourhood $[r_s, r_e]$. We could choose
$r_s, r_e$ such that $r_e - r_s \asymp n$ due to the assumption that $\tau^0_j / n \to q_j$.
Consider the quantity
$\hat{\sigma}^2_{\hat{N}_{SIC}}$, defined in (\ref{ch4:sigmahat}) and needed
in the evaluation of $\mbox{SIC}(\hat{N}_{SIC})$, defined in (\ref{ch4:sic}).
The presence of the change-point $\tau^0_r \in [r_s, r_e]$ would cause the contribution of the sum of squares
\[
\sum_{i=r_s}^{r_e} (X_i - \bar{X}_{r_s:r_e})^2
\]
to $\hat{\sigma}^2_{\hat{N}_{SIC}}$, see formulae (\ref{ch4:sn}) and (\ref{ch4:sigmahat}), to be, stochastically, much larger than $\sigma^2 (r_e - r_s + 1)$.
As
$r_e - r_s \asymp n$, this increased
contribution would take place over a relatively long interval, and therefore the remaining
components of the sum of squares (\ref{ch4:sn}) would not be able to compensate for
it and as a result, $\hat{\sigma}^2_{\hat{N}_{SIC}}$ would end up much larger than $\sigma^2$.
This would in turn mean that $\mbox{SIC}(N')$ could not be minimised at $\hat{N}_{SIC}$, which leads to a contradiction.

To formalise this reasoning, \cite{y88} shows two lemmas, which we give below without
proofs.

\begin{lemma}
\label{ch4:lem2}
As $n \to \infty$, we have $0 \le \sum_{i=1}^n  \varepsilon_i^2 - n \hat{\sigma}^2_{N_0} = O_p(\log\, n)$.
\end{lemma}

This lemma says that at the true number of change-points $N_0$, the estimator $\hat{\sigma}^2_{N_0}$ is close
to and bounded from above by $\frac{1}{n} \sum_{i=1}^n \varepsilon_i^2$ in probability.

\begin{lemma}
\label{ch4:lem3}
For every $N < N_0$, there exists $\nu > 0$ such that $P(\hat{\sigma}^2_N > \sigma^2 + \nu) \to 1$ as $n\to\infty$.
\end{lemma}

This lemma says that if $N < N_0$, the estimator $\hat{\sigma}^2_N$ is significantly larger than $\sigma^2$ in probability.
\cite{y88} then argues that by Lemma \ref{ch4:lem2}, $\hat{\sigma}^2_{N_0} \to \sigma^2$ in probability, so by Lemma \ref{ch4:lem3}
and the definition of $\mbox{SIC}(N)$, $P(\hat{N}_{SIC} \ge N_0) \to 1$.

Even though this is not discussed in \cite{y88}, it is worth taking a look at other penalty values that would lead to similar conclusions. Suppose that we extend SIC to the more general form
\begin{equation}
\label{ch4:sicext}
\mbox{IC}_{p(n)}(N') = \frac{n}{2} \log(\hat{\sigma}_{N'}^2) + N' p(n),
\end{equation}
where $p(n)$ is a general function of $n$ (not necessarily $\log\,n$). With $\mbox{IC}_{p(n)}$ in place of $\mbox{SIC}$, 
by the above discussion, for $N < N_0$ (for the $\nu$ from Lemma \ref{ch4:lem3} and with probability approaching 1), we would have
\begin{equation}
\label{ch4:gensic}
\mbox{IC}_{p(n)}(N) - \mbox{IC}_{p(n)}(N_0) \ge \frac{n}{2} \log\left(1 + \frac{\nu}{2\sigma^2} \right) + (N-N_0) p(n),
\end{equation}
which is guaranteed to be positive (uniformly over $N$, since $N_0$ is finite) as long as $p(n) = o(n)$.

\vspace{5pt}

\noindent {\bf Showing the impossibility of $\hat{N}_{SIC}  >  N_0$.} In typical proofs of the consistency
of SIC, showing the impossibility of this side inequality is usually more involved than the other
side. This is fundamentally because it can be difficult to exclude the possibility that $\hat{N}_{SIC}$ is ``close to
but larger than $N_0$''. This could potentially be caused by 
$\hat{\sigma}^2_{N}$ being so much below $\sigma^2$ for $N > N_0$ that
the penalty (which increases in $N$) would not be able to compensate
for this drop. Fortunately, with a high probability, this does not happen: if $N$ is larger than but close to $N_0$, 
$\hat{\sigma}^2_{N}$ still does not underestimate $\sigma^2$ by too much. Heuristically, this is for the following 
reason. 
Suppose we have already estimated the $N_0$ true change-points accurately, but we also add a small
number of spuriously estimated change-points, and denote the overall number by $N$. In these circumstances,
$\hat{\sigma}^2_{N}$ would be slightly smaller than $\hat{\sigma}^2_{N_0}$ (the latter being a good estimator of $\sigma^2$), but not by much,
as the extra spurious estimated change-points (being located, by definition, over a section of the data where
the signal is constant) would not induce much difference in the residual sum of squares, a key ingredient of
$\hat{\sigma}^2_{N}$.

More formally, to quantify the magnitude of $\hat{\sigma}^2_{N}$ for $N > N_0$ with the precision required to show the 
impossibility of $\hat{N}_{SIC}  >  N_0$, \cite{y88} introduces the Lemma below,
whose proof is considerably more involved than the proof of Lemmas \ref{ch4:lem2} and \ref{ch4:lem3}.

\begin{lemma}
\label{ch4:lem5}
For every $N$ ($N_0 < N \le N_U$) and for any $\nu > 0$, with probability approaching 1,
\[
0 \le \sum_{i=1}^n \varepsilon_i^2 - n\hat{\sigma}^2_N \le \{ \nu + (N - N_0 - 1) 2 (1 + \nu)   \}   \sigma^2 \log\,n.
\]
\end{lemma}

This lemma says that for $N > N_0$, as long as $N \le N_U$, a pre-set upper bound, the estimator $\hat{\sigma}^2_N$
is lower bounded by the quantity
\[
\frac{1}{n} \sum_{i=1}^n \varepsilon_i^2 - \{ \nu + (N - N_0 - 1) 2 (1 + \nu)   \} \frac{\sigma^2}{n} \log\, n.
\]
In other words, Lemma \ref{ch4:lem5} ensures that $\hat{\sigma}^2_N$, for $N_0 < N \le N_U$, is not too small, and therefore that there is
no danger of $\mbox{SIC}(N)$ being minimised at $N > N_0$.
More specifically, with Lemma \ref{ch4:lem5} in place, \cite{y88} shows the impossibility of $\hat{N}_{SIC}  >  N_0$ as follows.
Note that $\sum_{i=1}^n \varepsilon_i^2 > n(\sigma^2 - \nu)$ with probability approaching 1 for any $\nu > 0$. By Lemmas \ref{ch4:lem2} and
\ref{ch4:lem5}, for $N_0 < N \le N_U$, for any $\nu > 0$, with probability approaching 1,
\begin{eqnarray*}
2\{  \mbox{SIC}(N) - \mbox{SIC}(N_0)   \} & \ge & n\,\log\,\hat{\sigma}_N^2 - n \log \left\{  \sum_{i=1}^n \varepsilon_i^2 /n    \right\} + 2(N - N_0) \log\,n\\
& = & n \log\left\{  1 - \frac{\sum_{i=1}^n \varepsilon_i^2 - n \hat{\sigma}^2_N}{\sum_{i=1}^n \varepsilon_i^2}     \right\} + 2(N - N_0) \log\,n\\
& \ge & n \log\left\{   1 - \frac{\{ \nu + (N - N_0 - 1) 2 (1 + \nu)   \}   \sigma^2 \log\,n}{n(\sigma^2 - \nu)}      \right\}\\
& + & 2(N - N_0) \log\,n,
\end{eqnarray*}
which, using $\log(1-x) > (1 + \nu)(-x)$ for small $x > 0$, is greater than
\begin{equation}
\label{ch4:lbsic}
-(1+\nu)\frac{\{ \nu + (N - N_0 - 1) 2 (1 + \nu)   \}   \sigma^2 \log\,n}{\sigma^2 - \nu} + 2(N - N_0) \log\,n,
\end{equation}
for large $n$.
Since (\ref{ch4:lbsic}) is positive for sufficiently small $\nu > 0$ (as it is positive for $\nu=0$ and continuous at that point), we have $P(\mbox{SIC}(N) - \mbox{SIC}(N_0) > 0) \to 1$, completing the proof.

Let us again pick up the discussion of the more general penalty $\mbox{IC}_{p(n)}$ defined in (\ref{ch4:gensic}). For what other values of $p(n)$ could we obtain the conclusion that the equivalent of (\ref{ch4:lbsic}) is positive, and therefore that $\hat{N}_{SIC_{p(n)}} > N_0$ is impossible?

Clearly, from (\ref{ch4:lbsic}), $p(n) = \log\,n$ is practically the smallest penalty that works, but it is interesting to observe that anything larger than that would also do the job. Therefore, combining this discussion with the restriction on $p(n)$ obtained through formula (\ref{ch4:gensic}), we can see that $p(n)$ in the region
\begin{equation}
\label{ch4:eq:penrange}
p(n) \in [\log\, n, \bar{p}(n)],
\end{equation}
where $\bar{p}(n)$ is any sequence satisfying $\bar{p}(n) = o(n)$, permits the same proof of consistency for $N_0$ of the $\mbox{IC}_{p(n)}$ criterion as the one we saw in this section for the classical $\mbox{SIC}$ criterion. From this discussion, we can also see that the classical SIC penalty $p(n) = \log n$ is fundamentally the smallest penalty for which this argument for consistency works, and, from this point of view, the classical SIC leads to the smallest permitted amount of penalisation.

It is interesting to observe that the smallest permitted penalty (here: $p(n) = \log n$) is also the most desirable, because it gives us the highest power in detecting change-points. This is despite the fact that often, consistency proofs for the number of change-points can be substantially simplified for penalties higher than the lowest permitted one. For an example of a simpler consistency proof than that of Theorem \ref{ch4:prop:yao1988} -- because it uses a larger penalty -- see \cite{f14a}, Theorem 3.3.

\cite{ya89} show how to adapt the penalty to estimate $N_0$ consistently for noise for which only a finite number of moments are finite.
\cite{l95} generalises SIC to situations in which the minimum spacing between consecutive change-points is smaller than in \cite{y88}.

\subsection{When \texorpdfstring{$N_0$}{the Number of Change-points} Is Unknown: Penalised Approaches Derived from Other Information Criteria}
\label{ch4:sec:other}

Unless specified otherwise, estimation of the relevant quantities throughout this section is performed
via the minimisation of the criteria with respect to their parameters.

\vspace{5pt}

{\em Akaike Information Criterion.}
Besides the Bayesian (Schwarz) Information Criterion, the other arguably the most commonly used model selection
criterion in statistics generally is Akaike Information Criterion (AIC). To estimate the number $N_0$ of change-points via AIC, define
\begin{equation}
\mbox{AIC}(N') = \frac{n}{2}\log(\hat{\sigma}^2_{N'}) + 2N',
\end{equation}
which leads to the AIC estimator
\[
\hat{N}_{AIC} = \arg \min_{N' \in \{0, \ldots, N_U \}} \mbox{AIC}(N').
\]
For the purpose of model selection in change-point problems, AIC appears to have been used less extensively
than SIC in the literature.
A contributing factor to this relative lack of popularity may be the well-documented lack of statistical consistency of the AIC criterion in selecting the true model in other problems, such as linear
model choice in regression. Indeed, condition (\ref{ch4:eq:penrange}) excludes the AIC penalty from the range of penalties that lead to consistency for $N_0$. 
In practice, since AIC 
penalises models with more change-points less heavily than SIC, we can expect it to overestimate the number of changes -- and substantial over-estimation is often seen in practice.
For Gaussian models, AIC is equivalent to Mallows' $C_p$ \citep{bm01, bcfsw14}.

\vspace{5pt}

{\em Modified Bayesian Information Criterion.}
\cite{zs07} propose the so-called modified BIC (mBIC), constructed as follows.
Assume a Bayesian set-up in which we place the uniform prior on the number of change-points, on
their locations, on their sizes, and on the overall mean of the signal $\mathbf{f}$. Assume further that
$\mbox{Var}(\varepsilon_i)$ is known to be 1. Then Theorem 1 in \cite{zs07} shows that the log Bayes factor
of a model with $N'$ change-point versus the no-change-point model is asymptotically equal to the
maximised log-likelihood under a model with $N'$ change-points (that is, the log-likelihood of the most
likely model with $N'$ change-points), minus the following term, which can be interpreted as the mBIC
penalty.
\begin{equation}
\label{ch4:eq:mbic}
\mbox{mBIC}_{\mbox{pen}}(N', q_1, \ldots, q_{N'}) = \frac{1}{2}\left\{  3N' \log\,n + \sum_{j=1}^{N'+1} \log(q_j - q_{j-1})   \right\},
\end{equation}
where $0 = q_0 < q_1 < \ldots < q_{N'} < q_{N'+1} = 1$ are the postulated change-point locations rescaled
to lie in the interval $[0, 1]$, i.e. $q_j = \tau_j / n$. Therefore, unlike the standard BIC (SIC) penalty discussed 
in Section \ref{ch4:sec:sic},
the mBIC penalty penalises not just the number of change-points, but also
their relative locations.

As done in \cite{zs07}, it is interesting to investigate how the second, location-dependent term behaves in two
extreme cases: for rescaled change-point locations placed at equal intervals over $[0,1]$, and clustered together
as closely as possible (e.g. $\tau_1 = nq_1 = 1, \ldots, \tau_{N'} = n q_{N'} = N'$). Simple algebra shows that
in the former case, we have
\begin{equation}
\label{ch4:eq:mbicunif}
\sum_{j=1}^{N'+1} \log(q_j - q_{j-1}) = -(N'+1)\log(N'+1),
\end{equation}
whereas in the latter case, we have
\begin{equation}
\label{ch4:eq:mbicclust}
\sum_{j=1}^{N'+1} \log(q_j - q_{j-1}) \approx -N'\log\,n,
\end{equation}
which leads to, in this special case, $\mbox{mBIC}_{\mbox{pen}}(N', q_1, \ldots, q_{N'}) = \mbox{SIC}(N') = N' \log(n)$.
As, typically, $N'\log\,n > (N'+1)\log(N'+1)$, the mBIC penalty penalises change-points located close to each other less severely
than it penalises uniformly spread ones. This is somehow counter-intuitive. This particular aspect of the penalty is not discussed in \cite{zs07}. However, the authors mention that change-points
located too close to each other are precluded by their assumption regarding the minimum distance between each pair
of consecutive change-points, which is assumed to be $\asymp n$. This leads them to conclude that the case described in
formula (\ref{ch4:eq:mbicunif}) is the prevalent one, and therefore it is the first term in (\ref{ch4:eq:mbic}) that typically
dominates, which makes the mBIC penalty close to $\frac{3}{2}N' \log\,n$. However, they also remark that $\log\, n$ is such a slowly-growing
function that the second term in (\ref{ch4:eq:mbic}) ``can still make a significant contribution to the penalty".
An extension to the case of an unknown $\mbox{Var}(\varepsilon_i)$ is also described.
\cite{dkk16} propose a related, empirical Bayes approach to change-point estimation in a marginal 
likelihood framework, which, disregarding any information supplied by the prior, also leads to a penalty which 
prefers change-point estimates located close to each other.

In practice, \cite{zs07} propose to estimate a change-point solution path via an algorithm related to Circular Binary Segmentation \citep{ovlw04_4}, and only use mBIC to choose from among the candidate models along the thus-constructed solution path. This way of proceeding goes some way towards mitigating
the preference of mBIC for models with change-points located close to each other (among models with the same numbers of change-points): this potential issue will be of no consequence if the solution path includes no such models. Our experience is that, if we directly use the mBIC, that is we use it also to find the best segmentation with a given number of change-points, then it performs less well due to its preference to fit shorter segments to the data. 

\vspace{5pt}

{\em Minimum Description Length.}
\cite{dlry06} propose to use the Minimum Description Length (MDL) principle to find segments of homogeneity within a piecewise-stationary
autoregressive process. Based on this work, we examine how this model selection principle simplifies in our signal + Gaussian i.i.d. noise setting.
Heuristically speaking, the code length of a model is the number of bits used to encode the parameters, plus the number of bits used to
encode the residuals. We refer to \cite{r89, r07} for the basics of information-theoretic encoding of random variables and parameters, which we do not reproduce here. To encode the parameters of the model, we need to include the following components.
\begin{itemize}
\item
{\em The postulated number $N'$ of change-points.} To encode an unbounded integer $N'$, approximately $\log_2 (N'+1)$ bits are required.
We note that \cite{dlry06} propose to use $\log_2 N'$, but this technically cannot be correct if the search for the number of models is to permit
a model with $N' = 0$ (which it does in \cite{dlry06}), as $\log_2 0 = -\infty$, so the procedure would always end up selecting the null model.
Davis (2019, personal communication) clarifies that this term should equal, in our notation, $\log_2 (N'+1)$. We also note that despite being
naturally bounded by $n-1$, the postulated number $N'$ of change-points is treated as
unbounded here as in practice we usually have $N' \ll n$, so typical values for $N'$ are far from this upper bound.
\item
{\em The postulated segment lengths.} Each of the $N' + 1$ segment lengths is naturally bounded by $n$ (and this upper bound 
can easily be approached in practice). Therefore, each of the first $N'$ segments requires $\log_2 n$ bits to be encoded. We do not need to encode
the length of the final segment as the segment lengths add up to $n$. Therefore encoding the segment lengths requires $N' \log_2 n$ bits.
\item
{\em The maximum likelihood estimates of $\sigma^2$ and $f_{\tau^0_j + 1}$.} To encode the maximum likelihood estimates of the signal variance $\sigma^2$ and the
signal levels $f_{\tau^0_j + 1}$ for $j = 0, \ldots, N'$, we use a result by 
\cite{r89}
which states that maximum likelihood
estimates of a real parameter computed from $n'$ observations can be effectively encoded with $\frac{1}{2} \log_2 n'$ bits. The signal level $f_{\tau^0_j + 1}$
is estimated on a data stretch of length $l_{j+1} - l_j$, where $l_j$ are the postulated change-point locations. The signal variance $\sigma^2$ is estimated on
the entire dataset. Therefore to encode these parameters, we require $\frac{1}{2} \sum_{j=0}^{N'} \log_2 (l_{j+1} - l_j) + \frac{1}{2} \log_2 n$ bits.
\item
{\em The residuals from each fitted model.} From Shannon's results in information theory,
\cite{r89} shows
that the code length of the
residuals is given by the negative of the log-likelihood (where the base 2 logarithm is used) of the corresponding fitted model.
\end{itemize}
Putting together the above pieces, ignoring the additive constants in the same way as it is done in Schwarz Information Criterion \citep{y88}, and switching to base
$e$ for the logarithm, which is equivalent to just multiplying the criterion by a constant, we obtain the MDL criterion as
\begin{eqnarray*}
\mbox{MDL}(N', l_1 < \ldots < l_{N'}) & = & \frac{n}{2} \log(\hat{\sigma}^2_{N'}) + \log (N'+1) + N' \log\, n\\
& + & \frac{1}{2} \sum_{j=0}^{N'} \log (l_{j+1} - l_j)\\
& =: & \frac{n}{2} \log(\hat{\sigma}^2_{N'}) + \mbox{MDL}_{\mbox{pen}}(N', l_1 < \ldots < l_{N'}).
\end{eqnarray*}

As with the other penalties described in this section, we note that the MDL criterion can also be used for the selection
of $N'$ only, for any (other) solution path
algorithm that produces change-point location estimators for any given $N'$.

If the MDL criterion is used to select $N'$ as well as the change-point locations, it is interesting to observe that 
from the concavity of the logarithm, as with the mBIC criterion described earlier, the MDL also encourages estimated change-points that
cluster together, rather than being spread out.

Finally, we note that the thus-defined MDL penalty is uniformly larger than the SIC penalty defined earlier. If we compare it to mBIC, using $q_j = l_j / n$, it is easy to see that
\[
\mbox{mBIC}_{\mbox{pen}}(N', q_1, \ldots, q_{N'}) - \mbox{MDL}_{\mbox{pen}}(N', l_1 < \ldots < l_{N'}) = -\log(N'+1) + f(n),
\]
where the $f(n)$ term does not depend on the postulated number of change-points $N'$ -- so from the point of view of minimisation, the $\log(N'+1)$ term is the only term that separates the two penalties.

\vspace{5pt}
{\em Potts functional.}
\cite{bklmw09} consider minimisation of what they refer to as the Potts functional, defined by
\begin{equation}
\label{ch4:eq:potts}
H_\gamma(\mathbf{f}', \mathbf{X}) = \frac{1}{n} \sum_{i=1}^n (X_i - f'_i)^2 + \gamma |J(\mathbf{f}')|,
\end{equation}
where $J(\mathbf{f}') = \{  i\,:\, 1\le i \le n-1,\, f'_{i} \neq f'_{i+1}     \}$ is the set of change-points in the candidate fit $\mathbf{f}'$, and $|\cdot|$ is
the cardinality of its argument (denoted by $N'$ earlier in this chapter). Theoretically, one of the requirements on $\gamma$ is that $\gamma n / \log\,n \to \infty$; however, in practical
terms, the authors recommend that $\gamma$ be set to $2.5\, \hat{\sigma}^2 \log\,n / n$, where $\hat{\sigma}^2$ is a consistent
estimator of $\sigma$. This, however, means that this approach is very closely related to SIC: by the arguments outlined in Section \ref{ch4:sec:n0unknown},
if $\sigma^2$ were assumed known, in the notation of Section \ref{ch4:sec:n0known}, SIC for the estimation of the number and the locations of change-points
would be
\begin{equation}
\label{ch4:sicpotts}
\mathrm{SIC}(N', l_1 < \ldots < l_{N'}) = \frac{1}{n} S_n(l_1, \ldots, l_{N'}) + 2\sigma^2 N' \log\,n / n,
\end{equation}
which agrees with (\ref{ch4:eq:potts}) except the penalty uses the multiplicative constant of 2, rather than $2.5$ (if $\sigma^2$ were unknown, it would
be advisable to use the likelihood form of SIC as in formula (\ref{ch4:sic}), which does not require the knowledge of $\sigma^2$, or 
alternatively use a consistent estimator of $\sigma^2$ in (\ref{ch4:sicpotts}), which would bring (\ref{ch4:eq:potts}) and (\ref{ch4:sicpotts}) in even closer agreement).

We conclude this section with pointers to other material on penalised model selection in this monograph. A future chapter of this book will describe penalised model selection in parametric multiple change-point problems involving settings beyond the Gaussian i.i.d. model.
Section \ref{ch4:sec:l1} later in this chapter describes $L_1$-penalised methods for model selection in multiple change-point problems; the reason for the separate treatment of this material is the distinct computational nature of these methods, which tend to be closely linked to the lasso problem. Finally, Section \ref{ch4:sec:adapt} covers methods in which the penalties themselves are chosen from the data.

\subsection{Computation for Estimators Defined as Optima}
\label{ch4:sec:comp}

\noindent {\bf Segment Neighbourhood} \label{ch4:sec-SegNeigh}

\vspace{5pt}

We now turn to the problem of how we can calculate the estimators we have described. First we return to estimators that assume $N_0$ is known, such as the one based on minimising the residual sum of squares
\begin{equation} \label{eq:ch4-minimise}
(\hat{\tau}_1, \ldots, \hat{\tau}_{N_0}) = \arg\min_{l_1 < \ldots < l_{N_0}} S_n(l_1, \ldots, l_{N_0}),
\end{equation}
where the residual sum of squares function
$S_n(l_1, \ldots, l_m)$ is defined in formula (\ref{ch4:sn}).

A naive approach to calculating $(\hat{\tau}_1, \ldots, \hat{\tau}_{N_0})$ would be to evaluate $S_n(l_1, \ldots, l_{N_0})$ for all possible choices of $l_1<l_2<\cdots<l_{N_0}$. However, the computation involved is $O(n^{N_0})$, which is impracticable for moderate to large values of $N_0$. Fortunately there are dynamic programming algorithms that can solve the minimisation problem (\ref{eq:ch4-minimise}) with a cost that is $O(N_0 n^2)$.

These algorithms can be applied to a broader class of optimisation problems than just minimising the residual sum of squares (\ref{eq:ch4-minimise}), including many change-point estimators based on maximising some form of log-likelihood. To make this generality clear, we will introduce segment cost functions, $C({X}_{1:n},k,l)$, for $l>k$, which can be viewed as some measure of fit, such as based on a negative log-likelihood, for fitting data from time $k+1$ to time $l$ as a single segment. For the residual sum of squares,
\[
C({X}_{1:n},k,l)=\sum_{i=k+1}^l (X_i-\bar{X}_{(k+1):l})^2.
\]
In most cases these cost functions will depend on the data $X_{1:n}$ just through the data within the segment, $X_{(k+1):l}$; though this is not needed for the algorithms we will describe to work. 

We require that the minimisation problem we wish to solve involves minimising a sum of the segment costs. That is, we wish to minimise, for some specified $m$,
\[
Q_n(l_1,\ldots,l_{m})=\sum_{j=1}^{m+1} C({X}_{1:n},l_{j-1},l_j),
\]
where as above we require $0<l_1<l_2,\ldots<l_m<n$ and we set $l_0=0$ and $l_{m+1}=n$. 

It is helpful to consider a wider class of minimisation problems,
\[
Q_{t}(l_1,\ldots,l_{m})=\sum_{j=1}^{m} C({X}_{1:n},l_{j-1},l_j) + C({X}_{1:n},l_{m},t),
\]
where we now consider only segmenting the data up to time $t$ and require $l_m<t$.
 Denote the minimum of this function as 
 \[
Q_{t,m}=\min_{l_1 < \ldots < l_{m}<t} Q_{t}(l_1,\ldots,l_{m}).
\]
If $m\geq t$ then, as there is no possible segmentation of $t$ data points into $m+1$ segments, we set $Q_{t,m}=\infty$. 

Define these constants, $Q_{t,m}$, for all $t=1,\ldots,n$ and $m=1,\ldots,N_0$. The Segment Neighbourhood algorithm \citep{al89} is based on the idea that if we are forced to place the last change-point before $t$ at some location ($k$, say), then there is a simple recursion for the minimum cost of segmenting the data with this additional constraint. To see this let  
\[
Q_{t,m,k}=\min_{l_1 < \ldots < l_{m-1}<k} Q_t(l_1,\ldots,l_{m-1},k).
\]
Then we have for $m\geq 1$
\[
Q_{t,m,k}=Q_{k,m-1}+C(X_{1:n},k,t),
\]
the minimum cost of fitting the data ${X}_{1:k}$ with $m-1$ change-points, plus the segment cost for the data ${X}_{(k+1):t}$. Thus we can calculate $Q_{t,m}$ by considering each possible value for $k$ and taking the minimum of the costs
\begin{equation} \label{eq:ch4-SegNeigh} 
Q_{t,m}=\min_{k} Q_{t,m,k} = \min_{k} \{ Q_{k,m-1}+C(X_{1:n},k,t) \}.
\end{equation}
Finally, it is trivial to calculate $Q_{t,m}$ when $m=0$ as $Q_{t,0}=C(X_{1:n},0,t)$. Thus we can calculate $Q_{t,0}$ for $t=1,\ldots,n$, and then solve (\ref{eq:ch4-SegNeigh}) in turn for increasing $m=1,\ldots,N_0$ and, for each $m$, for $t=m+1,\ldots,n$.

Once we have solved the recursion it is simple to extract the optimal set of change-points. The optimal location of the last change-point for a segmentation with $m$ change-points is just
\[
\hat{\tau}_{m}=\arg \min_k  \{ Q_{k,m-1}+C(X_{1:n},k,n) \}.
\]
Then we can recurse backwards in time, so that if the optimal location of the $j$th change-point is $\hat{\tau}_j$ then, for $j>1$, the optimal location of the $(j-1)$th change-point is
\[
\hat{\tau}_{j-1}=\arg \min_k  \{ Q_{k,j-2}+C(X_{1:n},k,\hat{\tau}_j) \}.
\]

It is straightforward to calculate the computational complexity of this algorithm. One implementation is to calculate and store $C(X_{1:n},k,l)$ for all $0<k<l\leq n$ -- which for most models will have an $O(n^2)$ cost for both space and time.
For each value of $t$ and $m$, solving (\ref{eq:ch4-SegNeigh}) involves $O(t)$ calculations, thus solving the recursion for any $m$ will have an $O(n^2)$ computational cost. As the recursions need to be solved for $m=1,\ldots,N_0$ this leads to an overall computational cost of $O(N_0 n^2)$. Furthermore it is often possible to implement the algorithm with the same time complexity, but without storing all segment costs if needed \cite[see][]{mhrf14}.

One nice feature of the Segment Neighbourhood algorithm is that it not only gives the optimal location of the change-points for a segmentation with $N_0$ change-points, but also for those  with $1, 2, \ldots, N_0-1$. Thus, if the number of change-points is unknown, we can run Segment Neighbourhood to find the best segmentations with $1, 2, \ldots, N_U$ change-points for some suitable chosen upper bound, $N_U$, on the number of change-points we think are present. We can then estimate $N_0$ from this solution path of best segmentations that minimise our cost for differing numbers of change-points, using an appropriate criterion such as one of the information criteria discussed in Section \ref{ch4:sec:n0unknown}, or one of the adaptive procedures that we will introduce in Section \ref{ch4:sec:adapt}. However, if we intend to estimate $N_0$ using an information criterion that is linear in the number of change-points, there is a more efficient dynamic programming approach that can jointly estimate the number and locations of the change-points under such a criterion: Optimal Partitioning.

\vspace{10pt}

\noindent {\bf Optimal Partitioning} \label{ch4:sec-OP}

\vspace{5pt}

As described in Section \ref{ch4:sec:n0unknown}, when we do not know the number of change-points, $N_0$, we can estimate it by minimising an information criterion. For example, a common information criterion for the change-in-mean problem is Schwarz Information Criterion under a model where the variance is assumed known (which is equivalent to using the criterion defined in Equation \ref{ch4:sicpotts}). This is
\begin{equation} \label{eq:ch4-SIC_OP}
\mathrm{SIC}(m;l_1,\ldots,l_m) = \frac{1}{\sigma^2} S_n(l_1,\ldots,l_m) + m (2\log n),
\end{equation}
where $S_n(l_1,\ldots,l_m)$ is the residual sum of squares for the segmentation with change-points at $l_1<\cdots<l_m$. 

Using the notation introduced earlier in this section, $Q_{n,m}=\min S_n(l_1,\ldots,l_m) $, and we can re-write the criterion as estimating the number of change-points by the value that minimises $(1/\sigma^2)Q_{n,m} + m (2 \log n)$. As discussed above, we can calculate this by using the Segment Neighbourhood algorithm to find $Q_{n,m}$ for $m=0,1,\ldots,N_{\max}$ for some suitable choice of $N_{\max}$. However, it turns out that there is a faster algorithm that we can use to directly minimise (\ref{eq:ch4-SIC_OP}) with respect to both the number and location of the change-points. 

This algorithm, called Optimal Partitioning \citep{jsbaaggstt05}, is a dynamic programming algorithm that can be used to minimise functions of the form
\begin{equation} \label{ch4:eq-OP}
Q_{n,\lambda}(m;l_1,\ldots,l_{m})=\sum_{j=1}^{m+1} C({X}_{1:n},l_{j-1},l_j) +m \lambda
\end{equation}
with respect to both the number of change-points, $m$, and their locations, $l_1<\cdots<l_m$. This is often called a penalised cost, and the key is that it involves a term that is the sum of segment costs plus a penalty that is linear in the number of change-points. 

For any $t=1,\ldots,n$ and $\lambda>0$ it is helpful to define 
\[
Q_{t,\lambda}= \min_{m;l_1<\cdots<l_m<t} \left\{ \sum_{j=1}^{m} C({X}_{1:n},l_{j-1},l_j) + C({X}_{1:n},l_m,t) +m \lambda
\right\},
\]
the minimum value of the penalised cost for segmenting the data up to time $t$. This definition is a slight abuse of notation, given our earlier definition of $Q_{n,m}$, but the context should mean that this causes no confusion.

The idea of Optimal Partitioning is that we can derive, and solve, a recursion for $Q_{t,\lambda}$. As with Segment Neighbourhood, we first consider fixing the location of the most recent change-point prior to $t$. If this is fixed at $k$, it is immediate that the minimum penalised cost for segmenting the data to time $t$ given the last change is at $k$ will be $Q_{k,\lambda}+C({X}_{1:n},k,t)+\lambda$. Thus we have
\[
Q_{t,\lambda}=\min \left\{ C({X}_{1:n},0,t), \min_{k=1,\ldots,t-1} \left\{ Q_{k,\lambda}+C({X}_{1:n},k,t) +\lambda \right\}
\right\}.
\]
Here the first term on the right-hand side corresponds to no change-points, and the remaining terms correspond to one or more change-points with the most recent change being at time $k$. If we define $Q_{0,\lambda}=-\lambda$ this can be simplified to
\begin{equation} \label{eq:ch4-OP}
Q_{t,\lambda}= \min_{k=0,\ldots,t-1} \left\{ Q_{k,\lambda}+C({X}_{1:n},k,t) +\lambda \right\}.
\end{equation}

This recursion can be solved for $t=1,\ldots,n$. Once it has been solved it is straightforward to obtain the estimated change-point locations. To describe this it is simplest to denote the change-points in reverse order, so $\hat{\tau}_1>\hat{\tau}_2>\cdots>\hat{\tau}_m$. The location of the last change-point will be
\[
\hat{\tau}_1=\arg\min \left\{ Q_{k,\lambda}+C({X}_{1:n},k,t) +\lambda \right\}.
\]
Then if we have estimated the $j$th to last change-point, $\hat{\tau}_j>0$ we set
\[
\hat{\tau}_{j+1}=\arg\min_{k=0,\ldots,\hat{\tau}_j-1} \left\{ Q_{k,\lambda}+C({X}_{1:n},k,\hat{\tau}_j) +\lambda \right\}.
\]
We repeat this until we obtain $\hat{\tau}_{m+1}=0$, which corresponds to no earlier change-points.

The advantage of Optimal Partitioning over Segment Neighbourhood is that it has a lower computational complexity. 
For most models calculating $C({X}_{1:n},k,t)$ can be done with cost $O(1)$ regardless of $k$ and $t$ (for example after storing appropriate summaries of the data as discussed in Section 1.2 of \cite{fearnhead2022detecting}). Assuming this is the case, then solving the Optimal Partitioning recursion (\ref{eq:ch4-OP}) for one value of $t$ has complexity that is linear in $t$. As we need to solve the recursion for $t=1,\ldots,n$ this leads to an $O(n^2)$ computational complexity.

When minimising the penalised cost, the choice of $\lambda$ can greatly affect the number of estimated change-points and hence how accurate the estimated segmentation is. Whilst there are default choice for $\lambda$, for example based on various information criteria described in Sections \ref{ch4:sec:n0unknown}--\ref{ch4:sec:other}, these can perform poorly in real applications when the strong assumptions that are required by the theory supporting the information criteria does not hold. In such situations it is recommended to compare how the segmentations vary as we vary $\lambda$. This can be done efficiently using the CROPS algorithm of \cite{haynes2017computationally_4}.

\vspace{10pt}

\noindent {\bf Pruning}
\label{ch4:sec-PELT}

\vspace{5pt}

\begin{figure}[t]
\centering
\includegraphics[scale=.65]{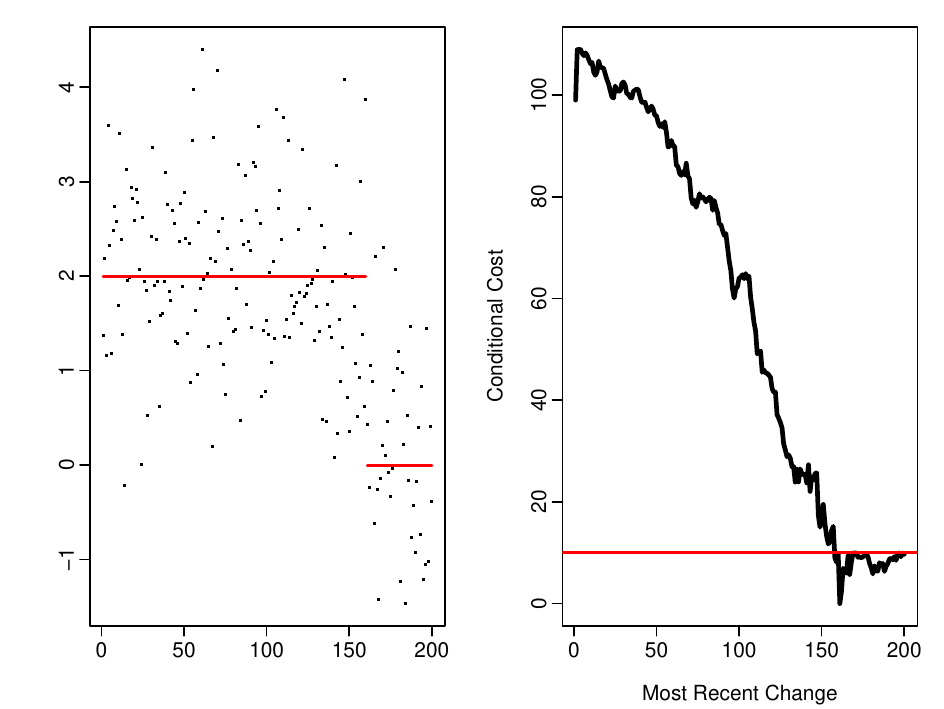}
\caption{ Plot of simulated data and underlying mean (left) and plot of conditional cost (see the text for the definition of this term) in the Optimal Partitioning recursion (right). The conditional cost has been shifted so its minimum is at 0, and the horizontal red line is at $\lambda$ and determines the condition for the PELT pruning.}
\label{Fig:ch4-PRUNE}       
\end{figure}

If we consider the Optimal Partitioning recursion, then at time $t$ we are minimising over terms, $Q_{k,\lambda}+C({X}_{1:n},k,t) +\lambda $, which we will call the conditional cost for the most recent change at $k$.  Figure \ref{Fig:ch4-PRUNE} shows example data and a plot of this conditional cost minus its minimum value. For this example, $t=200$ and the data has had a change at time 160. We notice that for locations of the most recent change-point, $k$, prior to 160 the conditional cost tends to increase as $k$ decreases. This is intuitive as the conditional cost includes a segment from $k$ to time 200, and these will ignore the change in the data. The increasing conditional cost is indicative of the increasing evidence that there has been a change-point since time $k$.

A natural question is whether we could use such information to speed up the Optimal Partitioning algorithm. That is, if the conditional cost at some location $k$ is sufficiently high, does that mean that $k$ can never be the best choice of the most recent change-point for future time-steps, regardless of what the future data is? If this was the case we could ``prune'' $k$ and reduce the set of possible most recent change-points we search over at future steps of the algorithm.

\cite{kfe12} show that in many instances this is the case. In particular if the segment cost is such that adding change-points can only ever reduce the unpenalised cost, then we can prune any $k$ for which the conditional cost is greater than $\lambda$ above the minimum conditional cost. This corresponds to all time-points, $k$, for which the conditional cost is above the red line in the right-hand plot of Figure \ref{Fig:ch4-PRUNE}. 

The intuition behind this pruning rule is that at any future time-point, regardless of the future data, there would be a smaller cost of adding the most recent change at the current time-point than having the most recent change at $k$. To prove this formally, let the current step of the algorithm be $t$, the value that we are considering pruning being $k$, and $T$ being any future time-point. Then the conditional cost at time $T$ with a most recent change-point at $k$ is
\[
Q_{k,\lambda}+C({X}_{1:n},k,T) +\lambda.
\]
Now our assumption on the cost means that $C({X}_{1:n},k,T)\geq C({X}_{1:n},k,t)+C({X}_{1:n},t,T)$ as adding any change-point, in this case at time $t$, can only reduce the unpenalised cost. Thus we have
\[
Q_{k,\lambda}+C({X}_{1:n},k,T) +\lambda \geq Q_{k,\lambda}+ C({X}_{1:n},k,t)+C({X}_{1:n},t,T) +\lambda.
\]
The conditional cost at time $T$ of a most recent change at $t$ is just
\[
Q_{t,\lambda}+C({X}_{1:n},t,T) +\lambda = \min_l\left\{ Q_{l,\lambda}+C({X}_{1:n},l,t)\right\} + \lambda + C({X}_{1:n},t,T) + \lambda,
\]
where we have just plugged in the expresssion for $Q_{t,\lambda}$. This will be smaller than the conditional cost at time $T$ of a change at $k$ if and only if
\[
 \min_l\left\{ Q_{l,\lambda}+C({X}_{1:n},l,t)\right\} + \lambda \leq Q_{k,\lambda}+ C({X}_{1:n},k,t).
\]
The left-hand side is the minimum conditional cost at time $t$ plus $\lambda$. Thus if the conditional cost at time $t$ of a most recent change at $k$ is above this value, $k$ can never be the best choice of most recent change-point at future time points, hence it can be pruned.

\cite{kfe12} introduce an algorithm, PELT, that implements this pruning rule. They show that it can give substantial improvements in computational speed over Optimal Partitioning, particular if the data has segments that are short relative to the length of data. In fact they show that if we consider a data generating mechanism where, on average, the segment lengths are constant as we increase $n$ (and hence the number of change-points increases linearly with $n$) then PELT can have an expected computational complexity that is linear in $n$. 

The intuition behind this result comes from Figure \ref{Fig:ch4-PRUNE}, as we consider most recent changes that are older than the actual most recent change-point, the conditional cost tends to increase linearly. Thus the PELT pruning step will
often only keep candidate values for the most recent change-point that are close to or more recent than the actual most recent change-point. Thus, on average the computational cost of one step of the algorithm will be of the order of the length of a segment rather than proportional to $t$.

\section{Binary Segmentation and Other Hierarchical Methods}
\label{ch4:sec:bs}

In the preceding sections, we described an approach to change-point detection in which the estimator was formulated as the theoretical minimum of a criterion. An algorithm was then needed to compute the minimum. By contrast, the estimators described in this section will already be defined as algorithms to begin with.

This section focuses on {\em hierarchical} methods for change-point detection, i.e. methods for which the
corresponding algorithms either:
\begin{itemize}
    \item first look at the data as a whole to identify change-point candidate(s) considered to be the most prominent, and then gradually narrow their focus to smaller and smaller sections of the data to propose less prominent change-point candidates -- we will refer to these types of techniques as {\em top-down}; or
    \item start at the finest level of resolution of the data, where they remove from consideration locations considered to be the least likely change-points, and then gradually broaden their focus to bigger and bigger sections of the data to propose ever more prominent change-point candidates -- these types of techniques will be referred to as {\em bottom-up}.
\end{itemize}

In either case, hierarchical methods gradually add ever less likely change-point candidates (top-down methods), or remove ever more likely ones (bottom-up methods), and therefore naturally arrange the change-point candidates in their implied order of importance. This means that they tend to yield nested solution paths to the multiple change-point problem, in the sense defined in Section \ref{ch4:sec:n0unknown}.

Model selection, i.e. deciding which set of change-point candidates along the solution path should be preferred, will also be discussed for each method separately.

\subsection{Binary Segmentation}
\label{ch4:multi:bs}

Arguably the simplest hierarchical, top-down method for multiple change-point detection is binary segmentation.
Before we describe it in simple algorithmic terms, one necessary building block that needs introducing is a more general definition of the CUSUM statistics compared to that introduced in \cite{fearnhead2022detecting}. The reason for the generalisation is that the CUSUM statistic $C_t$ (and the signed CUSUM $C_t^*$) operated on the entire data sequence, i.e. on the time domain $[1, n]$. In this section, we will need CUSUMs that can be applied on arbitrary intervals $[s,e] \subseteq [1,n]$
(we will use the term interval to denote an interval subset of the integers $[1,n]$).

For any input sequence $\mathbf{X} = (X_1, \ldots, X_n)'$ and $e>b\geq s$ with $m = |[s,e]| = e-s+1$, we define the signed CUSUM statistic $C^*_{s,b,e}(\mathbf{X})$,
and the CUSUM statistic $C_{s,b,e}(\mathbf{X})$ for $\mathbf{X}$ on the interval $[s,e]$ by
\begin{eqnarray}
C^*_{s,b,e}(\mathbf{X}) & = & \sqrt{\frac{e-b}{m(b-s+1)}} \sum_{t=s}^b X_t - \sqrt{\frac{b-s+1}{m(e-b)}} \sum_{t=b+1}^e X_t\nonumber\\
\label{ch4:eq:signedcusum}
& = & \sqrt{\frac{(b-s+1)(e-b)}{m}} (\bar{X}_{s:b} - \bar{X}_{(b+1):e}),\\
\label{ch4:eq:cusum}
C_{s,b,e}(\mathbf{X}) & = & |C^*_{s,b,e}(\mathbf{X})|.
\end{eqnarray}
Remember that the CUSUM statistic is giving the evidence for a change at $b$, in this case based just on the data $X_{s:e}$. This evidence is in terms of a suitably re-scaled difference of the empirical mean of the data before and after $b$. Larger values correspond to greater evidence for a change.

The binary segmentation algorithm is best defined recursively and hence described by pseudocode. The main function is described as follows.

\begin{algorithmic}[1]
\Function{Binary Segmentation}{$s$, $e$, $\zeta_n$}
\If {$e-s < 1$}
\State STOP
\Else
	\State $b_0 := \arg\max_{b\in\{s, \ldots, e-1\}} C_{s,b,e}(\mathbf{X})$
	\If {$C_{s,b_0,e}(\mathbf{X}) \ge \zeta_n$}
		\State add $b_0$ to the set of change-point candidates
		\State \textsc{Binary Segmentation}($s$, $b_0$, $\zeta_n$)
		\State \textsc{Binary Segmentation}($b_0+1$, $e$, $\zeta_n$)
	\Else
		\State STOP
	\EndIf
\EndIf
\EndFunction
\end{algorithmic}

With the above definition in place, the standard binary segmentation procedure is launched
by the call \textsc{Binary Segmentation}($1$, $n$, $\zeta_n$), where $\zeta_n$ is a threshold
parameter. If $\zeta_n = 0$, then the above function acts in the ``solution path'' mode: no judgment is made as to whether the successively proposed change-points are significant or not. For this mode to be useful in change-point detection, a model selection procedure needs to be applied at a later stage. The same principle applies in the case in which $\zeta_n$ is positive but small -- smaller than the level required for consistent model selection.
If $\zeta_n > 0$ and has been set at an appropriate level, the binary segmentation routine can combine solution path generation and model selection.
We discuss model selection for binary segmentation and suitable choices of $\zeta_n$ later on.

The main idea of binary segmentation is very simple: at the start, we search for the single most prominent change-point, by proceeding in the same way as in the AMOC setting described in \cite{fearnhead2022detecting}. If
$\zeta_n = 0$, or if $C_{1,b_0,n}(\mathbf{X}) \ge \zeta_n$, we save the location of this change-point candidate and proceed in the same way to the left and to the right of it, thereby ``zooming in'' on smaller and smaller sections of the data. On any given current interval $[s,e]$, we stop if the interval is too short for us to proceed ($e-s < 1$) or if we do not discover significant change-points ($C_{s,b_0,e}(\mathbf{X}) < \zeta_n$). On completion, we execute one of the following steps.
\begin{itemize}
    \item In the solution path mode ($\zeta_n = 0$ or small), we sort the thus-obtained set of change-point candidates in decreasing order of their corresponding detection CUSUMs to create the solution path
    \[
    \mathcal{P}_{BS} = \left( (b_i), C_{(s_i),(b_i),(e_i)}(\mathbf{X})     \right)_{i=1}^{n-1},
    \]
    where the ordering $(b_i)$ is such that 
    \[
    C_{(s_1),(b_1),(e_1)}(\mathbf{X}) \ge \ldots \ge     C_{(s_{n-1}),(b_{n-1}),(e_{n-1})}(\mathbf{X}).
    \]
    The solution path $\mathcal{P}_{BS}$ is then passed onto a model selection routine (more details below).
    \item In the model selection mode ($\zeta_n > 0$ and set appropriately), we return the obtained set of change-point candidates as the final set of change-point location estimates; it is good practice to sort it in increasing order beforehand.
\end{itemize}


\vspace{5pt}

For the solution path mode, any existing model selection tool can, in principle, be used to choose the model based on the solution path $\mathcal{P}_{BS}$. This is done by choosing from among the nested sequence of models
\begin{eqnarray*}
\lefteqn{\mathcal{M}_0 = \emptyset}\\
\lefteqn{\mathcal{M}_1 = \{(b_1)\}}\\
\lefteqn{\cdots}\\
\lefteqn{\mathcal{M}_{n-1} = \{(b_1), \ldots, (b_{n-1})\}}\\
\end{eqnarray*}
One natural option would be to use SIC, i.e. choose the model that minimises
\begin{equation}
\label{ch4:eq:bsic}
\mbox{SIC}(N', (b_1), \ldots, (b_{N'})) = \frac{n}{2} \log\left\{ \frac{1}{n} S_n((b_1), \ldots, (b_{N'}))    \right\} + N' \log(n),
\end{equation}
over $N' \in \{ 0, \ldots, N_U    \}$, where $N_U$ is a pre-determined upper bound for the number of change-points.

It is easy to think of binary segmentation as an appealing algorithm: to start with, it is conceptually simple and easy to explain and to code -- its naturally recursive structure suggests recursion as an appropriate programming device. Binary segmentation breaks up the difficult multivariate problem of finding possibly multiple change-points into a series of one-dimensional (and hence easier) problems: observe that the only task of the \textsc{Binary Segmentation} routine is to find the {\em single} most prominent change-point on its current domain of execution, before calling itself on the two implied children intervals. (The property of binary segmentation whereby it fits the ``best'' single-change-point model on each current interval $[s,e]$, i.e. the model that maximises the CUSUM $C_{s,b,e}(\mathbf{X})$ over $b$, can be describes as its {\em greediness}, a term used in algorithmic science to describe algorithms that make locally optimal choices at each stage.) Another advantage of binary segmentation is that it is relatively fast: a typical execution, in which successive change-point candidates are found in the middle part of each current interval (i.e. away from its edges) passes the entire dataset $O(\log(n))$ times and hence takes $O(n \log(n))$ operations.
Finally, an appealing feature of binary segmentation is the generality of its principle -- it is possible to replace the CUSUM test with any other test appropriate for a given type of change. However, care is needed as early iterations of binary segmentation involved detecting a single change for intervals of data which may contain multiple changes. As we saw, for example, in Section 2.4 of \cite{fearnhead2022detecting} for detecting a change in slope, some methods for detecting a single change can perform poorly when there are actually multiple changes -- for example, detecting the change in the wrong position even for data where the signal for the true change locations is strong. A future chapter of this book will return to this issue, and describe an approach that overcomes this drawback of binary segmentation.



This latter issue shows that binary segmentation should only be used for change-point detection methods that if applied to data without noise but with multiple change-points, would detect one of these change-points. That is that the maximum of the CUSUM statistic will be at a value $b$ that corresponds to a true change-point. This can be shown to be the case for detecting changes in mean with the CUSUM statistic. However even in this case, the fact the binary segmentation can involve detecting a single change-point on intervals of data which may contain multiple changes can affect its accuracy. We now illustrate this. 

\begin{figure}[t]
\centering
\begin{minipage}{.5\textwidth}
  \centering
  \includegraphics[width=\linewidth]{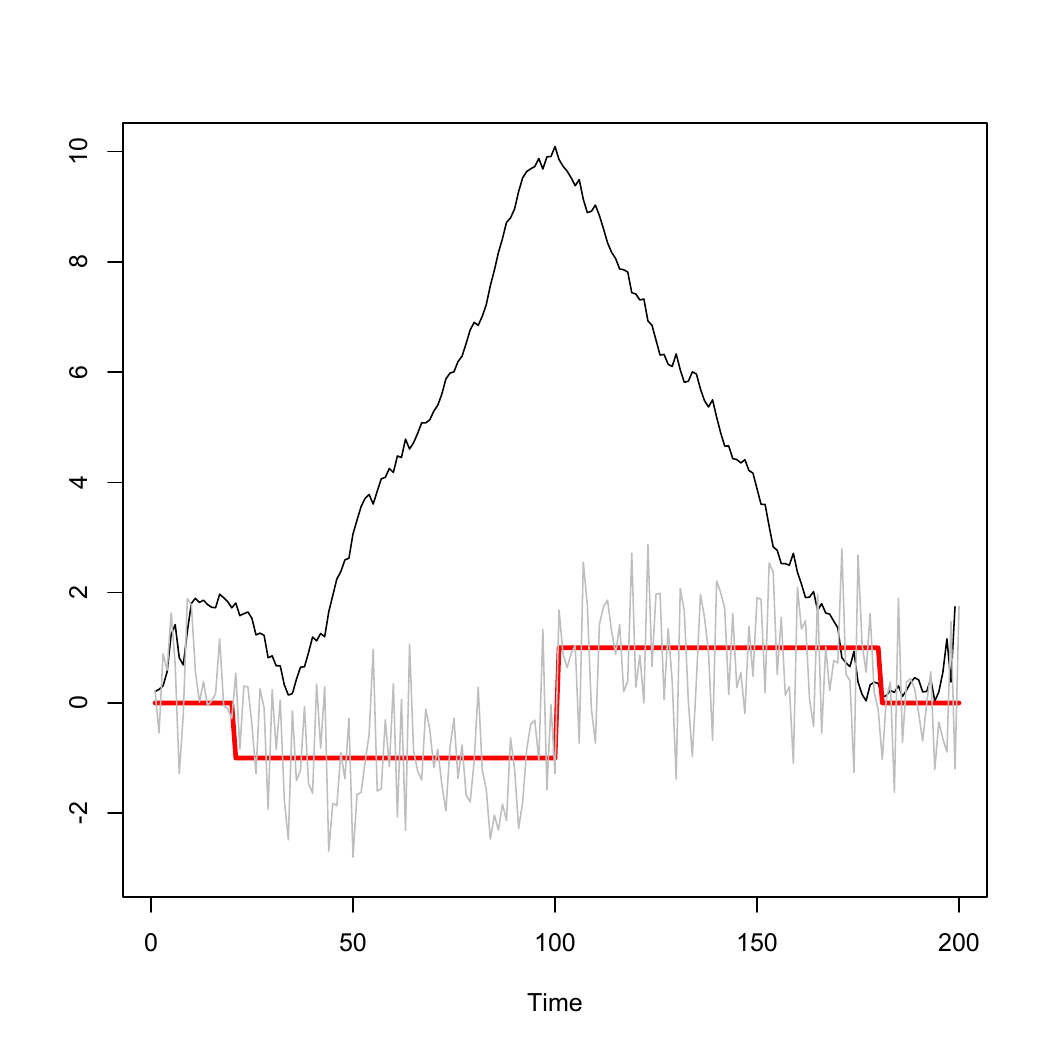}
\end{minipage}%
\begin{minipage}{.5\textwidth}
  \centering
  \includegraphics[width=\linewidth]{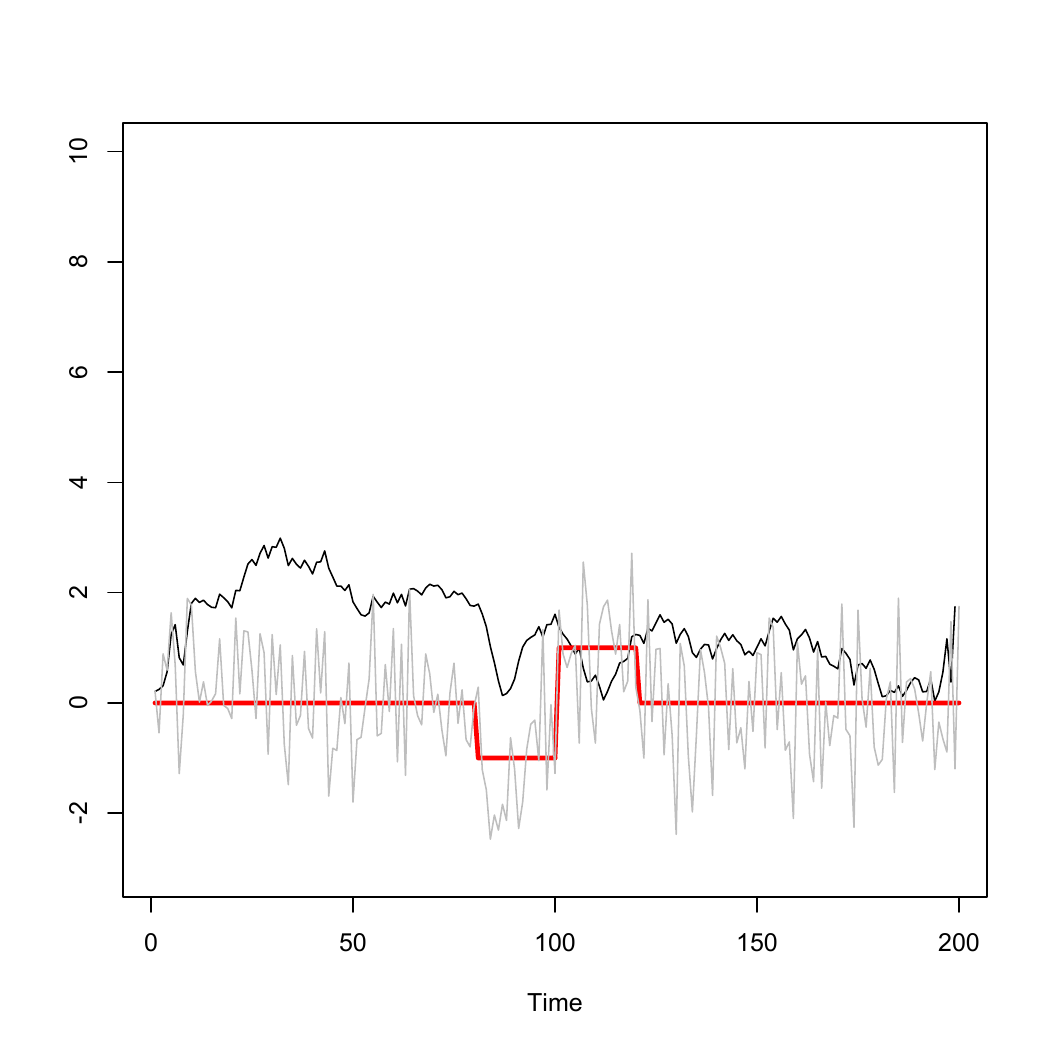}
\end{minipage}\\

  \caption{Both panels: signal (red), its noisy realisation $\mathbf{X}$ (grey), corresponding CUSUM statistic $C_{1,b,200}(\mathbf{X})$ as a function of $b$ (black).
  \label{ch4:fig:01}}
\end{figure}

To begin with, consider the left panel of Figure \ref{ch4:fig:01}. Starting with the whole data domain $[1,200]$, the CUSUM fitted by binary segmentation on $[1,200]$ clearly indicates $b = 100$ as the first change-point candidate. This is unsurprising as the two change-points at times $20$ and $180$ appear to be secondary in importance to the change-point at time 100, so overall the signal is, in a way, not far from a signal with just one change-point at 100; hence fitting the single-change-point model to the entire data would not be ``very wrong'' in this particular instance.

Let us contrast this with the situation in the right panel of Figure \ref{ch4:fig:01}. Here, visually, there does not appear to be one dominant change-point, and it is unclear how to approximate the signal well by a signal with only a single change-point. This, coupled with the fact that the noise level is relatively large, causes the CUSUM statistic to miss all of the change-points: the maximum of the CUSUM is attained around time $t = 30$, far from any of the actual change-points. In this instance, fitting the best single-change-point model to the entire data is not a good idea, and provides an example of how binary segmentation can be led onto the wrong track. As we will see in Section \ref{ch4:multi:wbs}, the detection failure here is not because the problem may be inherently difficult for any method to detect the change-points correctly (it is not the case); the root of the failure of binary segmentation is that it considers the whole data sample at once, which ``averages away'' the interesting features of the data. As we will see in Section \ref{ch4:multi:wbs}, in this example it is a better idea to focus on smaller sections of the data with more prominent presence of the features of interest, for example $X_{1:100}$ or $X_{81:120}$, which binary segmentation is unable to do.

Authors such as \cite{v81_4} and \cite{v93} provide theoretically interesting consistency results regarding binary segmentation, but we do not reproduce them here: the most important messages that we want to convey are that it is a particularly simple and fast algorithm, but that it can fail unless the change-points have a natural ordering in terms of prominence/importance, i.e. unless there is one dominant change-point on every current interval under consideration.
However, our practical experience with problems of this type suggests that whether or not this holds for a given dataset may be difficult to determine a priori. Methods described in the following sections have been designed to mitigate, in one way or another, this weakness of binary segmentation.

The fact that binary segmentation discovers change-points one by one has one other important consequence: the best binary segmentation fit with $N'$ change-points (that is, the fit corresponding to the first $N'$ elements of the binary segmentation solution path) is not guaranteed to be the same as the least-squares fit to the data of a piecewise-constant function with $N'$ change-points. However, the least-squares fits to the data are not nested in the sense that the best $N'+1$ change-point locations are not guaranteed to be a superset of the best $N'$ change-point locations, which may have computational speed consequences when one wishes to compare least-squares model fits for a range of proposed change-point numbers. By contrast, most algorithms that add change-points one by one (such as binary segmentation) produce nested solution paths which can by construction be traversed quickly when one wishes to compare different model fits.

\subsection{Wild Binary Segmentation}
\label{ch4:multi:wbs}

\begin{figure}[th]
\centering
  \includegraphics[width=0.9\linewidth]{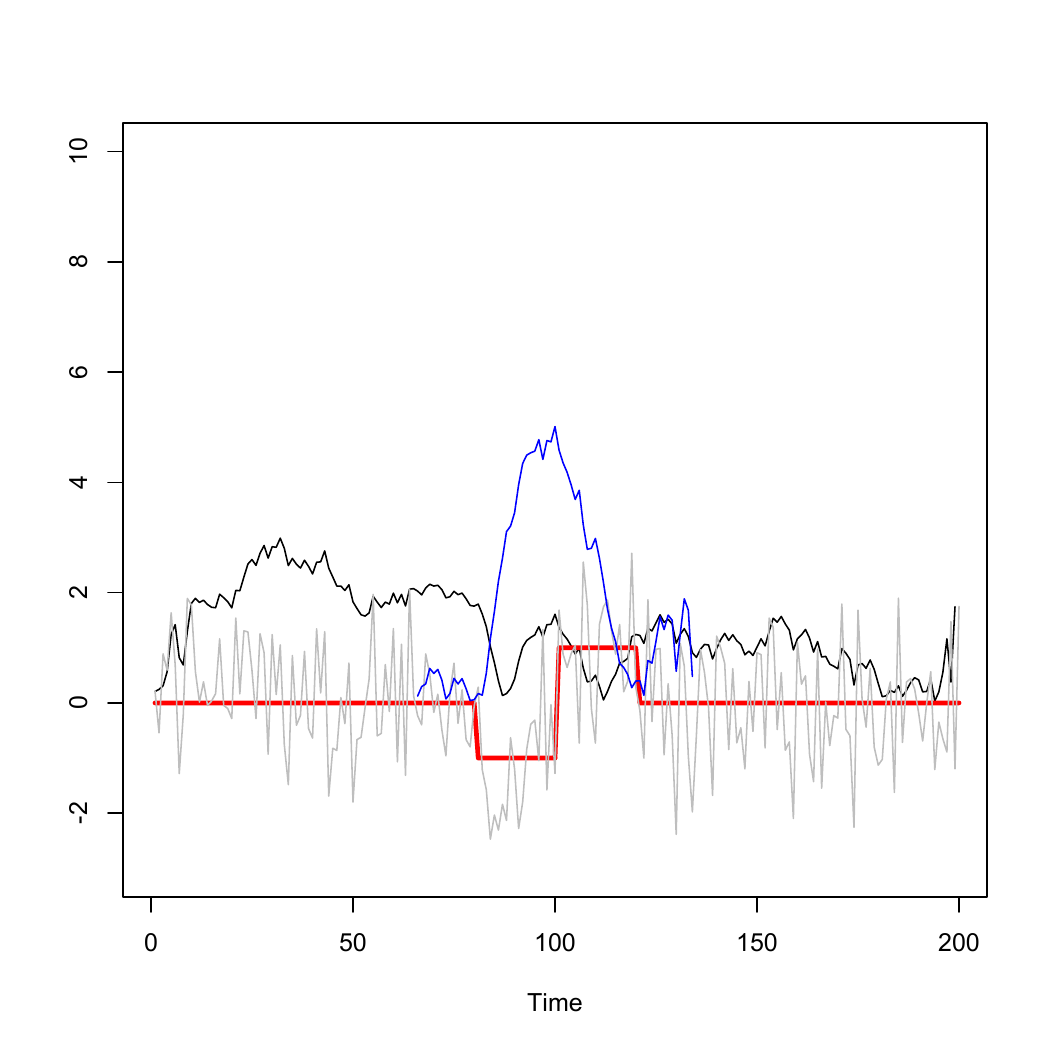}
  \caption{Signal (red), its noisy realisation $\mathbf{X}$ (grey), corresponding CUSUM statistic $C_{1,b,200}(\mathbf{X})$ as a function of $b$ (black) and the CUSUM statistic $C_{66,b,135}(\mathbf{X})$ as a function of $b$ (blue).
  \label{ch4:fig:02}}
\end{figure}

In the situation from the right-hand plot of Figure \ref{ch4:fig:01}, rather than starting with a CUSUM taken over the entire data sample, it would have been more advisable to ``zoom in'' on the three-change-point feature of interest in the middle part of the signal, by taking the CUSUM statistic over a shorter interval containing this feature. Figure \ref{ch4:fig:02} shows the CUSUM statistic taken over the interval $[66, 135]$, its endpoints being much closer to the change-points at 80, 100 and 120 than those of the original CUSUM over $[1,200]$. As the section of the signal on the interval $[66, 135]$ can be approximated relatively well by a signal with a single change-point, it is not surprising to see that the CUSUM computed over $[66, 135]$ brings out the location of one of the change-points, the middle one, very well.

If this observation were to be built into a change-point detection algorithm, the obvious first issue to face is that the analyst (more often than not) has little knowledge of where any features of interest may be located, so it is unclear which parts of the signal to zoom in on. The main idea underlying Wild Binary Segmentation (WBS; \cite{f14a}) is to take the CUSUM statistic over each of a ``representative sample'' of intervals of varying lengths and locations within $[1, n]$, in the hope that some of these intervals will align well with the features of interest in the signal (as illustrated in Figure \ref{ch4:fig:02}) and therefore the CUSUMs computed on those will be informative for change-point locations.
At each stage of the procedure, WBS works with the largest CUSUM (over the collection of all intervals drawn) and chooses the next change-point candidate as the argument-maximum of this largest CUSUM. 

To define WBS, we denote by $F_n^{M_1}$ a set of $M_1 \ge M$ intervals
$[s_m, e_m]$, $m = 1, \ldots, M_1$, produced in either of the two following ways.
\begin{enumerate}
    \item (Random grid.)
    Their start- and endpoints are drawn uniformly at random, independently with replacement, from the set $\{1, \ldots, n\}$. In this random setting we draw $M$ intervals and therefore $M_1 = M$.
    \item (Deterministic grid.)
    Construct an equispaced grid $\mathcal{K} = \{i/(K-1)\,\,:\,\,i=0,\ldots,K-1\}$ on $[0,1]$ such that the number of gridpoints is the smallest integer $K$ for which $M_1 = K(K-1)/2$, the number
    of all intervals with start- and endpoints in $\mathcal{K}$, is such that $M_1 \ge M$. Transform the grid $K$ to an integer grid on $[1,n]$ through the transformation $i/(K-1) \to [(n-1)i/(K-1) + 1]$, where $[\cdot]$ is rounding to the nearest integer. Take $[s_m, e_m]$ to be all intervals with start- and endpoints on the thus-transformed grid.
\end{enumerate}
Other ways of choosing the intervals -- mixed random plus deterministic, random but drawn from a distribution different than uniform, a different deterministic scheme such as that of Seeded Binary Segmentation \citep{klbm20} -- are also acceptable as long as they ensure the property that, with a high probability, each true change-point is well covered by at least one interval which does not contain any other true change-points. This can typically be achieved by (a) ensuring that the number of intervals is sufficiently large and (b) that the start- and end-points of the intervals cover the domain of the data sufficiently densely.

In either of the above cases (random or deterministic grid), we are guaranteed at least $M$ intervals, where 
$M$ is decided by the user. Version 2.2 of the
R package \verb+breakfast+ \citep{breakfast2} uses $M=10000$ by default. An appropriate choice of $M$ is key: on the whole, the larger the value of $M$, the more accurate but slower the procedure. Theoretical lower bounds on $M$ are available, but, unsurprisingly, they depend on the minimum spacing between neighbouring change-points, which the analyst typically does not have access to. Naturally, if $n$ is so small that the $M$ requested by the user is larger than or equal to the number of all intervals on $[1,n]$, then WBS simply uses all those.

Using pseudocode, the main function of the WBS algorithm, with model selection
performed via thresholding with threshold $\zeta_n$ (as with binary segmentation, if $\zeta_n = 0$ or is positive but small, then the procedure acts in the solution path mode), is defined in the following way.

\vspace{10pt}

\begin{algorithmic}[1]
\Function{WBS}{$s$, $e$, $\zeta_n$}
\If {$e-s < 1$}
\State STOP
\Else
	\State ${\mathcal M}_{s,e} := $ set of those indices $m$ for which $[s_m, e_m] \in F_n^{M_1}$ is such that
	$[s_m,e_m]\subseteq [s,e]$
	\State $(m_0, b_{m_0}) := \arg\max_{m\in {\mathcal M}_{s,e}, b\in\{s_m, \ldots, e_m-1\}}
	C_{s_m,b,e_m}(\mathbf{X})$
	\If {$C_{s_{m_0},b_{m_0},e_{m_0}}(\mathbf{X}) \ge \zeta_n$}
		\State add $b_{m_0}$ to the set of change-point candidates [with the additional side information $(m_0, s_{m_0}, e_{m_0}, C_{s_{m_0},b_{m_0},e_{m_0}}(\mathbf{X}))$]
		\State \textsc{WBS}($s$, $b_{m_0}$, $\zeta_n$)
		\State \textsc{WBS}($b_{m_0}+1$, $e$, $\zeta_n$)
	\Else
		\State STOP
	\EndIf
\EndIf
\EndFunction
\end{algorithmic}

The WBS procedure is launched by the call \textsc{WBS}($1$, $n$, $\zeta_n$).

The recursive calls in lines 9 and 10 of the WBS algorithm may create an impression that the intervals on which WBS performs computation have to be re-drawn, but this is not the case: all recursive calls use the same set of intervals drawn prior to the launch of the WBS routine, and the same set of CUSUM statistics computed (once, prior to the start of the recursive mechanism) on these interevals. Therefore, WBS can equally be implemented non-recursively as follows.

\begin{algorithmic}[1]
\Function{WBS.NONREC}{$\zeta_n$}
\While{$F_n^{M_1}$ non-empty} 
	\State ${\mathcal M} := $ set of those indices $m$ for which $[s_m, e_m] \in F_n^{M_1}$
	\State $(m_0, b_{m_0}) := \arg\max_{m\in {\mathcal M}, b\in\{s_m, \ldots, e_m-1\}}
	C_{s_m,b,e_m}(\mathbf{X})$
	\If {$C_{s_{m_0},b_{m_0},e_{m_0}}(\mathbf{X}) \ge \zeta_n$}
		\State add $b_{m_0}$ to the set of change-point candidates [with the additional side information $(m_0, s_{m_0}, e_{m_0}, C_{s_{m_0},b_{m_0},e_{m_0}}(\mathbf{X}))$]
		\State remove from $F_n^{M_1}$ those $m$ for which $b_{m_0}\in\{s_m, \ldots, e_m-1\}$
	\Else
		\State STOP
	\EndIf
\EndWhile
\EndFunction
\end{algorithmic}

Suppose we wish to use WBS to estimate the locations of significant change-points, as opposed to merely producing a solution path of all possible models. It is clear that we should use $\zeta_n > 0$, but what values of $\zeta_n$ are appropriate? One natural requirement is that $\zeta_n$ is large enough to control the probability of false alarms, in the sense that if the signal $\mathbf{f}$ contains no change-points, the WBS estimator also returns no change-points with a suitably high probability. Let us examine how high $C_{s_{m_0},b_{m_0},e_{m_0}}(\mathbf{X})$ can possibly get (with a high probability) given that $\mathbf{f} \equiv 0$ (the same bound will apply in the more general case $\mathbf{f} \equiv$ const, as the CUSUM statistic ignores additive constants).
Trivially, it is clear that
\[
C_{s_{m_0},b_{m_0},e_{m_0}}(\mathbf{X}) \le
\max_{1\le s \le b < e \le n} C_{s,b,e}(\mathbf{X}).
\]
We wish to choose $\zeta_n$ such that
\[
P\left(\max_{1\le s \le b < e \le n} C_{s,b,e}(\mathbf{X}) > \zeta_n\right) 
\]
is small. Perhaps the simplest, albeit suboptimal, bound is provided by the Bonferroni correction. We first observe that for a fixed triple $(s, b, e)$, the quantity $C_{s,b,e}(\mathbf{X})$ is distributed as $|N(0, \sigma^2)|$. This is easy to establish directly: $C^*_{s,b,e}(\mathbf{X})$ is clearly normally distributed as a linear combination of the data points, and considering its expectation and variance yields the above fact. Using the Bonferroni correction and Mill's ratio inequality for the Gaussian distribution, we can therefore bound
\begin{eqnarray*}
P\left(\max_{1\le s \le b < e \le n} C_{s,b,e}(\mathbf{X}) > \zeta_n\right) & \le & \sum_{1\le s \le b < e \le n} 
P(C_{s,b,e}(\mathbf{X}) > \zeta_n)\\
& \le & n^3 2(1 - \Phi(\zeta_n/\sigma))\\
& \le & C n^3 \exp\{-\zeta_n^2 / (2 \sigma^2)\},
\end{eqnarray*}
where $\Phi()$ is the standard normal cdf, and $C$ is a certain positive constant. This upper bound converges to zero as $n\to\infty$ as long as $\zeta_n = \sigma \sqrt{C_0 \log\,n}$ with $C_0 > 6$. As the Bonferroni correction is, in this case, very loose (because of the strong dependence between $C_{s,b,e}(\mathbf{X})$ for neighbouring values of $(s, b, e)$), the constant $C_0 = 6$ can be expected to be too high in practice. On the other hand, the discussion in Section \ref{ch4:sec:sic} implies that the $C_0$
in $\zeta_n = \sigma \sqrt{C_0 \log\,n}$
must exceed 2 for $P\left(\max_{1\le s \le b < e \le n} C_{s,b,e}(\mathbf{X})\right)$ to have any hope of tending to 0. To summarise, it is a good strategy to look for a suitable value of $C_0$ in the region $(2, 6]$. In \cite{f14a}, in which $\sigma$ is estimated via the Median Absolute
Deviation (MAD) estimator with
$\{2^{-1/2}(X_{t+1} - X_{t})\}_{t=1}^{T-1}$ on input (see Section \ref{ch4:sec:noise_variance} for more on the MAD estimator and on estimating the noise variance in general),
two particular values of $C_0$ are considered and tested numerically on a variety of signals: $C_0 = 2$ (it is concluded that this constant value is too low in practice and tends to lead to the overestimation of the number of change-points) and $C_0 = 3.38$ (which appears too high in practice and tends to underestimate the number of change-points). A good universal choice of $C_0$ is, at the time of writing, still an open question. For more details on this problem, and partial solutions, we refer the reader to \cite{fls20}, and to Corollary 1 in \cite{af18}.


\cite{f14a} also explores the use of WBS with an information criterion referred to as the strengthened SIC. This is close to the standard SIC, and we do not elaborate on its form here. Conceptually, WBS with an information criterion as the model selector proceeds similarly to binary segmentation (see Section \ref{ch4:multi:bs}): first, a solution path is executed via the call \textsc{WBS}($1$, $n$, 0).
Next, the change-point candidates are sorted in decreasing order of their corresponding detection CUSUMs to create the solution path
    \[
    \mathcal{P}_{BS} = \left( (b_i), C_{(s_i),(b_i),(e_i)}(\mathbf{X})     \right)_{i=1}^{M_2},
    \]
where $M_2 \le M$. Finally, SIC is applied on the resulting nested sequence of model candidates in the same way as in formula (\ref{ch4:eq:bsic}). A different model selection procedure for WBS, termed Steepest Drop to Low Levels \citep{f20}, directly processes the WBS solution path and is covered in Section \ref{ch4:sec:sdll}.

In WBS, if $M$ is large enough, each true change-point is guaranteed (with probability
approaching one in $n$) to be contained
within at least one of the intervals $[s_m, e_m]$ which contains only that single change-point, and to be sufficiently distanced from either $s_m$ or $e_m$, i.e. to sit sufficiently deeply within the interior of $[s_m, e_m]$.
This alone provides suitable guarantees for the size of the largest maximum of the CUSUM
statistics, and is enough to yield theoretical and practical advantages of WBS
over binary segmentation, as similar guarantees are unavailable in the latter method. However, WBS, as defined here, is still not rate-optimal in terms of the permitted minimum distance between change-points. The fundamental reason for this is the fact that the interval of the largest CUSUM, selected by WBS at each stage, is not necessarily guaranteed to contain a single change-point only, and therefore the CUSUM spanned on that interval is not necessarily the best possible estimator for the change-point location. The Narrowest-Over-Threshold (NOT) localisation approach described in \cite{bcf19} overcomes this issue by preferring the shortest intervals on which the contrast statistics (such as CUSUMs) are deemed to be significant. However, NOT applies to more general models than the multiple mean shift model considered in this chapter and will therefore be covered in a future chapter.

In WBS, the collection $F_n^{M_1}$ is drawn only once at the start of the procedure. In a signal with many change-points, this may potentially mean that WBS can ``run out of intervals'', if there are still undetected change-points but no intervals left to operate on. This can be referred to as data-nonadaptivity of WBS with respect to interval choice, and there is no mechanism in WBS itself to remedy this.

A solution to this problem is proposed in \cite{f20} by the method termed ``Wild Binary Segmentation 2'' (WBS2), in which intervals are drawn afresh on each interval on which the procedure currently operates (which can be seen as a data-adaptive interval drawing scheme). This ensures that there is always a supply of intervals to work with. While this can in principle lead to an increase in the computational cost over WBS, \cite{f20} notes that each interval draw in WBS2 requires of the order of 100 or 1000 intervals, which is 100-10 times fewer than the corresponding default value in WBS used in the R package \verb+breakfast+. 

Conceptually, WBS is a modular procedure in the sense that it is relatively straightforward to see how it can be extended to the multiple mean-shift problem in noise settings other than i.i.d. Gaussian: the CUSUM statistic $C_{s,b,e}$ can be replaced, for example, by a suitable generalised likelihood ratio test. This preserves the construction principle of WBS whereby it fits simple single-change-point models on subsamples of the data, and combines the results to create a multi-change-point fit on the entire sample. Plain binary segmentation is equally easy to extend conceptually to more complex settings; therefore, part of the appeal of the methods belonging to the binary segmentation family is that they can potentially be applied to multiple change detection e.g. in high-dimensional regression models or network data -- topics we will return to in a future chapter of this book.

\subsection{From the Bottom Up: Tail-Greedy Unbalanced Haar}
\label{ch4:sec:tguh}

Both binary segmentation and WBS are top-down hierarchical methodologies, successively subdividing the data by the next most likely change-point candidate. An alternative direction may be to proceed from the bottom up, by successively merging regions of the data considered the least likely to contain change-points. This principle is formalised in the Tail-Greedy Unbalanced Haar (TGUH) approach of \cite{f18_4}, which can be described as a hierarchical agglomeration technique. As the generic notation for TGUH can get a little involved, we introduce this algorithm on an example, rather than in its generality.

Suppose that $n = 6$, and our input data vector is $(X_1, X_2, \ldots, X_6)$. We initially look at all pairwise differences between consecutive data points, i.e. $2^{-1/2}(X_i - X_{i+1})$ for $i = 1, \ldots, 5$. The $2^{-1/2}$ is a normalisation factor which helps ensure that the data transformation we have just begun ends up being an orthonormal transformation (rotation) of the input data, conditioning on the order of merges taking place.

Having formed the pairwise differences, we then look for a certain number of the pairwise differences that are the smallest in absolute value. In general, at this first stage of the procedure, we would be looking for $\lceil \rho n \rceil$ smallest differences, where $\rho$ is a given fixed constant $\in (0, 1)$ (best thought of as being small), $n$ is the sample size (in our example, we have $n = 6$) and $\lceil \cdot \rceil$ is the ceiling operator.

The reason we are looking for the smallest differences (and this principle will continue throughout the algorithm) is to be able to identify the regions in the data that are the least likely to contain change-points. Of course, the fact that, for example, 
$2^{-1/2}|X_3 - X_{4}|$ is small does not mean there is no change-point at time $t = 3$: the test we are carrying out, involving just one observation to the left ($X_3$) and to the right ($X_4$) of the postulated change-point ($t = 3$) has very little power. However, as we will see later, as the algorithm progresses, we will be obtaining more and more powerful tests for the presence of change-points in the data.

For the sake of the example, suppose that $\rho = 0.25$ so that $\lceil \rho n \rceil = 2$ and that the two smallest differences in absolute value are, in this order,
$2^{-1/2}|X_3 - X_{4}|$ and $2^{-1/2}|X_4 - X_{5}|$. Note that the second smallest difference, $2^{-1/2}|X_4 - X_{5}|$, uses one of the data points ($X_4$) already used by the first smallest difference. For this reason, we do not accept $2^{-1/2}|X_4 - X_{5}|$ and look for the next smallest difference: suppose this is given by $2^{-1/2}|X_5 - X_{6}|$. We save those two smallest differences (now without the absolute values) as the so-called ``detail'' coefficients: they explain the deviations from constancy between, respectively, $(X_3, X_4)$ and $(X_5, X_6)$. We use the $d$ notation to denote detail:
\begin{eqnarray*}
    d_{3,3,4}^{(1,1)} & = & 2^{-1/2}(X_3 - X_{4}),\\
    d_{5,5,6}^{(1,2)} & = & 2^{-1/2}(X_5 - X_{6}).
\end{eqnarray*}
In the notation $d_{p,q,r}^{(j,k)}$, $j$ is the index of the stage, or one complete pass through the data, that the given detail coefficient is computed at; $k$ indexes the detail coefficients (starting from the smallest in absolute value) extracted within a single stage; $p$ is the start of the support of the data that enters $d_{p,q,r}^{(j,k)}$, $r$ is its end, and $q$ is the index of the final data point that enters $d_{p,q,r}^{(j,k)}$ with the positive sign (as opposed to negative).

We now ``merge'' the corresponding portions of the data: $(X_3, X_4)$ and $(X_5, X_6)$, that is we replace them by the scaled averages of the data at those locations:
\begin{eqnarray*}
s_{3,4} & = & 2^{-1/2}(X_3 + X_4),\\
s_{5,6} & = & 2^{-1/2}(X_5 + X_6).
\end{eqnarray*}
The letter $s$ has been chosen to represent the ``smoothing'' of the data provided by the averages. In the $s_{p,r}$ notation, $p$ and $r$ represent the start and the end of the relevant support of the data, respectively.
Note the orthonormality (the property of being orthogonal vectors of Euclidean length 1) of the filter used to compute each detail coefficient, $2^{-1/2}(1, -1)$, and the filter for the corresponding smooth coefficient, $2^{-1/2}(1, 1)$. This orthonormality will hold throughout the algorithm and is important both for the theoretical analysis of the algorithm, and so that the detail coefficients can be interpreted as (signed) CUSUM statistics; we expand on both these points below. In the next stages of the algorithm, $X_3$ and $X_4$ will always be considered together as part of a single merged region, and the same will be true of $X_5$ and $X_6$.

Summarising, after the first stage (a stage is understood as a complete pass through the data which computes all possible scaled pairwise differences) we have the following.

\begin{description}
    \item[{\bf After Stage 1}]
    
    \vspace{5pt}

    \item[{\bf detail coefficients}] $d_{3,3,4}^{(1,1)}$, $d_{5,5,6}^{(1,2)}$
    \item[{\bf remaining data}] $\{X_1, X_2, 2^{-1/2}(X_3 + X_4), 2^{-1/2}(X_5 + X_6)\}$
\end{description}

We are now ready for the second stage of the algorithm: the second pass through the data in which we again compute all the scaled pairwise differences of the remaining data, and retain the smallest ones.

Do we need to adjust some of our differencing weights to account for the fact that we now compare data averages taken over regions of different sizes? The answer is yes, and to understand why this is so, we now introduce the general principle of constructing an arbitrary scaled difference. We will use the following example to illustrate this general principle.

At some point in the second pass through the data, we will need to compute the scaled difference between $X_2$ and its right-hand neighbour, $2^{-1/2}(X_3 + X_4)$. The difference will take the general form
\begin{equation}
\label{ch4:eq:det2ndstage}
a X_2 + b 2^{-1/2}(X_3 + X_4),
\end{equation}
and we will show how to find the values of $a$ and $b$. The following general guiding principles will enable us to determine $a$ and $b$.
\begin{description}
    \item{{\bf Principle 1.}} If the data values over the section over which the scaled difference is being computed are all constant, the value of the scaled difference should be zero.

    \vspace{10pt}

    For us, this means that if $(X_2, X_3, X_4) = (1, 1, 1)$ (or any other sequence of three equal values bar zero), the linear combination in (\ref{ch4:eq:det2ndstage}) should return zero. From this, we obtain the condition

\[
a + \sqrt{2} b = 0.
\]

    The thought behind this principle is that the scaled differences record local deviations of the data from constancy. If the data is locally constant, the corresponding scaled difference should be zero.

    \vspace{10pt}
    
    \item{{\bf Principle 2.}} $a > 0, b < 0$. This is merely a convention: the left-hand neighbour in each pairwise scaled difference enters with the positive sign. Equally, we could use the opposite rule (as long as we did so consistently), but this would in no way change our findings.
    
    \vspace{10pt}

    \item{{\bf Principle 3.}} $a^2 + b^2 = 1$. This is to preserve the orthonormality of the data transformation we are building, conditional on the order of the merges taking place. The importance of this conditional orthonormality was briefly mentioned earlier and we will return to it later on.
\end{description}

Principles 1, 2 and 3 always uniquely determine the values of $a$ and $b$; in our particular example, we obtain
\begin{eqnarray*}
    a & = & \sqrt{2/3},\\
    b & = & -1/\sqrt{3}.
\end{eqnarray*}

In the second pass of the data, using Principles 1, 2 and 3, we construct the scaled differences between the following pairs of neighbouring entries of the remaining data:
\begin{itemize}
\item $X_1$ and $X_2$,
\item $X_2$ and $2^{-1/2}(X_3 + X_4)$,
\item $2^{-1/2}(X_3 + X_4)$ and $2^{-1/2}(X_5 + X_6)$.
\end{itemize}
Recall that $\rho = 0.25$ and there are 4 remaining data regions, so we only need to record $\lceil 0.25 \cdot 4 \rceil = 1$ smallest (in absolute value) scaled difference. For the sake of the example, suppose the smallest difference is given by
\[
\left|\sqrt{\frac{2}{3}} X_2 - \sqrt{\frac{1}{6}} (X_3 + X_4)\right|.
\]
We therefore record
\[
d_{2,2,4}^{(2,1)} = \sqrt{\frac{2}{3}} X_2 - \sqrt{\frac{1}{6}} (X_3 + X_4).
\]
How to compute the corresponding smooth coefficient $s_{2,4}$? To this end, recall that we need to take the inner product of the data $\{ X_2, 2^{-1/2}(X_3 + X_4)   \}$ with a filter orthonormal to the detail filter $(\sqrt{2/3}, -\sqrt{1/3})$. This orthonormal filter (we take the version with positive signs) is given by $(\sqrt{1/3}, \sqrt{2/3})$. We therefore compute
\[
s_{2,4} = \sqrt{\frac{1}{3}}X_2 + \sqrt{\frac{2}{3}} 2^{-1/2} (X_3 + X_4) = \frac{X_2 + X_3 + X_4}{\sqrt{3}}.
\]
The form of the smooth coefficient is not a coincidence: by simple algebra, it is possible to show that we always have
\[
s_{p,r} = \frac{X_p + \ldots + X_r}{\sqrt{r-p+1}},
\]
for all values of $p < r$.

In summary, we are in the following situation after two stages of the algorithm.

\begin{description}
    \item[{\bf After Stage 2}]
    
    \vspace{5pt}

    \item[{\bf detail coefficients}] $d_{3,3,4}^{(1,1)}$, $d_{5,5,6}^{(1,2)}$, $d_{2,2,4}^{(2,1)}$
    \item[{\bf remaining data}] $\{X_1, 3^{-1/2} (X_2 + X_3 + X_4), 2^{-1/2}(X_5 + X_6)\}$
\end{description}

Fast-forwarding through the next stage, suppose the next merge to take place is between $X_1$ and $3^{-1/2} (X_2 + X_3 + X_4)$, after which we have the following.

\begin{description}
    \item[{\bf After Stage 3}]
    
    \vspace{5pt}

    \item[{\bf detail coefficients}] $d_{3,3,4}^{(1,1)}$, $d_{5,5,6}^{(1,2)}$, $d_{2,2,4}^{(2,1)}$, $d_{1,1,4}^{(3,1)}$
    \item[{\bf remaining data}] $\{4^{-1/2} (X_1 + X_2 + X_3 + X_4), 2^{-1/2}(X_5 + X_6)\}$
\end{description}

We now only have two regions to merge, so, naturally, the last merge must be between the data regions $4^{-1/2} (X_1 + X_2 + X_3 + X_4)$ and $2^{-1/2}(X_5 + X_6)$. Effectively, the top-down algorithm is estimating that if there is only one change-point in the data, it will be at location $\hat{\tau} = 4$, as the data points $X_1, \ldots, X_4$ have merged prior to the final outstanding merge, and so have $X_5, X_6$. Of course, whether or not there is a real change-point at $\tau = 4$ is a matter for significance testing, and we return to this question later. We therefore end up with the following decomposition.

\begin{description}
    \item[{\bf After [final] Stage 4}]
    
    \vspace{5pt}

    \item[{\bf detail coefficients}] $d_{3,3,4}^{(1,1)}$, $d_{5,5,6}^{(1,2)}$, $d_{2,2,4}^{(2,1)}$, $d_{1,1,4}^{(3,1)}$, $d_{1,4,6}^{(4,1)}$
    \item[{\bf remaining data}] $6^{-1/2} (X_1 + X_2 + X_3 + X_4 + X_5 + X_6) =: s_{1,6}$
\end{description}

It is, of course, not a coincidence, that we end up with six coefficients (five detail coefficients plus a single smooth), for a data input for length six. This is because the transformation we have carried out is, conditional on the order of merges, a rotation of the data, and preserves the entirety of the information contained in the original sequence $(X_1, \ldots, X_6)$. Expressing this mathematically, we have
\[
\left[ \begin{array}{c}
d_{3,3,4}^{(1,1)}\\
d_{5,5,6}^{(1,2)}\\
d_{2,2,4}^{(2,1)}\\
d_{1,1,4}^{(3,1)}\\
d_{1,4,6}^{(4,1)}\\
s_{1,6}
\end{array}   \right] = A \left[ \begin{array}{c}
X_1\\
X_2\\
X_3\\
X_4\\
X_5\\
X_6
\end{array}   \right],
\]
where $A$ is an orthogonal (orthonormal) matrix, i.e. a matrix for which $A^T = A^{-1}$. If $(X_1, \ldots, X_6)^T$ were seen as a point in a six-dimensional Euclidean space, then $A(X_1, \ldots, X_6)^T$ would be its rotation about the origin. Note, however, that the matrix $A$ was not given to us at the start of the transformation, but got constructed in the process of the transformation, as dictated by the choice of the merges. Therefore, the bottom-up transformation just introduced can be seen as a conditionally orthonormal transformation of the data, in which the rotation $A$ is chosen in a data-adaptive way.

Because the detail coefficients are chosen ``adaptively'' to be always as small as possible, and recalling that rotations preserve the $L_2$ norm of the input vector (in the sense that $\|x\|_2^2 = \|Ax\|_2^2$ where $\|x\|_2$ is the $L_2$ norm of $x$), it must be the case that the largest detail coefficients tend to appear towards the final stages of the transformation, hopefully encoding the locations of the change-points in the data, if any are present. In a way, this bottom-up transformation can be thought of as a localised, multiscale version of the principal component analysis: they both share the aim of compressing the information contained in the data in as few coefficients as possible.

Following Principles 1, 2 and 3 introduced earlier, it is easy to derive the explicit form of the final detail coefficient $d_{1,4,6}^{(4,1)}$:
\begin{equation}
\label{ch4:eq:d}
d_{1,4,6}^{(4,1)}  = \sqrt{\frac{1}{12}} \sum_{i=1}^4 X_i - \sqrt{\frac{1}{3}} \sum_{i=5}^6 X_i.
\end{equation}
Using this representation, from formula (\ref{ch4:eq:signedcusum}), it is straightforward to see that $d_{1,4,6}^{(4,1)}$ is simply equal to 
$C^*_{1,4,6}(\mathbf{X})$, the signed CUSUM statistic spanned on the entire dataset, with a break at index 4. In fact, more is true: we always have
\begin{equation}
\label{ch4:eq:dC}
d_{p,q,r}^{(\cdot,\cdot)} = C^*_{p,q,r}.
\end{equation}
In other words, the detail coefficients all take the form of a signed CUSUM statistic. However, unlike in top-down approaches, in (\ref{ch4:eq:d}) we did not arrive at the decision to test $\sqrt{\frac{1}{12}} \sum_{i=1}^4 X_i$ versus $\sqrt{\frac{1}{3}} \sum_{i=5}^6 X_i$ by maximising $|C^*_{1,b,6}(\mathbf{X})|$ over $b$; we obtained $b=4$ through the bottom-up merging process described earlier. While the top-down maximisation of $|C^*_{1,b,6}(\mathbf{X})|$ over $b$ would have been optimal (in the Gaussian model (\ref{ch4:eq:univ_mult})) in the presence of a single change-point, intuitively, the bottom-up merging approach, being more localised, could be preferable in the presence of multiple change-points delineating more local, finer features in the signal.

Having introduced the bottom-up transformation, we now discuss some of its aspects, starting with its name. The term ``tail-greedy'' originates from the fact that we are simultaneously extracting (up to) $\lceil \rho \alpha_{j,n} \rceil$ detail coefficients that are the smallest in magnitude (where $\alpha_{j,n}$ is the number of data regions remaining after $j-1$ passes through the data), and hence target the entire lower tail of the
distribution of their magnitudes. We discuss the importance of tail-greediness below. The term ``unbalanced Haar'' comes from the fact that the detail coefficients $d_{p,q,r}^{(j,k)}$ resemble the Haar wavelet coefficients of the data; however, the difference is that the classical Haar wavelet coefficients are balanced in the sense that $q - p + 1 = r - q$. As we saw above, this is not necessarily true of $d_{p,q,r}^{(j,k)}$, and hence the label ``unbalanced''.

Returning to the identity (\ref{ch4:eq:dC}), the fact that the detail coefficients $d_{p,q,r}^{(j,k)}$ have the form of signed CUSUM statistics
provides a direct route to change-point model selection with TGUH. \cite{f18_4} suggests thresholding the $d_{p,q,r}^{(j,k)}$ coefficients with a suitable threshold of magnitude $O(\sqrt{\log\,n})$. Other model selection techniques, such as SIC, involving $|d_{p,q,r}^{(j,k)}|$ sorted in decreasing order, can also be used by following the workflow described in Sections \ref{ch4:multi:bs} and \ref{ch4:multi:wbs}.

The tail-greediness of the TGUH algorithm brings in two ingredients. One is an upper bound on the computational complexity of the TGUH decomposition, coming from the fact that there are at most a logarithmic number of stages $j$, which leads to the $O(n\, \log^2(n))$ upper bound on the computational complexity of the TGUH decomposition. The other automatic implication of the tail-greediness of TGUH is the $L_2$ consistency of the estimator of $\mathbf{f}$ constructed by thresholding the detail coefficients $d_{p,q,r}^{(j,k)}$ with an appropriate threshold and taking the inverse TGUH transformation. This is an important result as it directly implies that it is possible to post-process this $L_2$-consistent estimator into one that is consistent for the number $N_0$ and locations $\tau^0_1, \ldots, \tau^0_{N_0}$ of change-points in $\mathbf{f}$. Details are provided in \cite{f18_4}.

\section{Moving Sum and Isolation Approaches}

In the presence of multiple change-points, one natural idea is to sequentially consider smaller sections of the data,
on a moving-window basis, in the hope that each of them only contains at most a single change-point. The main motivation
behind this is that as we saw in \cite{fearnhead2022detecting}, it is usually easier to detect and estimate the location of a change-point if there is only
one present in the data than if there are more than one. We will refer to approaches that align with this way of thinking as isolation-based
methodologies, as the key idea here is to attempt to isolate potential change-points from their neighbours (if any). We review two such approaches in some detail: one based on ``moving sums'' (MOSUM), and the other on the idea of ``isolation followed by detection'' (Isolate-Detect).

\subsection{MOSUM}
\label{ch4:sec:mosum}

\cite{km11} describe the use of the Moving SUM (MOSUM) methodology to the problem of multiple change-point detection in the mean.
In its simplest form, it consists of repeated application to the data, on a moving window-basis, of a contrast statistic of the form
\begin{equation}
\label{eq:mosum}
M_{\tau, b} = (2b)^{-1/2}\left( \sum_{i=\tau-b+1}^\tau X_i - \sum_{i=\tau+1}^{\tau+b} X_i   \right),
\end{equation}
where $b$ represents the bandwidth parameter.
The MOSUM statistic $M_{\tau,b}$ is applied for $\tau = b, \ldots, n-b$. In \cite{mkc21}, a CUSUM-type boundary extension is proposed for $\tau = 1, \ldots, b-1$ and $\tau = n-b+1, \ldots, n-1$. For $\tau = 1, \ldots, b-1$, this takes the form
\[
M_{\tau, b} = \sqrt{\frac{2b}{\tau(2b - \tau)}}\sum_{i=1}^\tau (\bar{X}_{1:2b} - X_i)
\]
and is mirrored for $\tau = n-b+1, \ldots, n-1$.

As with the (signed) CUSUM statistic,
large positive or negative values of $M_{\tau,b}$ are indicative of the presence of a change-point in the interval $[\tau-b+1, \tau+b]$. In the following, we will call $M_{\tau, b}$ the signed MOSUM statistic, and $|M_{\tau, b}|$ the MOSUM statistic. Thus tests for changes will be based on whether the MOSUM statistic exceeds a given threshold $\lambda$, i.e. whether $|M_{\tau, b}|>\lambda$.

Unlike in the CUSUM statistic, for a fixed interval $[\tau-b+1, \tau+b]$, the simplest version of MOSUM only does a single test, one in which the mean of the data
over $[\tau-b+1, \tau]$ is compared to that over $[\tau+1, \tau+b]$; these two intervals are of the same length. The reason why it is not considered essential to
perform more tests over $[\tau-b+1, \tau+b]$ (as would be the case if CUSUM were to be applied over this interval) is that the testing interval 
$[\tau-b+1, \tau+b]$ will be moved along the domain of the data, which will lead to more tests. (However, there is no reason why the CUSUM statistic could not be applied in a similar fashion on a moving-window basis, leading to more tests on any one interval.)

\begin{figure}[t]
\centering
  \includegraphics[width=0.75\linewidth]{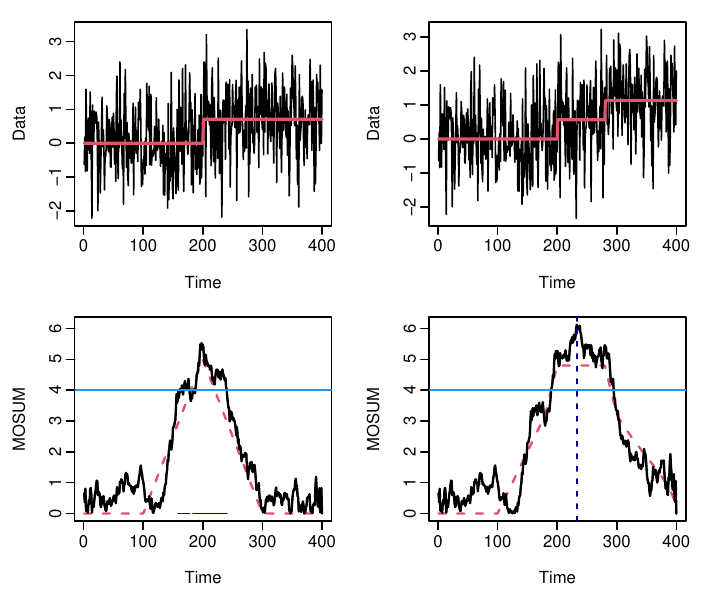}
  \caption{Example simulated data and associated MOSUM statistics. Data (black line) and true mean function (red line) simulated with one change-point (top left) and two change-points (top right). The corresponding MOSUM statistics, for a bandwidth $b=100$, are shown below each data plot. The MOSUM statistics for the data (black line) and the mean function (dashed red line) and an example threshold (horizontal blue line). For the data with one change-point, there are two contiguous regions where the MOSUM statistics exceed the threshold (dark blue lines close to $x$-axis). For the data with two change-points we would estimate a single change-point at the maximum of the MOSUM statistics (vertical blue dashed line).
  \label{ch4:fig:MOSUM1}}
\end{figure}

To gain intuition behind the MOSUM method, it is helpful to look at MOSUM statistics for some data. First consider the left-hand plots in Figure \ref{ch4:fig:MOSUM1}, corresponding to data with a single change at $t=200$. We plot the MOSUM statistics for the mean function (i.e. the ideal case where there is no noise) and for the data. The former is non-zero for values where the MOSUM window contains that change, and has its maximum at the location of the change-point. The MOSUM statistics for the data are then a noisy version of this.

The idea of the MOSUM test is that we can set a threshold for the MOSUM statistics, and we would detect a change-point if the MOSUM statistics exceeds the threshold. However a single change will appear in the windows of multiple MOSUM statistics, and would lead to multiple statistics exceeding the thresholds. If we observed the mean without noise, then we see that a change would lead to a single contiguous region of time where the statistics exceeded the threshold. This suggests constructing such regions of time, estimating a single change within each such region, with the change-point location at the value of time within that region where the MOSUM statistic is largest.

However, if we look at the MOSUM statistics for the data in this example we can see that such an approach may lead to false positives. Due to the noise in the data, and hence the corresponding noise in the MOSUM statistics, a single change may lead to multiple disjoint regions where the statistics exceed the threshold: in the bottom-left plot of Figure \ref{ch4:fig:MOSUM1} we see that the single change-point results in two contiguous regions of time where the MOSUM statistics for the data exceed the threshold. To avoid this, MOSUM algorithms have heuristic rules to prune regions. These are discussed in \cite{mkc21}. One approach is to use the property that multiple regions associated with the same change-point are likely to be very close together. This suggests merging regions that are within a certain distance of each other -- and then estimating a single change within each remaining region.

Thus the general idea is to monitor the MOSUM statistics, calculate regions where these statistics are greater than some threshold, and then estimate a single change-point within each such region. One drawback with this approach is shown in the right-hand plots of Figure \ref{ch4:fig:MOSUM1}, where we show MOSUM statistics for data with two changes. In this case the changes are closer together than the bandwidth. This means that we get just one contiguous region where the MOSUM statistics are greater than the threshold (both for the statistics calculated for the mean function and for the data). Thus we would underestimate the number of changes. Furthermore in this case it also means that the estimated change-point location is poor: for this particular realisation, it is estimated roughly half way between the two changes. This problem is related to the choice of bandwidth, and shows that if we have too large a bandwidth so that two change-points lie within the same window used by the MOSUM statistics, we may not be able to detect both changes. Below, we will see this issue appear in the conditions for the theory of the MOSUM approach. We can avoid this problem by choosing a small enough bandwidth, but that can lose power at detecting small changes. Thus for MOSUM methods, the choice of bandwidth is of particular importance, and we discuss this in more detail later.

Based on the above intuition,
\cite{km11} describe the following algorithm for the estimation of the number $N_0$ and locations $\tau^0_1, \ldots, \tau^0_{N_0}$ of change-points 
via the MOSUM method.
\begin{enumerate}
\item
Fix the bandwidth parameter $b$, a significance threshold $\lambda$ and a gap-correction parameter $0 < \eta < 1/2$.
\item
Calculate $M_{\tau, b}$ for $\tau = b, \ldots, n-b$.
\item
Identify $T := \{  \tau\,:\, |M_{\tau, b}| > \lambda    \}$.
\item
Represent $T = T_1 \sqcup \ldots \sqcup T_K$, where $T_j$ are intervals in $\{1, \ldots, n\}$ such that their lengths are maximum possible, and where $\sqcup$ denotes
disjoint union.
\item
For any pair of consecutive intervals $T_j = [s_j, e_j], T_{j+1} = [s_{j+1}, e_{j+1}]$, if there is a gap between them that is of length $< \eta b$, merge them into a single interval $T_j = [s_j, e_{j+1}]$, set $K := K-1$ and relabel the resulting intervals as $T_1, \ldots, T_K$. Do this iteratively until there is no longer such a pair of neighbouring intervals.
\item
Set $\hat{N} = K$.
\item
For each $j = 1, \ldots, \hat{N}$, define
\[
\hat{\tau}_j = \arg\max_{\tau \in T_j} \,\, |M_{\tau, b}|.
\]
\end{enumerate}

Step 4 of the algorithm is calculating the contiguous regions of time where the MOSUM statistics exceed the threshold, and Step 5 performs the merging of nearby regions. Steps 6 and 7 then estimate the number of change-points as the number of regions, with their locations set to the location of the largest MOSUM statistic within each region.

\cite{km11} are using the following statistic
\[
M_b = \max_{\tau = b, \ldots, n-b} |M_{\tau, b}|
\]
to test for the presence of change-points in $\mathbf{f}$. They show the following result, which we give in a simplified version (only for i.i.d. Gaussian
noise with a known variance).

\begin{theorem}
Let the null hypothesis hold, i.e. $N_0 = 0$. If $b$ is such that $n/b \to \infty$ and $\log^3 n /b \to 0$, then
\[
\alpha(n/b) M_b - \beta(n/b) \to \Gamma
\]
in distribution, where $\Gamma$ is the Gumbel distribution with cdf $F(x) = \exp(-2\exp(-x))$ and
\begin{eqnarray*}
\alpha(x) & = & \sqrt{2 \log(x)},\\
\beta(x) & = & 2 \log(x) + \frac{1}{2}\log\log(x) + \log(3/2) - \frac{1}{2}\log\pi.
\end{eqnarray*}
\end{theorem}

To investigate the power of their test under the alternative $N_0 \ge 1$, \cite{km11} require that true change-points are separated at least by a distance of $2b$, so that each
statistic $M_{\tau, b}$ is computed over a portion of the data containing at most one true change-point. Under this and some other assumptions
related to the minimum jump size, the authors show that their test has asymptotic power one.

We now summarise a consistency result from \cite{km11} regarding the estimated number and locations of change-points, again simplified for Gaussian i.i.d. noise with a known variance.

\begin{theorem}
\label{ch4:th:mosum}
Let the true change-point locations be separated from the adjacent locations, and from the edges of the data, by the distance larger than $2b$, where $b$ is as in
formula (\ref{eq:mosum}). Let the true jump sizes satisfy
\begin{equation}
\label{ch4:eq:a3a}
\frac{1}{\min_{1\le j\le N_0} (f_{\tau_j^0+1} - f_{\tau_j^0})^2} = o_p(b \log^{-1}(n/b)).
\end{equation}
Let $\lambda$, the significance threshold, be of the form
\[
\lambda = \sigma D_n(b, \zeta_n),
\]
where $D_n(b, \zeta_n)$ is a particular sequence in which the leading term is $\sqrt{2\log(n/b)}$ (we refer the reader to the paper for details).
We then have
\begin{eqnarray*}
P(\hat{N} = N_0) & \to & 1,\\
P\left( \max_{1\le j\le N_0} |\hat{\tau}_j \mathbb{I}(j \le \hat{N}) - \tau_j^0| \ge b     \right) & \to & 0.
\end{eqnarray*}
\end{theorem}

\cite{km11} also provide additional results regarding the accuracy of estimated change-point locations. For example, if the number of change-points does not increase with
the sample size, and if $\delta_n^{-2} b^{-1} \to 0$, they show that 
\[
\max_{1\le j\le N_0} |\hat{\tau}_j \mathbb{I}(j \le \hat{N}) - \tau_j^0|  = O_p(\delta_n^{-2}),
\]
where $\delta_n$ is the lower bound on the jump sizes; this rate in general cannot be improved as explained in \cite{fearnhead2022detecting} in the AMOC setting.

To complete the summary of the theoretical analysis in \cite{km11}, we quote their non-asymptotic result given specifically for the i.i.d. noise setting, again with a known variance $\sigma^2$. The result is below.

\begin{theorem} \label{ch4:thm:mosum2}
Let the true change-point locations be separated from the adjacent locations, and from the edges of the data, by the distance larger than $2b$, where $b$ is as in
formula (\ref{eq:mosum}). Let the threshold $\lambda$ be of the form $\lambda = \sigma c_n$, where
\begin{eqnarray*}
c_n & > & \sqrt{4 \log(n)}\\
\sqrt{2}c_n & \le & \eta \delta_n \sqrt{b} - \sqrt{8 \log(n)}.
\end{eqnarray*}
Then, for some constant $C > 0$,
\[
P\left(\hat{N} = N_0;\quad \max_{1\le j\le N_0} |\hat{\tau}_j - \tau_j^0| < b\right) \ge 1 - C/n.
\]
\end{theorem}

The above result provides a relatively clear recipe for the choice of the threshold $\lambda$. 

\begin{figure}[p]
\centering
  \includegraphics[width=0.8\linewidth]{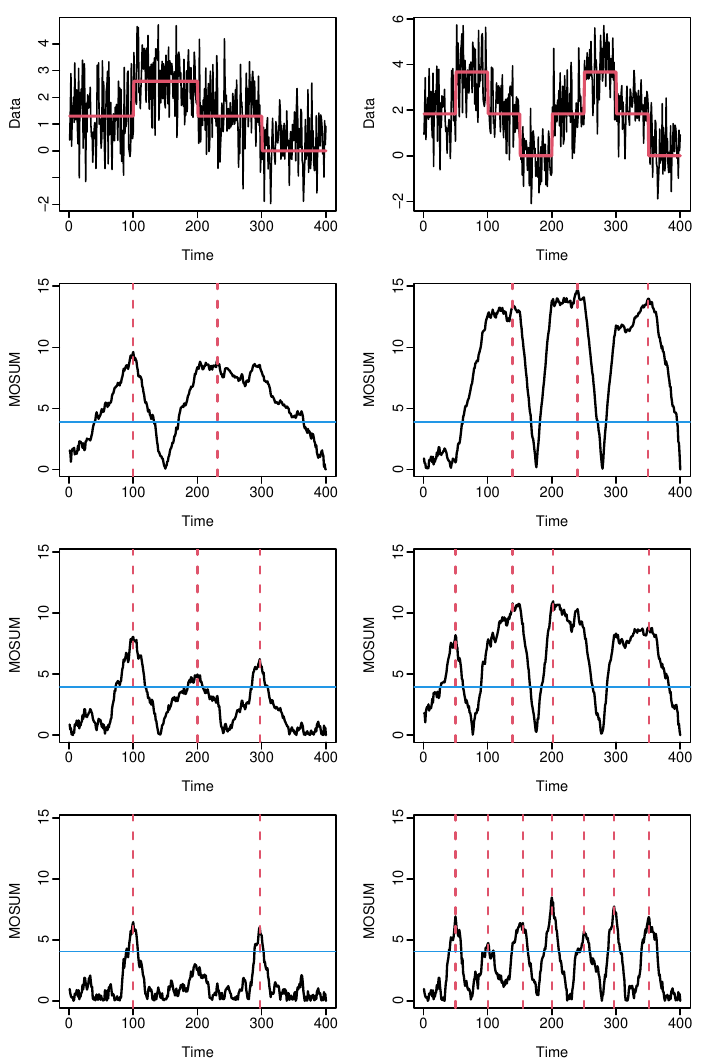}
  \caption{The MOSUM statistic, and associated estimated change-points, for different choices of $b$ and for two datasets: three equally spaced change-points (left-hand column) and seven equally spaced change-points (right-hand column) in data of length $n=400$. Data (black line) and mean function (red line) is shown in the top-row. Other rows show the MOSUM statistics (black line), estimated change-point location (red vertical dashed lines) and threshold used (blue horizontal line), for $b=100$ (second row), $b=50$ (third row) and $b=25$ (fourth row).
  \label{ch4:fig:MOSUM_bw}}
\end{figure}
As mentioned above, the choice of bandwidth, $b$, is particularly important. To further investigate the impact of this choice, we show the results of the MOSUM procedure on two simulated data sets in Figure \ref{ch4:fig:MOSUM_bw}. We consider two data sets, one with 3 change-points each 100 observations apart, and the other with 7 changes each 50 observations apart. We show the MOSUM statistics for $b=100$, $b=50$ and $b=25$ in each case.

There are a couple of observations to make. First we see the potential for loss of power as we reduce the bandwidth, in that the magnitudes of the MOSUM statistics are lower as we reduce $b$. Against this, we see the problem if we have windows that include multiple changes: we see evidence for changes, but it becomes difficult to correctly identify how many change-points there are. In this case we see that the bandwidth suggested by the conditions on e.g. Theorem \ref{ch4:th:mosum}, of half the segment length, produces the best results. 

\begin{figure}[t]
\centering
  \includegraphics[width=0.75\linewidth]{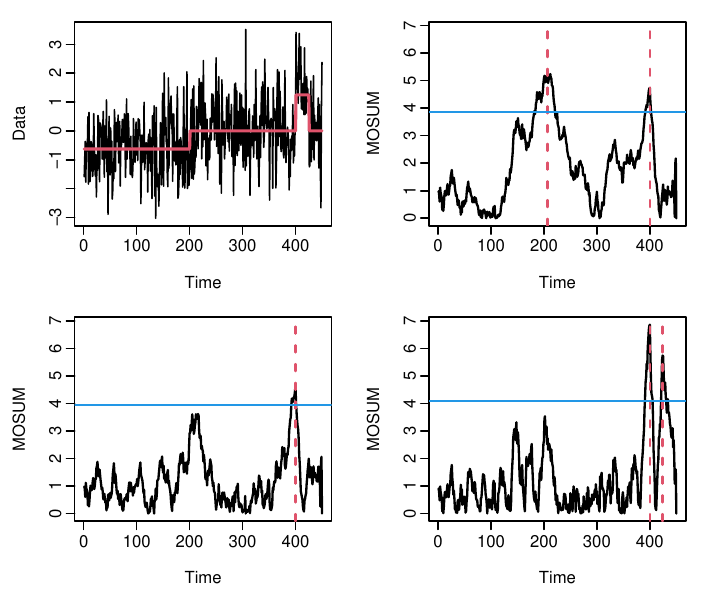}
  \caption{The MOSUM statistic, and associated estimated change-points, for different choices of $b$. Top left plot shows data (black line) and mean function (red line) is shown in the top-row. Other plots show the MOSUM statistics (black line), estimated change-point location (red vertical dashed lines) and threshold used (blue horizontal line), for $b=100$ (top right), $b=50$ (bottom left) and $b=25$ (bottom right).
  \label{ch4:fig:MOSUM_bw2}}
\end{figure}

In practice this can lead to two practical issues. First we will rarely know what is the minimum segment length, and thus what is an optimal choice of $b$. Second, in most applications, segments will vary in length, possibly substantially. Thus choosing a bandwidth that is optimal for one change-point, may mean that the method has little power to detect other changes. A large bandwidth may help us detect a small change that is surrounded by large segments, but will make it hard to detect multiple change-points that are close together; or a small bandwidth may help us to detect large changes that are close together but will lose power for detecting smaller changes. This can be seen in Figure \ref{ch4:fig:MOSUM_bw2}, where we implement the MOSUM statistic with different values of $b$ for data with a small change at $t=200$ and two bigger changes at $t=400$ and $t=450$. A large bandwidth is unable to detect both of the latter changes, whereas a small bandwidth is unable to detect the first change. 

To overcome the challenges of choosing a good bandwidth, and that different changes may need different bandwidth choices, we could try and combine results from applying the MOSUM method for a number of different $b$ values. See \cite{ck20}, \cite{lm21} and \cite{mkc21}. Also there are alternative ways of estimating change-points from the MOSUM statistic, that may overcome some of the problems with detecting multiple changes when we use too large a value for $b$, see \cite{mkc21}.

There are two further comments to make on MOSUM, with regard to how it compares to other methods. First, the choice of bandwidth sets an implicit minimum segment length for the method -- in the sense that the MOSUM approach will not be able to detect changes within a certain distance of each other. If the true segments are all greater than this minimum length, then this implicit assumption can be beneficial, as it can reduce false positives that are close to true changes. This feature is important when interpreting simulation results comparing with methods that do not impose a minimum segment length.

The second is that the seemingly optimal choice of bandwidth for MOSUM is half the segment length. However this choice is not using all the information in the data about the change -- which would come from a window that uses all data from the segments on either side of the change. Other methods have the opportunity to use all this information. For example methods that estimate changes by minimising a penalised cost jointly estimate all changes, and in some sense are deciding on whether there is a change at some time $t$ based on all relevant data given the estimates of where other changes occur. Binary Segmentation also adapts the windows it looks at based on previously estimated changes. Similarly, Wild Binary Segmentation is able to sample windows of data that include all, or most of the segments either side of the change. However, these alternative methods have their own drawbacks: binary segmentation can struggle due to analysing windows of data that contain multiple changes; and Wild Binary Segmentation's performance can depend on how many and which windows of data are chosen.

\begin{figure}[t]
\centering
  \includegraphics[width=0.75\linewidth]{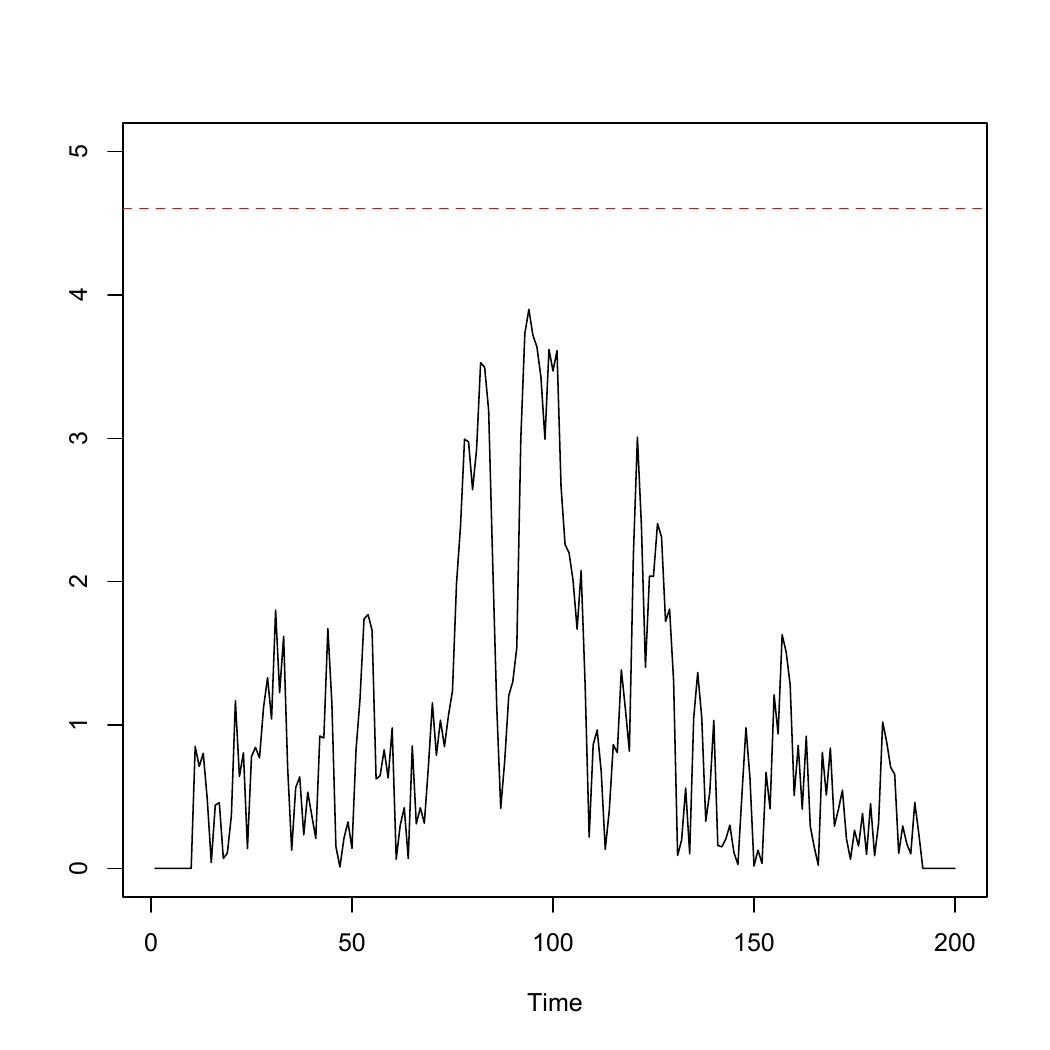}
  \caption{The absolute MOSUM statistic of the noisy signal from Figure \ref{ch4:fig:02}, with bandwidth $b=10$, padded with zeros at both edges. Red: the significance threshold $\sqrt{4\log(200)}$.
  \label{ch4:fig:03}}
\end{figure}

To see the potential loss of power, consider performance of MOSUM on the noisy signal from Figure \ref{ch4:fig:02}. We choose the most favourable value of $b$, equal to half the minimum distance between each pair of consecutive change-points; therefore $b = 10$. Figure \ref{ch4:fig:03} shows that the MOSUM statistic on this signal appears to correctly peak around times $80, 100$ and $120$, which correspond to the true change-point locations. However, even the lowest permitted significance threshold is too high for the procedure to judge the peaks as significant. This is in contrast to the performance for e.g. Wild Binary Segmentation, which was able to detect the changes, with high probability, provided it sampled appropriately large windows around each change.



\subsection{Isolate-Detect}
\label{ch4:sec:id}

The Isolate-Detect method of \cite{af18} is a pseudo-sequential procedure whose basic idea is that the CUSUM statistic is successively computed on $X_{1:k\lambda_n}$ for 
some integer $\lambda_n$, for $k = 1, 2, \ldots$, until detection takes place or until the entire dataset has been considered in this way. Detection is defined as the CUSUM exceeding a certain threshold $\zeta_n$. On detection, the procedure restarts anew
\begin{description}
    \item[(Version 1)] either from the location just after the CUSUM-maximising estimated change-point location, or
    \item[(Version 2)] from the end of the interval on which the detection took place.
\end{description}

Both versions have their pros and cons. In version 1, the procedure restarts from as early as it is able to. This gives it as much data as possible to carry out further scanning of the dataset. However, the price to pay is that if the estimated location of the first change-point falls before its true location, the procedure will then again examine a section of the data containing the same true change-point. This may lead to double detection, or may impact the accuracy in estimating the change-point that follows.

By contrast, in version 2, the procedure deliberately ``cuts out'' from further scanning the portion of the data between the estimated location of the first change-point and the end of the interval on which it was detected. This is to ensure that the same change-point does not get detected again, and (under suitable assumptions on the signal strength, on an event with a high probability) to guarantee that the detection of the following change-point (if any) takes place on an interval that does not contain any other change-points. It is this ``isolate and detect'' mechanism that gives the procedure its name.

The complete Isolate-Detect methodology is more complex: the authors advocate alternate application from the left and from the right, and various model selection criteria beyond thresholding are proposed. However, the main message is that  under certain assumptions on the signal strength and the minimum spacing between change-points, it is possible to guarantee, with probability tending to one with the sample size, that with a suitable threshold $\zeta_n$, each detection interval (in version 2 above) contains only one true change-point, and that the number and locations of change-points get estimated consistently.

\begin{figure}[t]
\centering
  \includegraphics[width=0.75\linewidth]{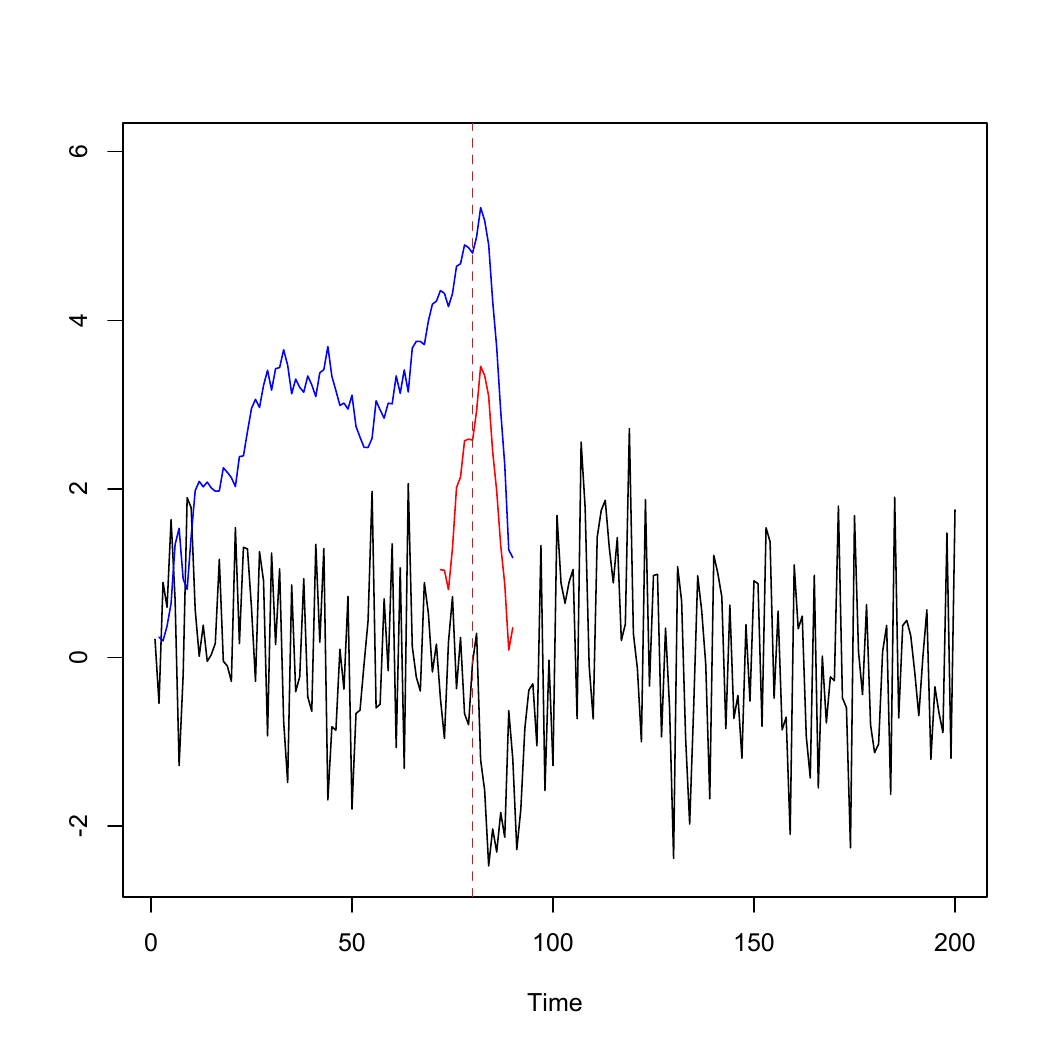}
  \caption{Black: noisy signal $\mathbf{X}$. Blue: $C_{1,b,90}(\mathbf{X})$. Red: $C_{71,b,90}(\mathbf{X})$. Dashed brown: 80, the location of the first true change-point.
  \label{ch4:fig:04}}
\end{figure}

To contrast Isolate-Detect and MOSUM, we return to the signal example from Figure \ref{ch4:fig:02}, and plot, in Figure \ref{ch4:fig:04}, the CUSUM statistic on $X_{1:90}$ as well as the CUSUM statistic on $X_{71:90}$; note that $C_{71,80,90}(\mathbf{X}) = M_{80,10}$. With the location of the first true change-point at 80, observe that the maximum of $C_{1,80,90}(\mathbf{X})$, occurring at $\hat{\tau}_1 = 82$, is much larger than that of $C_{71,80,90}(\mathbf{X})$, due to the fact that the inspection starts much earlier. While this in itself does not mean it is a ``better'' detection statistics (because significance thresholds will be different in the Isolate-Detect and MOSUM methods), Table 1 in \cite{fls20} suggests that $C_{1,b,90}(\mathbf{X})$ would have led to detection at 5\% significance level.

It is simplest version, the Isolate-Detect methods would then continue its inspection by successively considering CUSUM statistics as follows:

\begin{description}
    \item[(Version 1)] on $X_{83:(82+k\lambda_n)}$, or
    \item[(Version 2)] on $X_{90:(89+k\lambda_n)}$.
\end{description}

The Isolate-Detect methodology also applies to other signal classes (e.g. piecewise polynomials); see \cite{af18} for details.

\section{\texorpdfstring{$L_1$}{L1}-penalised Methods}
\label{ch4:sec:l1}

Given the multiple change-point model (\ref{ch4:eq:univ_mult}) and the problem of estimating $\mathbf{f}$, 
let $\mathbf{f}'$ be a candidate solution. In this section, we will need to quantify the behaviour of the first difference
of $\mathbf{f}'$, and for this purpose we introduce some notation first. For any vector $\mathbf{v} = (v_1, \ldots, v_n)^T$,
we define
\begin{eqnarray*}
\Delta\mathbf{v} & = & (v_2-v_1, \ldots, v_{n}-v_{n-1})^T\\
\| \mathbf{v}  \|_p & = & \left(\sum_{i=1}^n |v_i|^p\right)^{1/p},\quad 1 \le p < \infty\\
\| \mathbf{v} \|_\infty & = & \max_{i=1,\ldots,n} |v_i|\\
\| \mathbf{v} \|_0 & = & |\{ i\,\,: v_i \neq 0  \}|,\\
\end{eqnarray*}
with $|\cdot|$ denoting the cardinality of a set. For $p \ge 1$, not necessarily finite, $\| \mathbf{v}  \|_p$ is the $l_p$ norm of the
vector $\mathbf{v}$, while $\| \mathbf{v} \|_0$ is not a norm, but it is still a frequently used functional quantifying the magnitude (understood in the sense of non-zeroness)
of $\mathbf{v}$.  The residual sum of squares in (\ref{ch4:sn}) is defined
as a function of the postulated change-point locations, but we can alternatively define the residual sum of
squares with respect to any candidate fit $\mathbf{f}'$ as
\[
\bar{S}_n(\mathbf{f}') = \sum_{i=1}^n (X_i - f_i')^2.
\]
A similar reparametrisation of the minus log-likelihood function $-l(\cdot)$ is possible. Many of the penalised
estimators of the number $N_0$ and the locations $\tau^0_1, \ldots, \tau^0_{N_0}$ of change-points -- or, equivalently, of $\mathbf{f}$ -- introduced in Sections
\ref{ch4:sec:n0unknown}--\ref{ch4:sec:other} had one of the two forms,
\begin{eqnarray*}
\hat{\mathbf{f}}_{LS, 0} & = & \arg\min_{\mathbf{f}'} \{\bar{S}_n(\mathbf{f}') + 2\sigma^2 C_n \|\Delta \mathbf{f}'\|_0\}\quad \text{(penalised least squares), or}\\
\hat{\mathbf{f}}_{LL, 0} & = & \arg\min_{\mathbf{f}'} \left\{ \frac{n}{2}\log\left(\frac{1}{n}   \bar{S}_n(\mathbf{f}')  \right) + C_n \|\Delta \mathbf{f}'\|_0\right\}\,\,\, \text{(penalised minus log-likelihood)},
\end{eqnarray*}
for a certain sequence of constants $C_n$; for example, we had $C_n = \log\,n$ in SIC.

Before the recent computational advances, described in Section \ref{ch4:sec:comp}, which permitted fast computation
of estimators in which the penalty was a linear function of $\|\Delta \mathbf{f}'\|_0$,
the fastest available algorithms
were capable of computing estimators penalised in this way in computational
time $O(n^2)$. To reduce this complexity, several authors explored the idea of replacing the $\|\Delta \mathbf{f}'\|_0$ term in
the penalty with penalties linear in the $L_1$ norm of $\Delta \mathbf{f}'$, also referred to as the total variation \citep{rof92} or the fused
lasso penalty \citep{tsrz05}, the reason being that $\|\Delta \mathbf{f}'\|_1$ is a convex relaxation of 
$\|\Delta \mathbf{f}'\|_0$, which meant there was hope of achieving fast computation.
For example, \cite{fhht07} propose a fast descent-fusion-smoothing algorithm for finding solutions to the $L_1$-penalised problem
\begin{equation}
\label{ch4:eq:flasso}
\hat{\mathbf{f}}_{LS, 1} = \arg\min_{\mathbf{f}'} \{\bar{S}_n(\mathbf{f}') + 2\sigma^2 C_n \|\Delta \mathbf{f}'\|_1\}
\end{equation}
in the context of signal approximation (i.e. without the change-point detection context in mind).
This algorithm is
shown to offer substantial computational gains with respect to the quadratic programming approach considered
in \cite{tsrz05}.

We now briefly review the relevant literature related to $L_1$-penalisation in the context of change-point detection.
\cite{hll10} rewrite the problem (\ref{ch4:eq:flasso}) in the form
\begin{equation}
\hat{\mathbf{g}} = \arg\min_{\mathbf{g}'}  \sum_{i=1}^n \left(X_i - \sum_{k=1}^i g'_k\right)^2 + 2\sigma^2 C_n \|  \mathbf{g}'  \|_1,
\end{equation}
where $g'_1 = f'_1$ and $g'_i  = f'_i - f'_{i-1}$ for $i > 1$. This re-parametrisation means that
they move to the context of sparse regression in which most $g'_i$'s are zero and any that are not, with
the exception of $g'_1$, correspond to change-points in the candidate function $\mathbf{f}'$. This change of variables
enables them to use an adaptation of the LARS algorithm of \cite{ehjt04} to obtain a solution in 
computational time $O(N_U^3 + N_U n \log\,n)$, where $N_U$ is a known upper bound on the number of change-points.
\cite{hll10} provide partial consistency results for their
proposed estimator, which include uniform convergence of the estimated change-point locations to the truth
at near-optimal rates under the assumption that the number of change-points $N_0$ is estimated accurately with 
probability tending to one, as well as convergence to zero of an asymmetric version of the Haussdorf distance
between the estimated change-point locations and the truth if the number of change-points is over-estimated. Note, however, that the authors do not specify how to estimate the number of change-points in this framework; we argue later in this section why the $L_1$-penalised approach may be suboptimal for this task.
Related consistency results appear in the unpublished work \cite{rw14}.

The ``taut string'' method for finding the unique solution to (\ref{ch4:eq:flasso}) proceeds as follows. Consider the integrated
process
\[
X^0_t = 2 \sum_{s=1}^t X_s,\quad t = 0, \ldots, n,
\]
with the convention that the sum is zero if the lower summation limit is larger than the upper. Consider the upper and lower bounds
on $X^0_t$ defined by, respectively,
\begin{eqnarray*}
L_t & = & X^0_t - 2\sigma^2 C_n\\
U_t & = & X^0_t + 2\sigma^2 C_n.
\end{eqnarray*}
Take a piece of string attached at one end at the point $(0, 0)$ and at the other end at $(n, X_n^0)$, and constrained to lie between
$L_t$ and $U_t$. The string, pulled until it is taut, is denoted by $S_t^0$. \cite{mg97} show that the first difference of $S_t^0$
essentially (disregarding the edge effects) minimises criterion (\ref{ch4:eq:flasso}). \cite{dk01}, which is where the name ``taut string''
appears to originate from and where this algorithm is looked at from the perspective of local peak/trough detection, also provide a brief history
of related algorithms in other statistical contexts (albeit not in the context of change-point detection). Their work gives the computational
complexity of the taut string algorithm as $O(n)$.

There is a separate strand of literature which explores the conditions under which solutions to the problem (\ref{ch4:eq:flasso}) identify
the change-point locations exactly, i.e. with no error, which appears to be motivated by an earlier analogous discussion for the lasso.
Papers in this category include \cite{r09} (with a subsequent correction published on the author's web page) and \cite{qj16}. This necessarily
happens under rather strong conditions, as we saw earlier
that error-free recovery is in normal circumstances
impossible. For this reason, we do not discuss these results here.

\begin{figure}[t]
\centering
\includegraphics[scale=0.9]{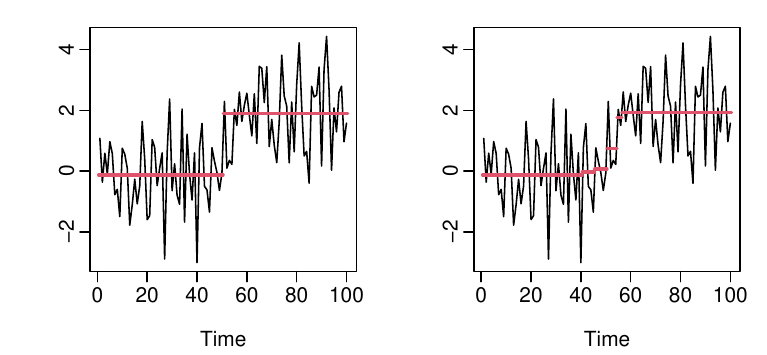}
\caption{The challenge of estimating the number of change-points with $L_1$ penalisation. Simulated data with a change in mean from 0 to 2 at time 50. The left-hand plot shows the best fitting model with a single change at the true change-point location. The right-hand plot shows a better fitting model with a ``staircase'' effect around the change-point: the fitted mean increases monotonically. The $L_1$ penalty for the models in both plots is the same, so the right-hand fit will be preferred as it has used its additional flexibility to fit to the noise around the change-point.
\label{ch4:fig:L1}}
\end{figure}

We now provide a simple heuristic argument for why the $L_1$-penalised approach may not be optimal
in the context of change-point detection. Consider again the problem (\ref{ch4:eq:flasso}), which minimises the
criterion
\[
{\mathcal F}(\mathbf{f}') = \bar{S}_n(\mathbf{f}') + 2\sigma^2 C_n \|\Delta \mathbf{f}'\|_1.
\]
Let the true signal $\mathbf{f}$ have a single change-point, e.g. $f_i = 0$ for $i = 1, \ldots n/2$ and $f_i = 1$ thereafter.
Note that the minimisation of ${\mathcal F}(\mathbf{f}')$ in no way discourages solutions with many change-points. For
example, consider a candidate signal $\mathbf{f}'$, which is non-decreasing (as is the true signal $\mathbf{f}$) but in such
a way that it forms a ``staircase'' starting at or around 0, and ending up at or around 1; if $\| \Delta \mathbf{f}'\|_1 \le \| \Delta \mathbf{f}\|_1 = 1$
(note that $\| \Delta \mathbf{f}'\|_1$ is simply the sum of all jump sizes for the non-decreasing signal $\mathbf{f}'$), then
chances are that we will be able to choose the locations of the change-points in $\mathbf{f}'$ so that it will provide a better fit to the data
than $\mathbf{f}$ itself, or in other words we will have $\bar{S}_n(\mathbf{f}') < \bar{S}_n(\mathbf{f})$. It is easy to imagine this
happening especially given that we do not have an a priori restriction on the number or locations of change-points in $\mathbf{f}'$. As 
a consequence, this will lead to ${\mathcal F}(\mathbf{f}') < {\mathcal F}(\mathbf{f})$ and we will end up preferring a solution with
multiple change-points, in a situation in which there is only one true one. An example of this problem is shown in Figure \ref{ch4:fig:L1}. The key point here is that the $L_1$ penalty $2\sigma^2 C_n \|\Delta \mathbf{f}'\|_1$ is insufficiently discouraging of solutions with many change-points. A related discussion under the heading of a ``staircase problem'' appears
in \cite{rw14}.

In light of this issue, consistent estimation of the number of change-points via $L_1$ penalisation necessarily requires 
further steps beyond merely minimising (\ref{ch4:eq:flasso}). With this in mind, a number of authors propose post-processing
algorithms for solutions to this minimisation problem, which hopefully sift out the spurious change-points and only retain the 
significant ones. 
\cite{lsrt17} propose a generic post-processing procedure for any $L_2$-consistent,
piecewise-constant estimator of $f_t$, including in particular the one discussed here, which yields consistency for the estimators of $\tau^0_i$ under certain assumptions
on the distances between the change-points and the jump sizes. Similar results for 
general ``trend filtering'' \citep{t14} estimators appear in \cite{glcs18}. A related post-processing
device is proposed in \cite{z19}.

\section{Estimating the Noise Variance} \label{ch4:sec:noise_variance}

All methods we have introduced involve some threshold or penalty that impacts on the number of change-points that we estimate. Apart from methods that choose the change-points based on the SIC under a model with unknown variance (see Section \ref{ch4:sec:sic}), these require an estimate of the noise variance of the data. 

\begin{figure}[t]
\centering
  \includegraphics[width=\linewidth]{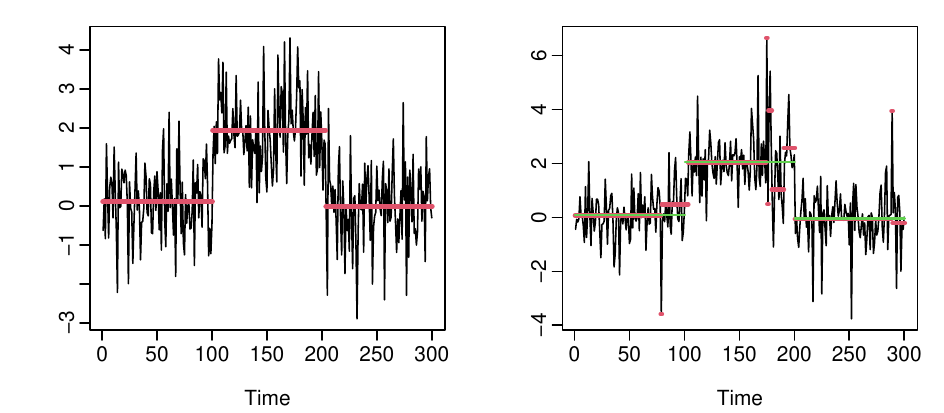}
\caption{Example of estimating change-points using the MAD estimator of the noise variance. The left-hand plot shows data simulated with independent standard Gaussian noise and two change-points. The MAD estimator, $\hat{\sigma}^2$ of the noise variance is $0.94$, and an estimator based on minimising the residual sum of squares plus a $2\hat{\sigma}^2\log(n)$ penalty for each change-point correctly identifies the changes (red line shows fitted mean). The right-hand plot shows data simulated with Laplacian noise with variance 1. In this case the MAD estimator, $\hat{\sigma}^2$ of the noise variance is $0.64$. The estimated change-points based on minimising the same penalised residual sum of squares leads to over-estimating the number of change-points due to the under-estimation of the noise variance (red line). In this example, estimating the noise variance after a preliminary estimate of the mean using a moving median filter with bandwidth $h=32$ gives an over-estimate of the noise variance of $1.14$. Using this value we correctly estimate the change-points (green line). \label{ch4:fig:sig_est}}
\end{figure}

How can we estimate the noise variance? If we knew the change-point locations this would be straightforward, as we could estimate it based on the variance of the residuals of the data from the segment means. Whilst we can estimate the change-point locations, this requires us to know the noise variance. Also an iterative approach where we guess the noise variance, estimate the change-points, used the fitted model to estimate the noise variance and then re-estimate the change-points with this new estimated noise variance, often works poorly. An incorrect initial guess can often lead to a fit to the data with the wrong number of change-points so as to get a residual variance that is close to our initial guess.

The common way around this is to consider the first difference of the data, that is $Y_i=X_{i+1}-X_i$, for $i=1,\ldots,n-1$. Unless there is a change-point between $i$ and $i+1$, then $X_i$ and $X_{i+1}$ are independent random variables with the same mean and variance $\sigma^2$. Thus $Y_i$ would have mean 0 and variance $2\sigma^2$. There will be some $Y_i$ values which have non-zero mean because there is a change-point at $i$, but assuming these are rare we can estimate the variance of $Y_i$ using some robust estimator. The most common estimate is the so-called MAD estimator \citep{rc93_4}. This calculates the Median Absolute Deviation of $Y_{1:n-1}$ from its median, and then re-scales this value to give an estimator of the standard deviation of $Y_i$ under the assumption that they are normally-distributed. We then re-scale this by dividing by $\sqrt{2}$ to get an estimate of $\sigma$. Thus our estimator is
\[
\hat{\sigma}= \frac{1}{\sqrt{2}} \textrm{MAD}(Y_{1:n-1}),
\]
where
\[
\textrm{MAD}(Y_{1:n-1}) = c\times\textrm{median}(Y_{1:n-1}-\textrm{median}(Y_{1:n-1})),
\]
with $c=1.483$, such that $1/c$ is the median of the absolute value of a standard Gaussian random variable.

This estimator works well providing the frequency of change-points is low, and that the i.i.d. Gaussian assumption for the noise holds. However, care should be taken as the latter assumption is often violated in practice. If the noise of the data is auto-correlated then, if there is no change-point between $i$ and $i+1$, 
\[
\mbox{Var}(Y_i)=\mbox{Var}(X_{i+1}-X_i)=\mbox{Var}(X_{i+1})+\mbox{Var}(X_i)-2\mbox{Cov}(X_{i+1},X_i)=2\sigma^2(1-\rho_1),
\]
where $\rho_1$ is the lag-1 autocorrelation of the noise. This will lead to under-estimation of the noise variance by a factor $(1-\rho_1)$. Similarly, even if the noise is independent, if it is heavier tailed than Gaussian the choice of $c=1.483$ in the definition of the MAD estimator will be inappropriate and will also lead to under-estimating the noise variance. In both cases this can lead to over-estimation of the number of change-points. 

To see this in practice we simulated data with $n=300$ and two change-points, at 100 and 200 respectively. We then considered two scenarios, one with independent standard Gaussian noise, and 
one with independent Laplacian noise with variance 1. In each case we estimated the noise variance using the MAD estimator, and then estimated the change-points by minimising a penalised version of the residual sum of squares with a penalty $2\hat{\sigma}\log (n)$. An example of the result is given in Figure \ref{ch4:fig:sig_est}. Whilst we correctly estimate the change-points in the Gaussian case, we under-estimate the noise variance substantially in the Laplacian noise case, leading to substantial over-estimation of the number of change-points. To check this is as a result of mis-estimation of the noise variance, as opposed to the heavier-tailed noise, we also estimated the change-points using the correct noise variance -- and correctly estimated the two change-points.

This example is consistent across multiple simulations. Over 100 simulations, for the Gaussian noise case we correctly estimated there were 2 change-points in 99 cases. For the Laplacian noise case we only correctly estimated there were 2 change-points in 29 cases, and on average estimated close to 5 change-points per dataset. Estimating the change-points with the Laplacian noise but using the true noise variance led to us estimating the correct number of change-points 63 times, with an average of just under 3 changes detected per dataset. This points to some of the loss of accuracy being due to the noise being heavier-tailed than our change-point method assumes, but that mostly is due to the tendency to under-estimate the noise variance.

An alternative approach, that avoids these problems, is to use a non-parametric method to estimate the mean function, and then to estimate $\sigma^2$ based on the variance of the residuals from the estimated mean. A natural non-parametric estimator is given by the moving median filter. That is we fix a bandwidth $h$ and estimate the mean at time $i$ by the median of the data within $h$ of $i$. Provided $h$ is smaller than the minimum segment length, the empirical median of the data window around $i$ should approximate the mean at $i$, as more than half the data-points considered when estimating the mean at $i$ will be from the segment containing $i$. The challenge is choosing $h$ small enough to satisfy this constraint, but large enough that the estimate of the mean function is reasonably accurate. For our simulation study above, estimating the variance for the Laplacian noise case by this approach with $h=32$ gave estimates of the change-points that were identical to those obtained using the true noise variance in all 100 simulations. 

We discuss this issue further in Section \ref{ch4:sec:well-log}.

\section{Adaptive Estimation of \texorpdfstring{$N_0$}{the Number of Change-points} Based on Output of Solution Path}
\label{ch4:sec:adapt}

The main model selection approaches described in this chapter so far can be referred to as non-adaptive, in the sense that their tuning parameters do not get chosen based on the data. For example, the thresholding approach, applicable (amongst others) in (wild) binary segmentation, used the ``universal'' threshold of the form $C\sigma\sqrt{\log(n)}$, in which the only data-dependent parameter is the data length $n$ (the constant $C$ does not change across datasets), and potentially an estimate of $\sigma$ as described in the previous section. Similarly, in the penalised approach, the SIC penalty is of the form $N' \log(n)$, where again the postulated number of change-points $N'$ in the functional form of the penalty is the same regardless of the dataset to which the penalty is applied.

However, there are some approaches in the literature, referred to as adaptive here, which attempt to infer the amount of regularisation in the multiple change-point problem from the data. More specifically, most of them work by examining the shape of the goodness-of-fit of each proposed change-point model as a function of the postulated number of change-points. In these approaches, briefly reviewed below, the goodness-of-fit of each postulated change-point model tends to be measured by either the residual sum of squares (or more generally: by the minus log-likelihood) or by a CUSUM-based measure of prominence of each additional change-point added to the model. In one way or another, these adaptive methods look for a ``drop'' or a ``bend'' in the thus-constructed goodness-of-fit as a function of the number of change-points, indicating that once all the relevant change-points have been added to the model, the increase in the goodness-of-fit coming from any additional change-point tails off, leading to a ``broken stick''/``bent elbow'' shape of the goodness-of-fit. Therefore, any model selection criterion that attempts to exploit this phenomenon can be informally referred to as an ``elbow criterion'', a term that occasionally appears in the literature in other contexts related to tuning parameter selection. We describe two automated implementations of this idea, but an example where we attempt to detect the ``elbow" by eye is given in Section \ref{ch4:sec:robust}.

\subsection{Dimension Jump, Slope Estimation and Lavielle's Formalisation}
\label{ch4:sec:dj}

\cite{bm01} and \cite{bm06} propose the adaptive choice of what they refer to as the minimal penalty, in the context of penalised Gaussian model selection in which the cost function
is the $L_2$ risk, via an algorithm referred to in later works as ``dimension jump''. The main idea of dimension jump is as follows. Consider
a family of penalties of the form $CD_m / n$, where $D_m$ is the dimensionality of the space of candidate models (in our context, this can simply be thought of as 
the postulated number of change-points). Heuristically, too small a choice of the constant $C$ when minimising the corresponding penalised $L_2$ risk criterion will lead 
to models with a lot of change-points, whereas a large choice (or indeed too large a choice) will reduce this considerably. Therefore the dimension jump 
approach looks at the graph of the function $C/\sigma^2 \mapsto D_{\hat{m}}$ and chooses the constant $C$ that corresponds to the largest jump in this
function, hoping that this will locate the ``sweet spot'' between penalties that are too small and those that are too large.

In the approach termed ``slope estimation" \citep{l05} the observation is that the optimal penalty for Gaussian model selection under the $L_2$ risk, in the sense
of \cite{bm06}, depends on the unknown parameter $\sigma^2$. For any postulated number of change-points $N'$, let $\hat{f}_t^{N'}$ be the best
fit to the data in the $L_2$ sense that contains exactly $N'$ change-points. For $N'$ large enough, the function
$N' \mapsto n^{-1}\sum_{t=1}^n (X_t - \hat{f}_t^{N'})^2$ is approximately linear with slope $-\sigma^2 / n$, which can in principle be used to estimate
$\sigma^2$ and then be plugged into the optimal penalty; \cite{l05} does this via the dimension jump approach, but \cite{a19} advocates robust regression.
More specifically, \cite{l05}'s criterion is
formed of the least-squares cost plus a penalty of the form
\[
\mbox{Leb}_{\mbox{pen}}(N') = \frac{N'+1}{n} \sigma^2 \left(  c_1 \log\left(  \frac{n}{N'+1}  + c_2 \right)    \right).
\]
It is interesting to observe that this penalty, in contrast to the majority of the literature, is not linear in the number of change-points. It is formulated for a known $\sigma^2$ and
uses two general constants $c_1, c_2$. Based on a simulation study, Lebarbier proposes the universal choice $(c_1, c_2) = (2, 5)$.

Besides \cite{l05}, specific algorithmic ideas regarding the implementation of dimension jump and slope estimation in the context of the multiple
change-point problem appear, amongst others, in \cite{bmm12}.
 \cite{a19} is a comprehensive review
article on minimal penalties and the slope heuristics, which describes these two approaches and their variants in detail,
including an unpublished two-stage refinement by Rozenholc, branded ``statistical base jumping".


\cite{l05a}, motivated by the problem of selecting a suitable penalty constant
from the data, proposes a heuristic algorithm for estimating the number of change-points,
which examines the second differential of the empirical loss of the piecewise-constant model
fit as a function of the number of change-points. The approach requires the provision of the
maximum number of change-points and a threshold parameter whose value appears critical
to the success of the procedure; the theoretical properties of the method are not investigated.

\subsection{Steepest Drop to Low Levels}
\label{ch4:sec:sdll}

Another formalisation of the ``elbow criterion'' is given by the Steepest Drop to Low Levels (SDLL) criterion of \cite{f20}. To undestand its mechanics, consider initially two noisy datasets simulated as below.

\begin{verbatim}
mscale <- c(rep(0, 150), rep(5, 30), rep(4, 120))
set.seed(1)
noise <- rnorm(300)
mscale.easy <- mscale + 2 * noise
mscale.difficult <- mscale + 3 * noise
\end{verbatim}

\begin{figure}[p]
\centering
\begin{minipage}{.49\textwidth}
  \centering
  \includegraphics[width=\linewidth]{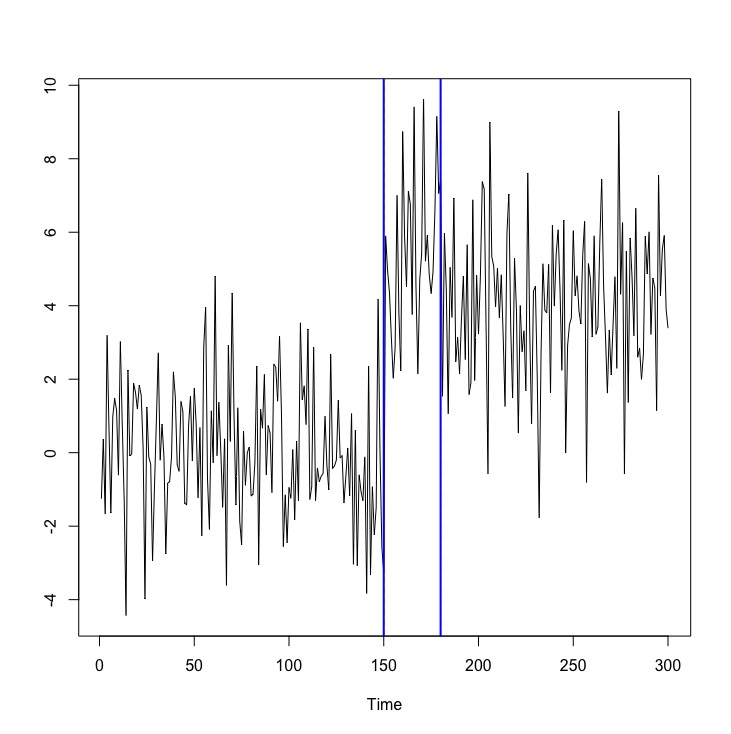}
\end{minipage}
\begin{minipage}{.49\textwidth}
  \centering
  \includegraphics[width=\linewidth]{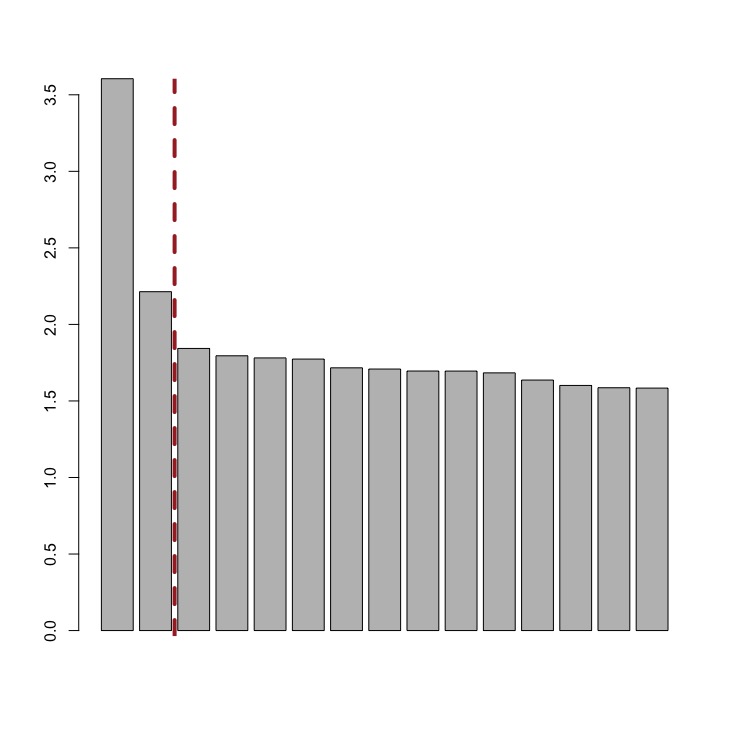}
\end{minipage}\\
\begin{minipage}{.49\textwidth}
  \centering
  \includegraphics[width=\linewidth]{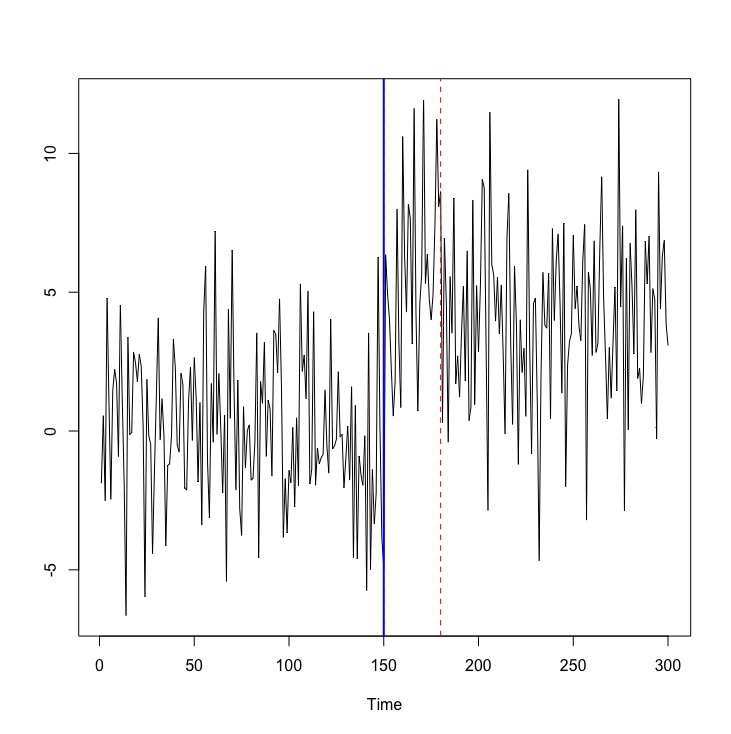}
\end{minipage}
\begin{minipage}{.49\textwidth}
  \centering
  \includegraphics[width=\linewidth]{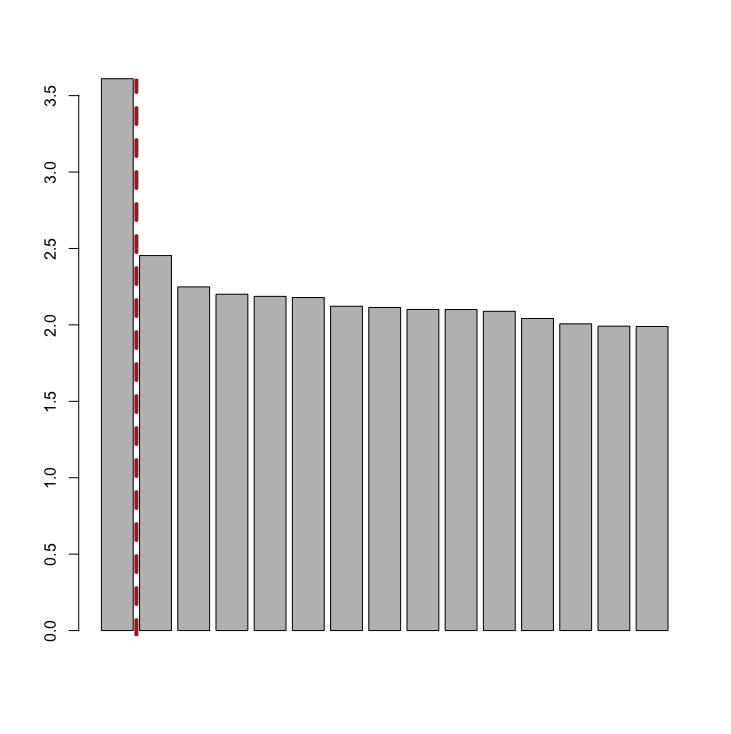}
\end{minipage}
\caption{Left column: a noisy signal $\mathbf{X}$ (black) generated as {\tt mscale.easy} (top) and {\tt mscale.difficult} (bottom; see definition within text). Change-points estimated via SDLL as implemented in {\tt breakfast::model.sdll}, acting on the WBS2 solution path, in blue. 
True change-points marked as dashed red lines, note each blue line overlaps with a dashed red.
Right column: logged CUSUMs corresponding to the 15 largest elements of the WBS2 solution path of $\mathbf{X}$. The brown dashed line shows where SDLL identifies the ``steepest drop to low levels'', thereby estimating the number of change-points.\label{ch4:fig:sdll_01}}
\end{figure}

In the above code snippet, {\tt mscale} is an abbreviation for `multiscale'; indeed, the change-points in the {\tt mscale} signal appear to live on two different scales -- the change-point at 150 is more prominent than the change-point at 180. One challenge for a change-point detection procedure here will be to choose between a model with 1 and 2 change-points, based on a noisy realisation of {\tt mscale}. The sample paths denoted by {\tt mscale.easy} and {\tt mscale.difficult} are shown in Figure \ref{ch4:fig:sdll_01}.

\begin{figure}[p]
\centering
\begin{minipage}{.49\textwidth}
  \centering
  \includegraphics[width=\linewidth]{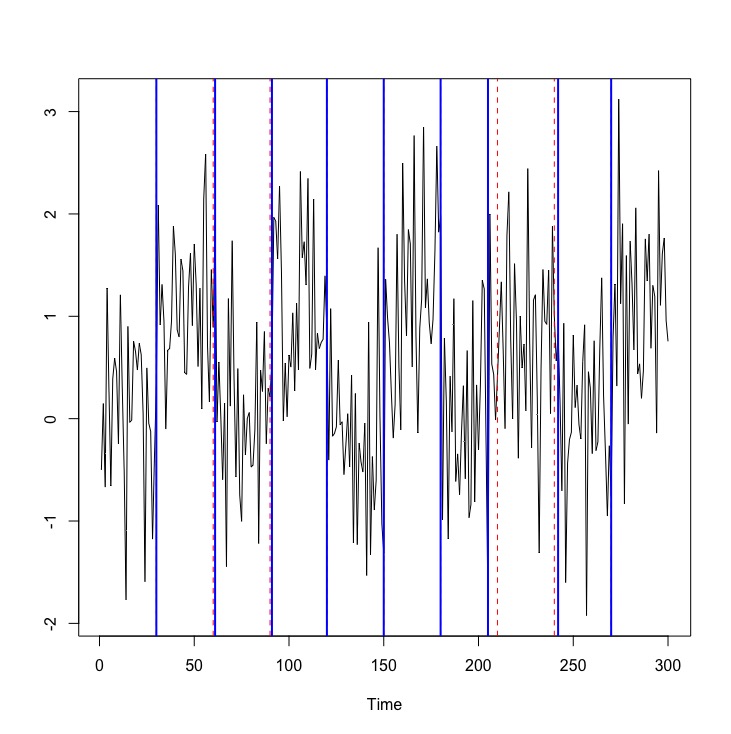}
\end{minipage}
\begin{minipage}{.49\textwidth}
  \centering
  \includegraphics[width=\linewidth]{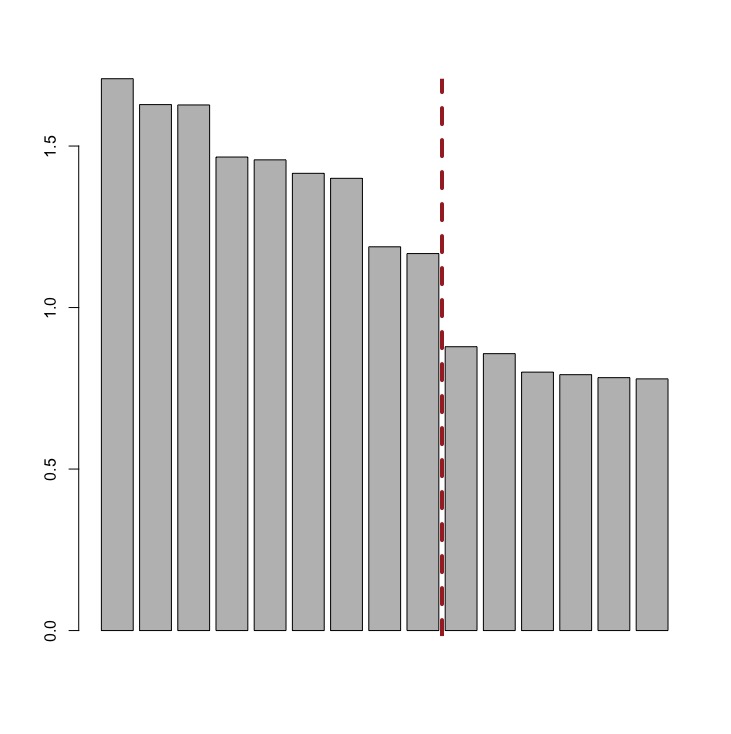}
\end{minipage}\\
\begin{minipage}{.49\textwidth}
  \centering
  \includegraphics[width=\linewidth]{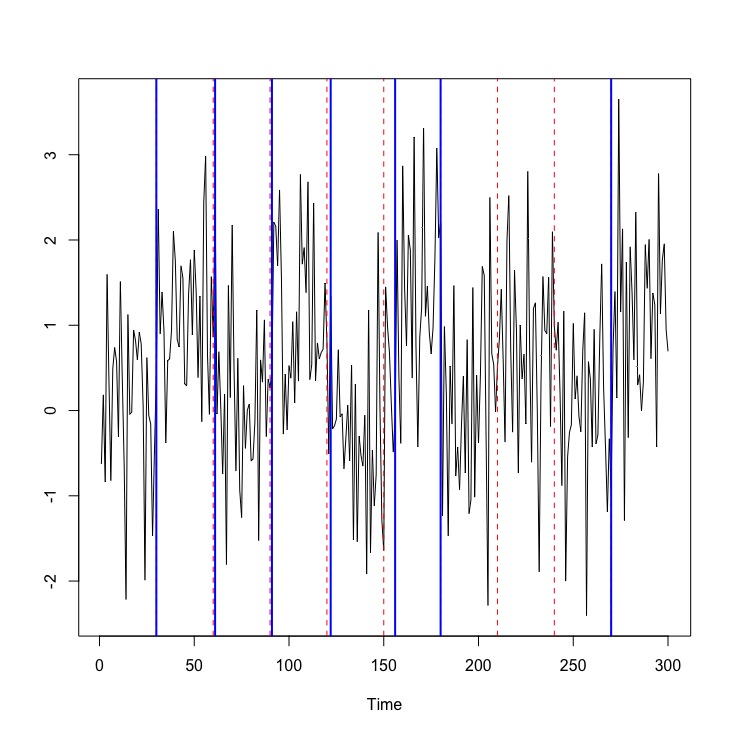}
\end{minipage}
\begin{minipage}{.49\textwidth}
  \centering
  \includegraphics[width=\linewidth]{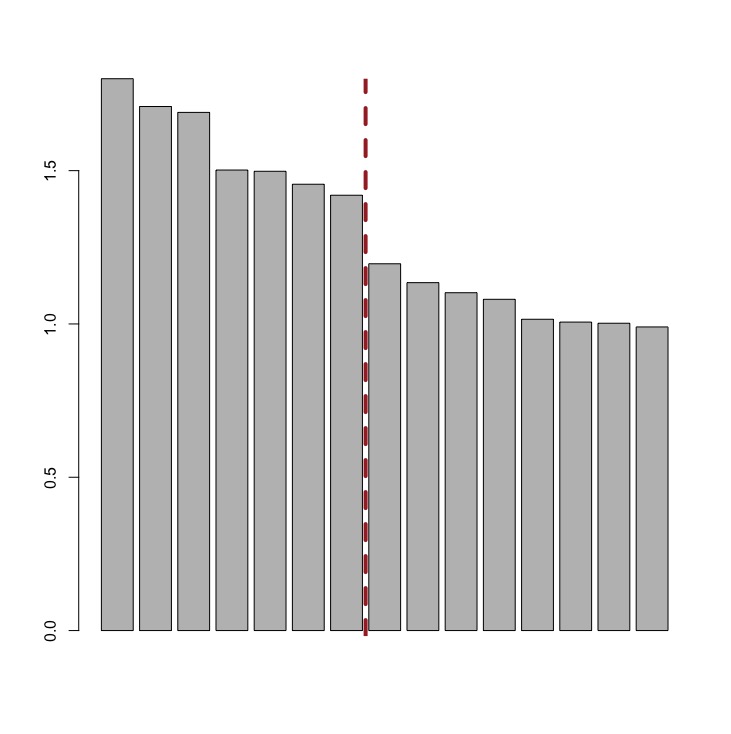}
\end{minipage}
\caption{Left column: a noisy signal $\mathbf{X}$ (black) in which the true piecewise-constant mean oscillates between 0 and 1 at nine change-points marked by the dashed red lines; the standard deviation of the noise is 0.8 (top) and 1 (bottom). Change-points estimated via SDLL as implemented in {\tt breakfast::model.sdll}, acting on the WBS2 solution path, in blue (nine estimated change-points in the top signal; seven in the bottom one). Right column: logged CUSUMs corresponding to the 15 largest elements of the WBS2 solution path of $\mathbf{X}$. The brown dashed line shows where SDLL identifies the ``steepest drop to low levels'', thereby estimating the number of change-points.\label{ch4:fig:sdll}}
\end{figure}

The input to the SDLL model selection procedure is a Wild Binary Segmentation or a Wild Binary Segmentation 2 solution path, as defined in Section \ref{ch4:multi:wbs}. Recall that 
a solution path is simply a sequence of CUSUM statistics, computed for each of a representative sample of intervals $[s_m, e_m] \subseteq [1, n]$, and sorted in decreasing order. In the first example, we compute the WBS2 solution path of {\tt mscale.easy} and {\tt mscale.difficult}, using a deterministic grid of intervals for our computation of the CUSUM statistics, as described in Section \ref{ch4:multi:wbs}, with $M = 1000$.

The right-hand plots of Figure \ref{ch4:fig:sdll_01} show the 15 largest elements of the WBS2 solution path of {\tt mscale.easy} and {\tt mscale.difficult} on a logarithmic scale. The logarithmic scale is appropriate here as CUSUMs are multiplicative in nature (in the sense that the CUSUM of twice the signal is twice the CUSUM of the original signal) and the SDLL criterion will be comparing the differences between them -- so we need to log them first.

Denote the sorted CUSUMs (in decreasing order) by $(C_{(s_i),(b_i),(e_i)})_{i=1}^{M}$, where $M$ is suitably large, as dictated by the WBS or WBS2 solution path algorithm. SDLL proceeds in two stages, described below.

\begin{enumerate}
    \item Testing for the presence of any change-points. To achieve this, $C_{(s_1),(b_1),(e_1)}$ is compared to a threshold of the form $\lambda_1 = C_1 \sqrt{\log n}$, for a suitably chosen $C_1$. If $C_{(s_1),(b_1),(e_1)} < \lambda_1$, the procedure estimates $\hat{N} = 0$ and stops. Otherwise, the procedure moves onto the next stage below.
    \item Estimation of the number of change-points. This proceeds in the following steps.
    \begin{enumerate}
        \item Take differences in the consecutive logged sorted CUSUMs, i.e. form
        \begin{equation}
        \label{ch4:eq:di}
        D_{i} = \log(C_{(s_i),(b_i),(e_i)}) - \log(C_{(s_{i+1}),(b_{i+1}),(e_{i+1})}).
        \end{equation}
        Under the assumptions outlined in \cite{f20}, those CUSUMs $C_{(s_i),(b_i),(e_i)}$ for which $(b_i)$ estimates a true change-point are of order $n^{1/2}$, whilst the remaining ones are $O(\log^{1/2}n)$. (This also suggests that we should only consider $D_i$ in (\ref{ch4:eq:di}) for which $C_{(s_{i+1}),(b_{i+1}),(e_{i+1})} \ge C_2 \log^{1/2} n$.) Therefore, as long as both logarithmic terms forming part of $D_{i}$ correspond to change-point-detecting CUSUMs, $D_i$ will be of order $O(1)$. If $\log(C_{(s_i),(b_i),(e_i)})$ is a change-point-detecting CUSUM but $\log(C_{(s_{i+1}),(b_{i+1}),(e_{i+1})})$ is not, then $D_i = O(\log\,n)$. Finally, if both logarithmic terms correspond to non-change-point detecting CUSUMs, then we again have $D_i = O(1)$.
        Hence, for $n$ large enough, the position $i$ of the largest difference $D_i$ should indicate the number of change-points $N_0$ as $D_i$ is only unbounded in $n$ when $C_{(s_i),(b_i),(e_i)}$ is a change-point-detecting CUSUM (i.e. is of order $n^{1/2}$) but $C_{(s_{i+1}),(b_{i+1}),(e_{i+1})}$ is not (i.e. is of order $\log^{1/2}n$).
        \item However, in practice, for a finite and not-necessarily-large $n$, it may be the case that $D_{N_0}$ is not the largest difference. For example, the inspection of the right-hand plots of Figure \ref{ch4:fig:sdll_01} shows that the largest difference in the logged CUSUMs is between the 1st and the 2nd CUSUM, which would point to a single change-point model, while we know that there are in fact two change-points in the mean of {\tt mscale}. Therefore, we will look for $D_{N_0}$ among the largest differences as outlined below. We first sort the differences $D_i$ in decreasing order to create $D_{(i)}$. In our running example, we have $D_{(1)} = D_1$ and $D_{(2)} = D_2$.
        \item Examine $D_{(i)}$ in the order $i = 1, 2, \ldots$. We know that $D_{(i)} = D_{N_0}$ would imply that the next CUSUM, $C_{(s_{i+1}),(b_{i+1}),(e_{i+1})}$ is not a change-point-detecting one, and is therefore of order $\log^{1/2}n$. Therefore, for each $i$, we check the ``low levels'' condition $C_{(s_{i+1}),(b_{i+1}),(e_{i+1})} \le C_1 \log^{1/2}n$. If it is not satisfied, by implication we cannot have $D_{(i)} = D_{N_0}$ for the current value of $i$, and we then move on to consider the next $i$, until the low levels condition is satisfied. In other words, we are not merely looking for the ``steepest drop'' in the sequence of sorted CUSUMs (which in our example is between CUSUMs indexed 1 and 2) but for the  ``Steepest Drop to Low Levels'', which in the {\tt mscale.easy} example is located between CUSUMs 2 and 3. We take the $i$ at which we stop to be the SDLL estimate of $N_0$. 
    \end{enumerate}
\end{enumerate}

It remains to choose the two constants $C_1$ and $C_2$.
\begin{description}
    \item[$C_1$:] A natural idea is to set $C_1$ so that type I error is controlled at the desired level, i.e. if the truth has no change-points, the procedure returns $\hat{N} = 0$ with a required high probability. The R package {\tt breakfast} implements this for significance levels 0.05 and 0.1.
    \item[$C_2$:] This is set to $0.3\sqrt{2}\hat{\sigma}_{MAD}$ by default, where $\hat{\sigma}_{MAD}$ is a Median Absolute Deviation estimate of $\sigma$, suitable for the Gaussian distribution, computed with $2^{-1/2}(X_{i+1}-X_i)$ on input. The constant determines the number of sorted CUSUMs considered by SDLL so provides an indirect upper bound on the number of change-points. This is the less important parameter to be chosen: it simply defines the range of operation of the SDLL procedure.
\end{description}

SDLL uses the threshold $\lambda_1$ in a different way than simply testing each CUSUM in the solution path against $\lambda_1$ and estimating $N_0$ as the number of CUSUMs that exceed $\lambda_1$. The key property of SDLL is that because it orders model candidates by the size of the ``drops'', i.e. differences between consecutive logged sorted CUSUMs, it often puts very different models in competition with each other. For example, SDLL may be considering models in the order: a model with 0 change-points, then a model with 2 change-points, then a model with 17 change-points. Therefore, the role of the threshold $\lambda_1$ is relatively ``easy'' as it frequently has to decide between models that are remote, e.g. in this case, perhaps between the model with 2 and the model with 17 change-points. This makes SDLL relatively robust with respect to the choice of $\lambda_1$.

The applicability of SDLL goes beyond solution paths generated by WBS or WBS2: it can be used with any other device for generating solution paths. The R package \verb+breakfast+ combines SDLL also with the Tail-Greedy Unbalanced Haar, the Narrowest-Over-Threshold and the Isolate-Detect solution paths. However, we note that WBS and WBS2 may be particularly suitable for SDLL as these two methods are based on the principle of explicitly selecting the largest available
CUSUM in each step. This means that the change-point-detecting CUSUMs are ``the largest possible'' in WBS(2), and because the non-change-point-detecting CUSUMs are uniformly bounded from above uniformly over all solution path approaches, the WBS(2) solution path promotes the maximisation of the gap between the change-point-detecting and the remaining CUSUMs, which is conducive to SDLL's success.

Continuing the running example involving the {\tt mscale} signal and its two noisy observed versions, we can see from Figure \ref{ch4:fig:sdll_01} that SDLL correctly identifies two change-points in the {\tt mscale.easy} signal, while mistaking the less prominent change-point for noise in the {\tt mscale.difficult} signal.

How does SDLL behave for signals without an obvious multiscale structure, in which all change-points have similar importance? To see an example of this, we consider the setting shown in Figure \ref{ch4:fig:sdll}, where the signal has a teeth-like structure: the $N_0 = 9$ change-points appear at equal intervals and the level of the signal oscillates between two values. We can see that SDLL is still able to identify the correct number of change-points in the less noisy setting. In the noisier setting, the two least prominent change-points (although in the noiseless signal, all change-points have equal importance, noise obscures some change-points more than others) fail to be identified by SDLL.


\section{Software: Speed, Design and Accuracy}
\label{ch4:sec:simstu}


It would be impossible for us to offer a thorough comparative numerical review of the state of the art in multiple change-point detection software, even for the simple piecewise constant mean plus i.i.d. Gaussian noise model studied in this chapter. Some methods may have less visibility than others, and we may be simply unaware of them. Some may be available in languages other than R, our preferred language for examples in this book. Different methods may have different objectives, making fair comparison difficult, or even impossible. Finally, any such review would quickly become out of date with the appearance of any new, better methods, or with tweaks to or re-implementations of the existing methodologies, which occur all the time.

At the time of writing, there are a number of recommendable comprehensive reviews on the multiple change-point detection problem, some of which devote more space to software than others. In particular, we would like to point the reader to
\cite{tov18}, \cite{bw20} and \cite{fr20}.

Rather than providing a similar review, our aim in this section is different: we would like to share with the applied user our practical experience with multiple change-point detection methods in the mean-shift model with i.i.d. Gaussian noise, by flagging up some important issues to bear in mind when using any method, and highlighting some of the characteristics of a selection of methods covered in this chapter. Throughout this section, we will occasionally be highlighting what we see as ``best practice'' -- in the design of multiple change-point detection methods, in their evaluation, and in their use. We will be mainly basing our illustrations on the R packages \verb+breakfast+ and \verb+changepoint+ (both available from the CRAN repository). This is, of course, a highly subjective and limited choice (these packages have been created in our respective research groups), but even this limited selection will serve to illustrate the issues we raise in this section. To keep discussion of software in one place, we include a couple of methods that will be introduced in later chapters.

\begin{figure}[t]
\centering
  \includegraphics[width=0.8\linewidth]{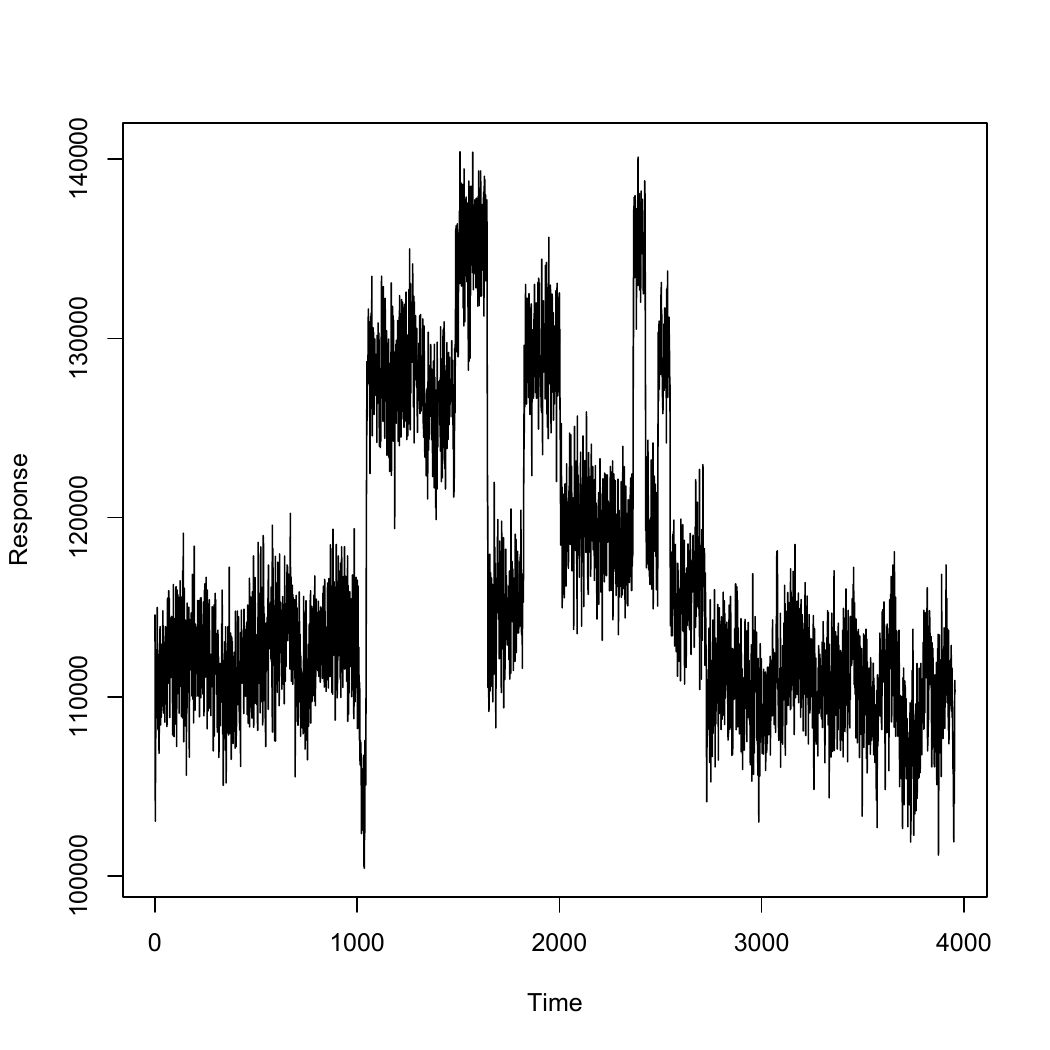}
\caption{The well-log data after removal of outliers -- see Section \ref{ch4:sec:well-log}.\label{ch4:fig:wl}}
\end{figure}

Our running examples in this section will include the well-log data, available from \url{https://github.com/pfryz/cpdds}. We use a version of the data after removing outliers (a future chapter of this book will discuss how to estimate change-points in the presence of outliers, and the original data will be analysed there). The time series plot of the subsequent data is shown in Figure \ref{ch4:fig:wl}. We will use this dataset (as well as some other, simulated ones) to illustrate some of the issues under discussion here. Our generic task in this section is to form a judgment, to the best of our ability, as to the existence and locations of (any) change-points in the input data, if modelled using the piecewise-constant mean plus i.i.d. Gaussian noise framework studied in this chapter. In the case of the well-log data, this will also lead us to reflect on the appropriateness of the i.i.d. assumption.

\subsection{Speed Considerations}
\label{ch4:sec:speed}

There are some speed-related issues to bear in mind when applying, or preparing to apply, a multiple change-point detection method; we list some of them below.
\begin{enumerate}
    \item While many methods come with theoretical bounds on their computational complexity, the importance of these bounds to the applied user will often be secondary to the actual computation times. With this in mind, before running a change-point study, it is worth checking the actual computation times for the longest signals in the study, for typical values of the tuning parameters of the given method. We do it for the well-log data, as well as for two longer simulated signals (one with no change-points and the other with frequent change-points) below.
    \item The language and style of implementation tend to have a significant impact on execution times. For example, implementations that use R only tend to be slower than those with the computationally intensive parts written in C and wrapped into R. Also, loop-rich implementations in R tend not to be as fast as those that rely on vector, matrix and list manipulation. If the speed of a method is unsatisfactory, it may be due to the language or programming style in which it was implemented, in which case it may with some effort be improved. Another possible route to improvement is parallelisation: some methods are more amenable to it than others.
    \item The speed of many methods depends on their tuning parameters. For example, in Wild Binary Segmentation (Section \ref{ch4:multi:wbs}) and related methods which combine change-point fits over many sub-samples of the data, the computational effort increases with the (minimum) number $M$ of the sub-samples drawn. Similarly, in the PELT algorithm (Section \ref{ch4:sec:comp}), higher penalty values tend to lead to longer execution times as they lead to fewer detections, so there are fewer opportunities to save time by pruning. By the same token, for the same penalty value, PELT will execute faster for signals with frequent change-points. From this point of view, it is useful to distinguish between the impact on the method's execution times of (a) its parameter values and (b) the input data. While the execution times of many methods display some degree of sensitivity to their tuning parameters, not all are sensitive, in this sense, to the input data other than through its length.
\end{enumerate}

To illustrate these and other speed-related issues, we run the following multiple change-point detection methods from the R packages \verb+breakfast+ (version 2.3), \verb+changepoint+ (version 2.2.4) and \verb+mosum+ (version 1.2.7). In each execution below, the input dataset is stored in \verb+data+.

\begin{enumerate}
    \item The PELT method, described in Section \ref{ch4:sec:comp}, from the R package \verb+changepoint+, version 2.2.4. The execution is
    \begin{verbatim}
    cpt.mean(data, pen.val = 2 * sigma.est^2 * log(length(data)),
      penalty = "Manual", method = "PELT")
    \end{verbatim}
    where \verb+sigma.est+ is the Median Absolute Deviation estimator of $\sigma$ described in Section \ref{ch4:sec:noise_variance}, that is \verb+sigma.est <- mad(diff(data))/sqrt(2)+.
    \item The CROPS algorithm, referenced in Section \ref{ch4:sec:comp}, from the same R package. This computes the segmentations for a range of penalties, here all penalties from $2\hat{\sigma}^2\log(n)$ to $40\hat{\sigma}^2\log(n)$. The execution is
    \begin{verbatim}
    cpt.mean(data, penalty = "CROPS",
      pen.value = 2 * sigma.est^2 * log(length(data)) * c(1,20),
      method = "PELT")    
    \end{verbatim}
    \item The MOSUM method, described in Section \ref{ch4:sec:mosum}, from the R package \verb+mosum+, version 1.2.7. The execution is
    \begin{verbatim}
    mosum(data, G = ceiling(log(length(data))))
    \end{verbatim}
    The \verb+G+ parameter specifies the bandwidth; \cite{km11} give $\log(n)$ as the lowest permitted order of magnitude of the bandwidth. Henceforth, we label this method MOSUM(1), to differentiate it from the next version of MOSUM below.
    \item The MOSUM method with execution
    \begin{verbatim}
    mosum(data, G = round(length(data)^(2/3)))
    \end{verbatim}
    henceforth labelled MOSUM(2).
    \item The WBS method, described in Section \ref{ch4:multi:wbs}, from the R package \verb+breakfast+, version 2.3. This computes the entire solution path of Wild Binary Segmentation, with $M = 10000$ systematically drawn intervals. The execution is
    \begin{verbatim}
    s.wbs <- sol.wbs(data)   
    \end{verbatim}
    \item The WBS2 method, described in Section \ref{ch4:multi:wbs}, from the same R package. This computes the entire solution path of Wild Binary Segmentation 2, with $M = 1000$ systematically drawn intervals at each recursive stage. The execution is
    \begin{verbatim}
    s.wbs2 <- sol.wbs2(data)
    \end{verbatim}
    \item The mean-shift version of the NOT method \citep{bcf19}, which will be introduced in a future chapter of this book (as it is appropriate for settings that go beyond the mean-shift setting of this chapter),
    from the same R package. This computes the entire solution path of the Narrowest-Over-Threshold method, with $M = 10000$ systematically drawn intervals. The execution is
    \begin{verbatim}
    s.not <- sol.not(data)
    \end{verbatim}
    \item The TGUH method, described in Section \ref{ch4:sec:tguh}, from the same R package. This computes the entire solution path of the Tail-Greedy Unbalanced Haar method, with the $\rho$ parameter (see Section \ref{ch4:sec:tguh} for details) set to $0.01$. The execution is
    \begin{verbatim}
    s.tguh <- sol.tguh(data)
    \end{verbatim}	
    \item The strengthened SIC model (sSIC) selection method applied on the NOT solution path (the same R package). How to apply an information-criterion-based model selection criterion on any nested solution path is described in Section \ref{ch4:multi:bs}; for details of sSIC see \cite{f14a}. The execution (which by default sets the maximum number of change-points to 25) is
    \begin{verbatim}
    model.ic(s.not)
    \end{verbatim}
    \item The thresholding model selection applied to the TGUH solution path (the same R package), which  retains those change-points for which the local CUSUMs in the nested solution path are (in absolute value) less than \\
    \verb+1.15 * sigma.est * sqrt(2 * log(length(data)))+. The execution is
    \begin{verbatim}
    model.thresh(s.tguh)    
    \end{verbatim}
    \item The SDLL model selection applied to the WBS2 solution path (the same R package). See Section \ref{ch4:sec:sdll} for details of the SDLL model selection. The execution is
    \begin{verbatim}
    model.sdll(s.wbs2)    
    \end{verbatim}
\end{enumerate}

We try each of these executions with three different choices of \verb+data+:
\begin{enumerate}
    \item[A.] The well-log data of Figure \ref{ch4:fig:wl}.
    \item[B.] A signal of length 10000, with no change-points: \verb+rnorm(10000)+.
    \item[C.] A signal of length 10000, with frequently occurring change-points:
    \begin{verbatim}
    rep(c(rep(c(0, 1), each = 10)), 10000/20) + rnorm(10000)/5
    \end{verbatim}
\end{enumerate}

Table \ref{ch4:tab:speed} reports execution times for each of these methods and choices of \verb+data+, but they should be read with a health warning: they will be different on a different machine, and they will likely change in any future versions of the two packages. However, their relative magnitudes will serve to illustrate some of the issues under discussion here.

\begin{table}
\centering
\begin{tabular}{ |c||c|c|c| } 
\hline
method & well-log & no change-points & frequent change-points \\
 \hline
PELT & 6 & 229 & 14 \\
CROPS & 225 & 552 & 3556 \\
MOSUM(1) &  1  &  2  &  3   \\
MOSUM(2) &  1  &  2  &  2   \\
WBS & 54 & 130 & 132 \\
WBS2 & 4618 & 13297 & 12114 \\
NOT & 69 & 134 & 135 \\
TGUH & 200 & 580 & 554 \\
\hline
sSIC & 80 & 77 & 74 \\
thresholding & 1 & 1 & 5 \\
SDLL & 1 & 1 & 5 \\
\hline
\end{tabular}
\caption{Execution times, in $10^{-3}$ sec., for the various method-signal combinations, averaged over 10 executions. For the simulated signals, the 10 executions are for 10 independently simulated realisations. R 4.2.3 on macOS Monterey, iMac from late 2015.\label{ch4:tab:speed}}
\end{table}

Although the results for both versions of the MOSUM method appear lightning-fast, we need to remember that they both compute the MOSUM solution for one particular, fixed bandwidth. The MOSUM method for any fixed bandwidth will naturally have reduced detection power for a change-point landscape that does not align with the bandwidth chosen. For example, a narrow bandwidth will struggle to detect weak change-points surrounded by long stretches of signal constancy (in this setting, a narrower bandwidth might be useful), whereas a wide bandwidth will struggle to detect frequent change-points (where a wider bandwidth might succeed). \cite{ck20} adapt MOSUM to combine detection results across a range of bandwidths.

The results for the PELT method are an example of how the character of the input data can impact execution times. The method is very fast for signals with frequent change-points, but the absence of change-points can slow it down. This is because its computational savings occur in pruning, which does not happen when change-points are not being detected. We return to this issue in a future chapter of this book, where we describe an alternative method, FPOP, that is an implementation of dynamic programming that is fast even in the no change-point case.

It is interesting to see that the timing of the CROPS method, relative to that of PELT, varies significantly with the data set. This depends on the number of different segmentations, i.e. corresponding to different numbers of change-points, that are optimal as we vary the penalty over the range we are considering. The cost of running CROPS will be roughly equal to the cost of running PELT times the number of different segmentations that are optimal over the range of penalty that is being considered. 

The WBS, WBS2, NOT and TGUH solution paths are examples of algorithms whose execution times are practically unaffected by the form of the input data other than through its length. WBS and NOT are much faster than TGUH and WBS2 in part because the latter two are implented in pure R, whereas the key components of the former two are implemented in C. In addition WBS2 is implemented via recursion, which leads to easy-to-understand code, but introduces computational inefficiencies, as parts of the data are unnecessarily processed anew at each recursive stage.

To obtain a single set of change-point estimates from the WBS, WBS2, NOT or TGUH solution paths, it is necessary to input them into a model selection procedure, and it can be seen that those based on thresholding or SDLL are much faster than that based on the sSIC information criterion, which is partly due to computational inefficiencies in the implementation of sSIC, and partly to its natural complexity.


\subsection{Design and Functionality}
\label{ch4:sec:design}

User preferences will inevitably vary when it comes to the functionalities of the different multiple change-point detection packages and routines available. The purpose of this section is to highlight some of the aspects of this side of change-point detection software that we found important in our own data-analytic work.

\vspace{10pt}

\noindent {\em Availability of reliable default parameters.} The availability of reliable default parameters in a multiple change-point detection routine may be of particular importance to non-expert users, or users who wish to gather initial experience with a new method quickly, or ``automated'' users (i.e. other programmes rather than humans). By way of examples, we saw in Section \ref{ch4:sec:speed} how the various methods of the \verb+breakfast+ package (version 2.2) can be used with only \verb+data+ on input; the same applies to the call
\begin{verbatim}
    breakfast(data)
\end{verbatim}
which computes a number of multiple change-point estimators at once, each with its own default parameters. The function \verb+cpt.mean+ from the \verb+changepoint+ package (version 2.2.4) is different as it defaults to the AMOC setting, as well as assuming $\sigma = 1$. To run it by default for the purpose of multiple change-point detection with an unknown $\sigma$, a suitable wrapper needs to be written first. An example of such a wrapper is below.

\begin{verbatim}
cpt.mean.pelt.mad <- function(data) {
  sigma.est <- mad(diff(data))/sqrt(2)
  cpt.mean(data/sigma.est, method = "PELT")
}
\end{verbatim}

\vspace{10pt}

\noindent {\em Interpretable parameterisation.} Some users will inevitably wish to have a full control over the change-point fit, and will therefore appreciate working with routines whose parameters play a clear and useful role, are easy to interpret, easy to adjust, are independent of each other, and cover the whole family of change-point fits available. We discuss below some commonly used parameters from this angle.
\begin{description}
    \item[{\bf The minimum segment length.}] This controls the length of the minimum interval between two consecutive change-points in a fit. Methods that implement this constraint can do so at the solution path or model selection stage. What is, however, frequently missing from help files is a precise description of how imposing this constraint  reduces the quality of the fit if there is strong evidence in the data against this constraint: e.g. will some ``obvious'' change-points have to be removed? If so, which ones?
    
\begin{figure}[t]
\centering
\includegraphics[width=0.95\linewidth]{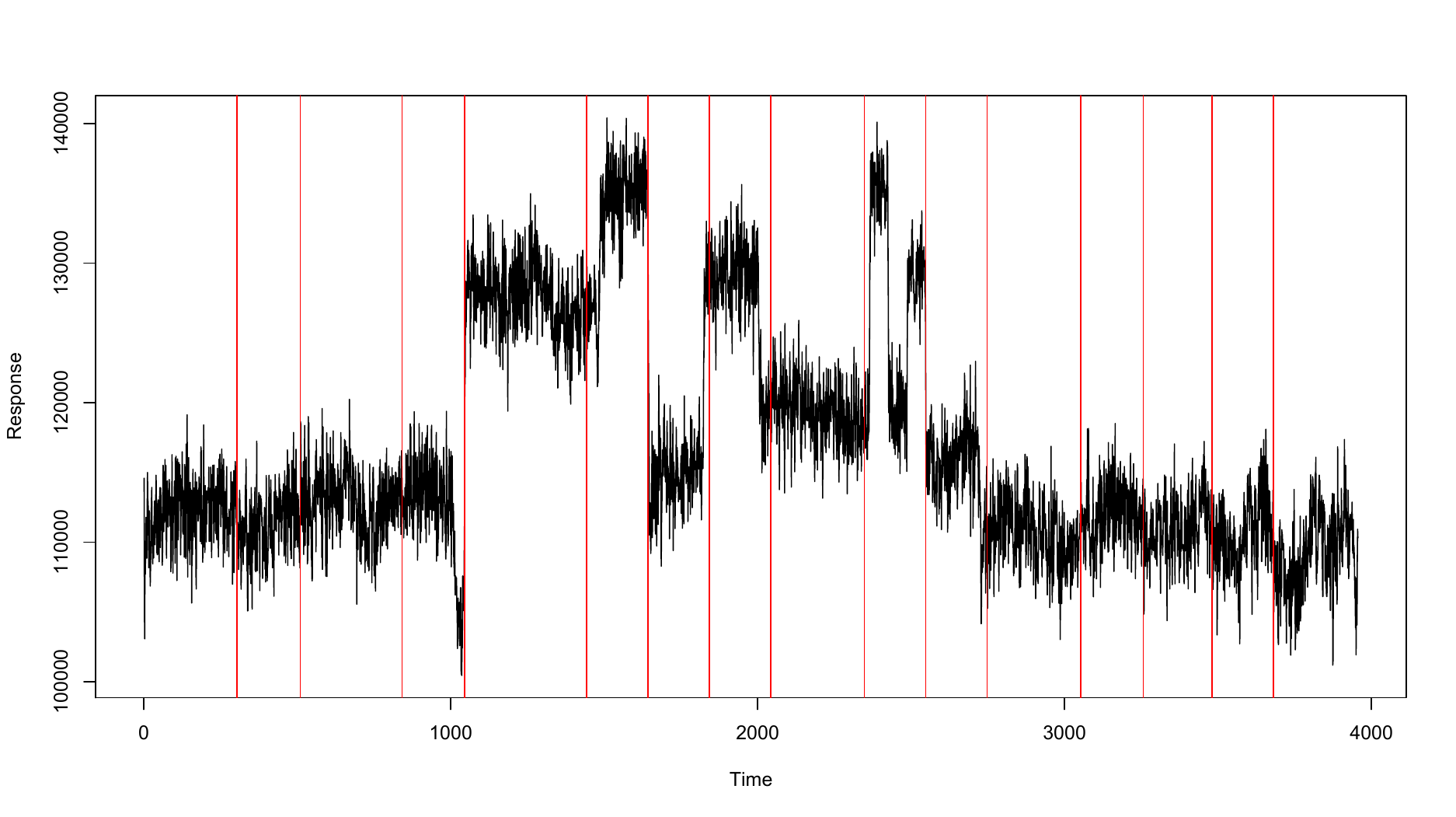}
\caption{The well-log data (black) and the least-squares change-point estimates with minimum segment length 200.\label{ch4:fig:wl200}}
\end{figure}

    As an example of its application, consider the well-log data of Figure \ref{ch4:fig:wl}. The execution
    \begin{verbatim}
    cpt.mean(data, penalty = "Manual",
      pen.value = 0, method = "PELT", minseglen = 200)
    \end{verbatim}
    uses the PELT algorithm to compute the least-squares fit under the sole constraint that the minimum segment length should be 200. The resulting change-point estimates are shown in Figure \ref{ch4:fig:wl200}. We can see that, visually, the change-points considered the most prominent by the algorithm align perfectly with their estimates, but any (also visually obvious) neighbouring change-points distanced by less than 200 time units from those estimated perfectly, are necessarily mis-estimated so that the minimum segment constraint is respected. Care is needed when comparing the statistical performance of methods with different minimum segment length parameters on simulated data -- as whether or not the minimum segment length restriction is appropriate for the simulated data will often be the primary driver as to which method is more accurate.
    
    \item[{\bf The penalty -- form or value.}] As described in detail in Section \ref{ch4:sec:optima}, penalisation is a common method of model selection in multiple change-point problems, and routines such as \verb+cpt.mean+ (R package \verb+changepoint+) or \verb+model.ic+ (R package \verb+breakfast+) enable the user to control, to some extent, the penalty value used. Despite the deserved popularity of penalisation, one possible warning to issue here is that the penalty parameter may not necessarily be an intuitive one to use by the non-expert analyst, as it may be difficult to visualise the effect of any particular penalty value. The reason for this is that the penalty only appears in an explicit form in the minimisation problem, which does not have an explicit, formulaic solution. We recommend looking at segmentations across different choices of penalty value if possible.
    \item[{\bf The threshold magnitude.}] Some multiple change-point approaches enable thresholding as a way of selecting the change-point model. Examples include binary segmentation, Wild Binary Segmentation (WBS) and WBS2, covered in Section \ref{ch4:sec:bs}. The criticism made of the penalty parameter above applies also to thresholding -- although the threshold is arguably a more explicit parameter than the penalty (as it applies directly to the relevant CUSUM statistics or similar contrast rather than being an ingredient of a minimisation problem), it may be unclear to the non-expert user what threshold values are suitable for the problem at hand.
    \item[{\bf The number of change-points (maximum or exact).}] In an extension of the AMOC setting, in some applications it may be desirable to limit the number of detected change-points to a given number -- either as an upper bound (when in addition model selection e.g. via penalisation is used) or on an exact basis. This functionality is present by default in approaches that return an entire solution path of change-point candidates sorted by the order of importance, as it is done e.g. in the solution path mode of the WBS method (see Section \ref{ch4:sec:bs}). As an example, the following code obtains the 10 most important change-point location (arranged in increasing order of time, not importance) for the WBS method in the \verb+breakfast+ package.
    \begin{verbatim}
    > sort(sol.wbs(data)$solution.path[1:10])
    [1] 1045 1485 1644 1823 2004 2365 2425 2487 2548 2724
    \end{verbatim}
    It is possible to achieve the equivalent functionality in the \verb+changepoint+ package as follows.
    \begin{verbatim}
    > cpts(cpt.mean(data, penalty = "Manual",
        method = "SegNeigh", Q = 11))
    [1] 1045 1485 1644 1823 2004 2365 2425 2487 2547 2724   
    \end{verbatim}
    In the above, \verb+Q+ is the number of segments.
    It is interesting and reassuring to see that the returned change-point locations show an almost perfect overlap between the two approaches.
    \item[{\bf The maximum probability of over-estimating $N_0$.}] This parameter comes in two different flavours: a type that can be referred to as ``strong'', where the user specifies a given small maximum probability of overestimating the number of change-points, and ``weak'', in which the user specifies the maximum probability of spuriously estimating any change-points if there are none. Control of the strong type is available in FDR-controlling approaches, which will be covered in a future chapter of this book. Control of the weak type is available, for example, in the \verb+model.sdll+ routine from the \verb+breakfast+ package (version 2.3). Consider the code
    \begin{verbatim}
    set.seed(1)
    for (i in 1:100)
      if (length(model.sdll(sol.wbs2(rnorm(1000)))$cpts > 0))
        print("Spurious change-points found!")
    \end{verbatim}
    This prints out the message in quotes each time any spurious change-points are estimated, out of 100 simulated sample paths of i.i.d. Gaussian noise of length 1000. The code prints out the message 10 times, which delivers on the promise of the default setting in the \verb+model.sdll+ routine, which guarantees to keep the probability of spurious detection to 10\%.
\end{description}

\vspace{10pt}

\noindent {\em Ability to supply own estimators of $\sigma$.} Many model selection methods for multiple change-point problems rely on the estimation of $\sigma^2$ to some degree of accuracy: this includes thresholding, SDLL, and penalisation of the least-squares fit. On the face of it, $\sigma$ or its estimators are absent in penalisation approaches that penalise the minus log-likelihood (rather than the residual sum of squares; see Section \ref{ch4:sec:n0unknown} for details), but this is only because the minus log-likelihood part of these penalised criteria implicitly estimates the residual variance for each candidate model fit, and the performance of the criterion may be poor if that estimate is unsatisfactory. For model selection methods that do not have a similar in-built mechanism for estimating $\sigma$, its estimate is required, and robust estimation methods may be preferred to stop the estimator being affected by the presence of change-points. Two options are the Median Absolute Deviation \citep{h74_4} or Inter-Quartile Range \citep{rc93_4} estimators suitable for Gaussian data on the input sequence $\{2^{-1/2} (X_{i+1} - X_i) \}_{i=1}^{n-1}$; see \cite{f20} and \cite{klb20} for a more detailed discussion. There is more on the estimation of $\sigma^2$ in Section \ref{ch4:sec:noise_variance}.

Some multiple change-point routines use their own estimator of $\sigma$, some permit an estimate or estimator of $\sigma$ as an argument, and some rely on the user to standardise the data so that the input data can be assumed to have a unit variance. We give below examples of each scenario.
\begin{enumerate}
    \item
    The \verb+breakfast+ function from the R package \verb+breakfast+ serves as an umbrella function for individual solution path and model selection functions within the package, and passes the input data onto them without worrying about the estimation of $\sigma$ -- this is left to the individual functions (if required). Therefore the \verb+breakfast+ routine does not permit the user to supply an estimator or estimate of $\sigma$.
    \item
    The \verb+model.*+ routines from the \verb+breakfast+ permit the user to specify their estimate of $\sigma$. For example, the \verb+model.thresh+ routine, which performs model selection via thresholding, is specified as follows.
    \begin{verbatim}
    model.thresh(
      cptpath.object,
      sigma = stats::mad(diff(cptpath.object$x)/sqrt(2)),
      th_const = 1.15)
    \end{verbatim}
    In the above, the default estimate of $\sigma$ is given, in \verb+sigma+ as the MAD estimator with $\{2^{-1/2} (X_{i+1} - X_i) \}_{i=1}^{n-1}$ on input.
    \item
    The \verb+cpt.mean+ does not explicitly permit an estimate of $\sigma$. This can be circumvented in one of two ways, as below.
    \begin{enumerate}
        \item As outlined earlier, an estimate of $\sigma$ can be passed as part of the penalty, for example
        \begin{verbatim}
        sigma.est <- mad(diff(data))/sqrt(2)
        cpt.mean(data, pen.val = 2 * sigma.est^2 * log(length(data)),
          penalty = "Manual", method = "PELT")
        \end{verbatim}
        \item
        The data can be divided by \verb+sigma.est+ prior to being fed into \verb+cpt.mean+.
        \begin{verbatim}
        cpt.mean(data/sigma.est, pen.val = 2 * log(length(data)),
          penalty = "Manual", method = "PELT")
        \end{verbatim}
    \end{enumerate}
\end{enumerate}
For transparency, it is good practice to allow an explicit estimate of $\sigma$ in change-point detection routines, but its absence can always be circumvented as in \verb+cpt.mean+.

\vspace{10pt}

\begin{figure}[t]
\centering
\includegraphics[width=\linewidth]{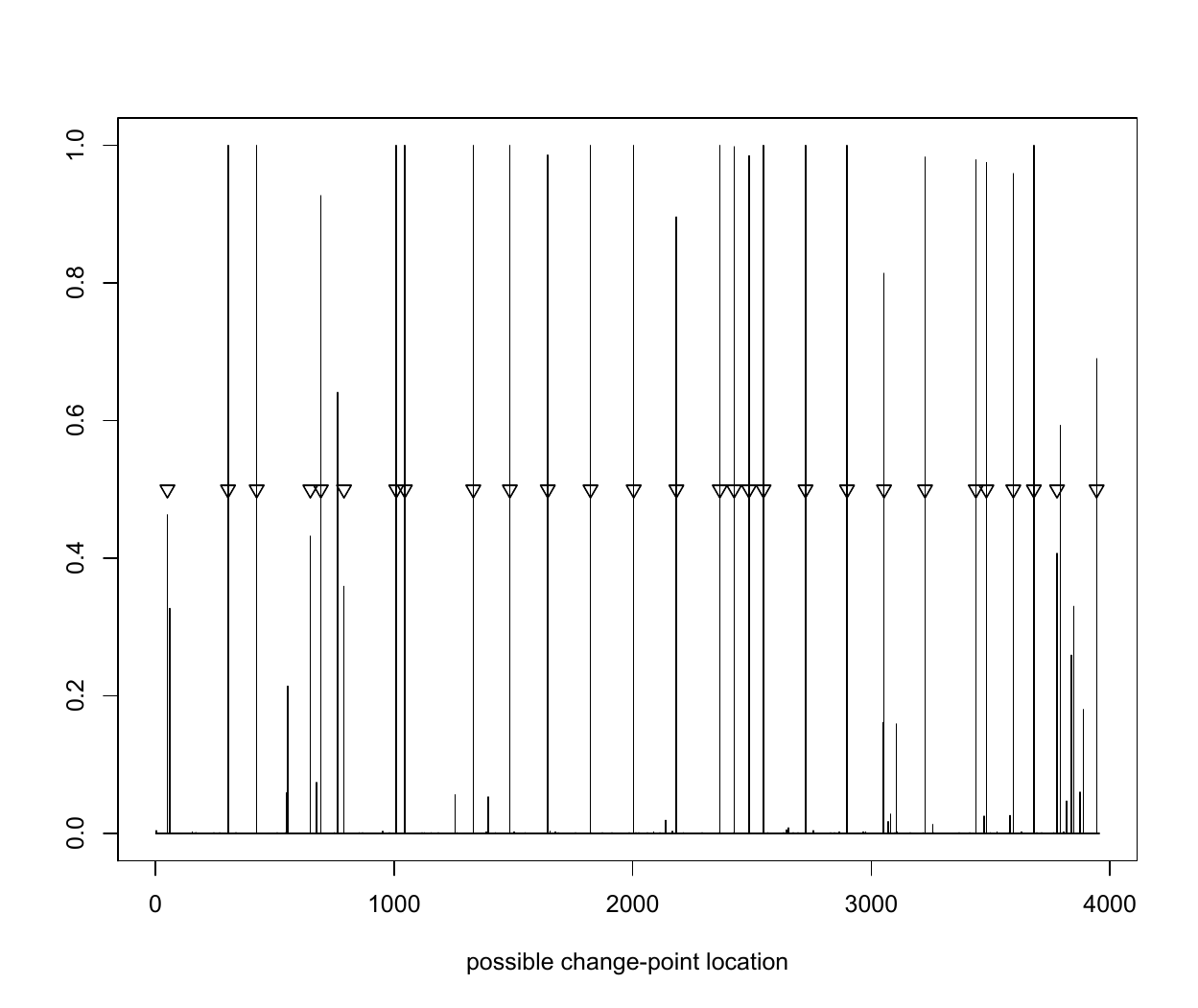}
\caption{Results for the well-log data. Vertical lines: the proportion of times, over 1000 runs, that the randomised WBS with the strengthened SIC model selection detected a change-point at each possible location. Triangle: the change-point locations estimated by the deterministic WBS with the same model selection criterion.\label{ch4:fig:randet}}
\end{figure}

\noindent {\em Is the procedure deterministic or random?} Versions of certain procedures can be randomised, in the sense of returning possibly different output when run repeatedly on the same data. An example is the Wild Binary Segmentation method when using the ``random grid'' option; see Section \ref{ch4:multi:wbs}. We now compare the deterministic and the random WBS on the well-log data; the results are in Figure \ref{ch4:fig:randet}. The deterministic WBS detects 28 change-points, while the lengths of the vertical lines sum up to 28.4, which can be interpreted to mean that the random WBS detects 28.4 on average. Their estimated locations now make up a probability distribution, in which the bar heights describe the estimated probability of a change-point occurring at each possible location in the data.

Executing a random procedure such as this one can serve as an informal way of assessing the uncertainty in estimating the change-point locations -- though here we are assessing the variability across the randomness in the algorithm rather than in the data. Looking across different analyses of the data can indicate what change-points are consistently identified, and which may be specific to a small subset of the analyses. 
We will discuss formal ways of evaluating statistical uncertainty in change-point problems in a future chapter of this book.

Randomised procedures such as this one are the most useful if they are executed multiple times and the outcomes pooled so the user can get an understanding of the entire sample space of possible outcomes. If there are time resources to only execute a procedure once, preference should be given to a deterministic procedure so avoid the unnecessary extra randomness in the output, which can sometimes be substantial.

\vspace{10pt}

\noindent {\em Nested or non-nested solutions?} Multiple change-point detection methods can be categorised into nested and non-nested ones. In a nested method, a candidate solution set $\{ \hat{\tau}_1, \ldots, \hat{\tau}_{N_1}  \}$ is necessarily a subset of a candidate solution set 
$\{ \hat{\tau}_1, \ldots, \hat{\tau}_{N_2}  \}$ if $N_1 \le N_2$, for all $N_1, N_2$. In a non-nested method, this does not necessarily hold, and estimated change-point locations can drop in and out of candidate solutions, as we increase/decrease the number of postulate change-points.

\begin{figure}[t]
\centering
\includegraphics[width=0.6\linewidth]{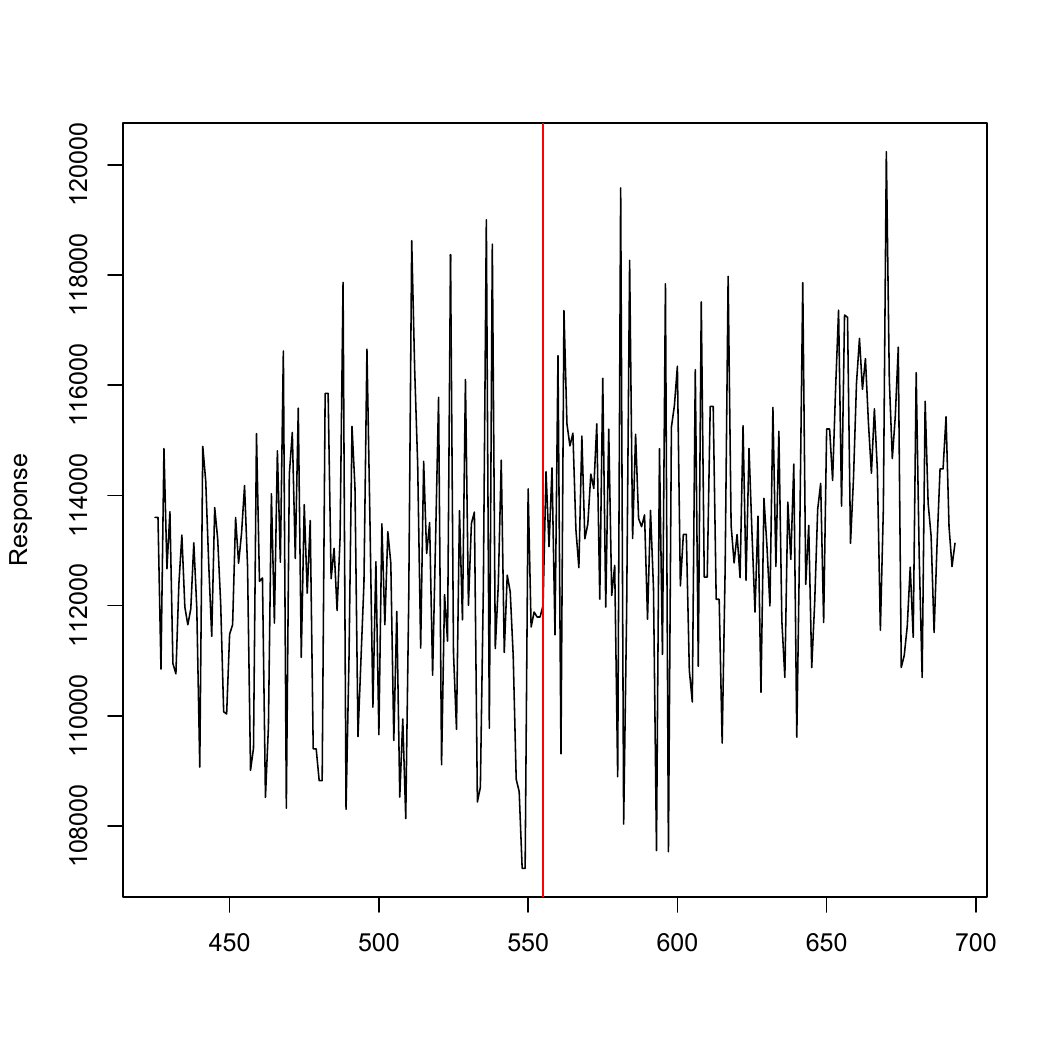}
\caption{Black: the well-log data between time indices 425 and 693. Red: location 555.\label{ch4:fig:555}}
\end{figure}

A prime example of a non-nested method is the least-squares fits to the data, indexed by the number of postulated change-points (or, equivalently, by the value of penalty), as computed, for example, by the PELT method. As an example, consider the following execution.
\begin{verbatim}
    cpts.full(cpt.mean(data, penalty = "CROPS",
      pen.value = 2 * sigma.est^2 * log(length(data)) * c(1,20),
      method = "PELT"))    
\end{verbatim}
This computes the least-squares fit to the well-log data for twenty different penalty values. Consider the first four columns of the output, which show the estimated locations of the first four detected change-points, as we increase the penalty.
\begin{verbatim}
[1,]   50  304  424  555 ...
[2,]   50  304  424  555 ...
...
[9,]   50  304  424  555 ...
[10,]   50  304  424  693 ...
[11,]   50  304  424  693 ...
[12,]   50  304  424  693 ...
[13,]  555  693  764 1009 ...
[14,]  555  693  764 1009 ...
[15,]  555  693  764 1009 ...
[16,]  304  424 1009 1045 ...
[17,]  555 1009 1045 1332 ...
[18,] 1009 1045 1332 1485 ...
...
\end{verbatim}
Consider the estimated location 555 (Figure \ref{ch4:fig:555}). It is interesting to see that it drops out of the solution set at the 10th penalty value (row 10 of the above matrix), to return in row 13, drop out again in row 16, return in 17 and drop out permanently in row 18. 


By contrast, nested procedures do not permit change-point candidates to drop out of/into solutions sets as the postulated number of change-points (or alternatively: the threshold magnitude or penalty value) is varied. Consider the WBS solution path on the well-log data, computed as below.
\begin{verbatim}
> sol.wbs(data)$solution.path -> swbs
> which(swbs == 424)
[1] 13
> which(swbs == 693)
[1] 17
> which(swbs == 555)
[1] 43
\end{verbatim}
It is interesting to see that location 555 also comes up in the WBS solution path, but relatively late, and much later than its neighbours 424 and 693. This may justify hesitation, on the part of the PELT fit, in including 555 in the model.

While it is natural that the least-squares fits do not result in nested change-point fits -- as there is nothing that links the solution sets for two different penalty values -- some analysts may prefer nested methods as they unambiguously order the change-point candidates in terms of their perceived importance/prominence, which may facilitate the interpretation of the output in some scenarios.

\vspace{10pt}

\noindent {\em Ability to handle missing data.} 

Often data is observed with missingness. For the change-in-mean model with i.i.d. data it is simple to adapt methods to deal with datasets which include missing data. One can remove the missing data and estimate the change-points for the remaining data points. Care is needed to map the estimated change-points back to the original time-scale. 

To make this precise, consider data $X_{1:n}$. Assume there are $n^*$ data-points which are not missing, and let $X^*_{1:n^*}$ be the set of observed data points. Define $j_1,\ldots,j_{n^*}$ such that $X^*_i=X_{j_i}$ for $i=1,\ldots,n^*$. Then we can estimate change-point locations $\tau^*_1,\ldots,\tau^*_m$ for $X^*_{1:n^*}$ using our favourite method. We then need to map these locations to the original time-scale to get our change-points for $X_{1:n}$. These will be $\tau_1,\ldots,\tau_m$ with $\tau_i=j_{\tau^*_i}$. Unfortunately, at the time of writing, neither of the two packages: \verb+breakfast+ or \verb+changepoint+, are able to do do this automatically.

\subsection{Accuracy and Performance Considerations}
\label{ch4:sec:acc}

This section covers a selection of accuracy-related aspects of multiple change-point detection in the i.i.d. Gaussian setting. We discuss the difficulty in measuring accuracy and the impact of the loss function, the importance of forming a view of the performance of a given method across signal classes, and possible ways of improving the accuracy if the execution with default parameter values is unsatisfactory.

\vspace{10pt}

\noindent {\em Difficulty in measuring the accuracy of change-point detection.} In addition to the statistical difficulty in detecting multiple change-points, one additional layer of difficulty is convincing oneself if the estimation result obtained is accurate. This is challenging partly by the nature of the problem, as the object returned by a change-point estimation procedure is fairly complex: it contains the estimate $\hat{N}$ of the number of change-points, and, conditional on it, the estimated locations $\hat{\tau}_1, \ldots, \hat{\tau}_{\hat{N}}$.
Different problems, and/or different users are likely to have a preference for solutions that ``get right'' different aspects of this solution object.

The lack of universal agreement as to what constitutes a successful estimate in a multiple change-point problem is reflected in the fact that many different accuracy measures feature in the existing literature. We briefly report on three styles of reporting accuracy in simulation studies below.
\begin{description}
    \item[{\bf Distribution of $\hat{N}$ plus an unconditional measure of location accuracy.}] A common way of reporting on the accuracy of a multiple change-point method is to provide a summary of the empirical distribution of $\hat{N} - N_0$, which describes the accuracy of estimating the number of change-points, supplemented by an unconditional measure of accuracy of the estimated locations. Examples of the latter include the mean-square or mean absolute error of the function estimate $\hat{f}$ obtained by taking the sample means of the data between each pair of consecutive estimated change-points. Although this has the advantage of attempting to separate the effects of number and location estimation, the disadvantage is that the location accuracy measure averages over all sample paths, therefore including sample paths with potentially very inaccurate estimates of $N_0$, which reduces its interpretability. Also, the accuracy of $\hat{N}$ can mask whether the change-points being found correspond to true changes. A method that misses two changes but has two false positives may appear better than one that misses the same two changes but has no false positives.
    \item[{\bf Distribution of $\hat{N}$ plus a conditional measure of location accuracy.}] To avoid the pitfalls associated with reporting an averaged-out measure of location accuracy (see directly above), some authors supplement the empirical distribution of $\hat{N} - N_0$ with a measure of location accuracy restricted to those sample paths for which $N_0 = \hat{N}$. However, the disadvantage is that for the more challenging signals, the number of sample paths for which $N_0 = \hat{N}$ may be small or even zero, which reduces the usefulness of this combined measure. Also different methods being compared may have different datasets for which $\hat{N}=N_0$, which can cause a selection bias if one tends to estimate the correct number of change-points on datasets for which estimating their locations is harder. One simple alternative, not often seen in the literature, may be to provide a separate measure of location accuracy for each value (or group of values) of $N_0 - \hat{N}$ separately, i.e. fully condition the location estimation accuracy of the number estimation accuracy. Alternatively we can separately compare location accuracy of some methods by running them conditional on knowing the true number of change-points. This latter approach can separate accuracy of estimating the number of change-points, which often relates to threshold or penalty choices, from accuracy of estimating their location.
    \item[{\bf A single measure combining number and location accuracy.}] Some authors depart from reporting the accuracy of number and location estimation separately, in favour of a single measure describing both. One of the most popular single measures is the Hausdorff distance, defined as follows. Denote $S_0 = \{\tau^0_1, \ldots, \tau^0_{N_0}  \}$ and $\hat{S} = \{ \hat{\tau}_1, \ldots, \hat{\tau}_{\hat{N}}    \}$.
    \[
    d_H(S_0, \hat{S}) = \max\left( \max_{\tau \in S_0} \min_{\hat{\tau}\in\hat{S}} |\tau - \hat{\tau}|,   \max_{\hat{\tau}\in\hat{S}} \min_{\tau \in S_0} |\tau - \hat{\tau}|  \right).
    \]
    Effectively, the Hausdorff distance finds, for each true change-point, the closest estimated change-point, and for each estimated change-point, the closest true change-point, and returns the largest of their distances. As well as penalising location mis-estimation, this also penalises under-estimation of the number of change-points (as then at least one of the true change-point does not have a close estimate) and their over-estimation (as then at least one of the estimated change-points does not have a true one nearby). While appealing, the Hausdorff distance does not work well in situations in which, for example, a true change-point has more than one close estimate, as then $d_H(S_0, \hat{S})$ can be deceptively low.
    
    There are also measures in this category that involve an extra parameter, for example by counting the proportion of true change-points that contain an estimated change-point within a certain distance, and/or vice versa. This is more subjective than measures that do not need an extra distance parameter (such as the Hausdorff distance), but may be justified by the context; e.g. in certain applications an estimated change-point may only be valuable if it falls within a certain prescribed distance from a true change-point.

\end{description}

\vspace{10pt}

\noindent {\em Evaluating a new multiple change-point detection method.} Suppose the user would like to apply a particular, new to them, multiple change-point detection method to a real-life problem in which the truth (i.e. the number and the locations of change-points) is unknown. The user may have a particular loss function in mind (e.g. they may want to be able to detect as many change-points as possible, but avoid overestimating their number, and hope to the detect them as close as possible to the true locations). To obtain a better understanding of the performance of a new method, if resources permit, we would like to emphasise the importance of running multi-tiered simulation studies before its application.

Ideally, such multi-tiered simulation studies would include signals of increasing complexity, up to and including the kind of complexity the user expects to see in their real dataset. Here is one possible workflow, which is naturally non-exhaustive, but we have found it helpful in our own work. This assumes that the user has set the parameters of the method to the best of their ability (or that the default values are used); we discuss this important point later in this section.
\begin{enumerate}
    \item {\em Checking if the method performs as expected in trivial cases.} For example, does the method return the correct number and locations of change-points in piecewise-constant signals without noise? Are all locations returned as change-points in linear signals with a non-zero slope and no additional noise? Does the method execute correctly on extremely short signals? Does it return correct results for signals with, visually, extremely obvious change-points? Such pre-tests are useful for identifying programming errors in methods, and help identify ``hidden'' issues that are not well-documented (e.g. an unreported upper bound on the estimated number of change-points).
    \item {\em Checking performance on constant signals.} A good multiple change-point detection method should be able to return $\hat{N} = 0$ if $N_0 = 0$, with a high frequency. What constitutes a high frequency is not always prescribed -- some methods, which we describe in more detail in a future chapter on change-point inference, have a tuning parameter that permits the user to specify the nominal size, i.e. the probability that $\hat{N} \ge 1$ given $N_0 = 0$ (however, this does not always imply correct finite-sample control of the empirical size). Others are not parameterised in this way, but stem from theory which guarantees that the probability that $\hat{N} \ge 1$ given $N_0 = 0$ tends to zero as $n\to\infty$. This, however, says nothing about the practical performance for a given fixed $n$. These uncertainties make it important to evaluate empirically the behaviour of a method of signals of varying lengths with $N_0 = 0$, as a proxy of the method's tendency to over-estimate the true number of change-points. Another reason why such an investigation may be useful is if the frequency with which $\hat{N} \ge 1$ is too high, it may be possible to correct it by adjusting the tuning parameter of the method, such as the penalty or threshold value.
    \item {\em Does the method have particular strengths?} Not all methods are equally appropriate for all types of signals, and discovering the strengths and weaknesses of a method may help in optimising its application. For example, some methods may be systematically better than others in some of the following signal classes: signals with high or low signal-to-noise ratios, signals with frequent or infrequent change-points, signals with teeth- or stair-like structures, signals with a mix of prominent and weak change-points. To briefly illustrate this point, we apply two different model selection methods to the ``teeth'' signal shown in Figure \ref{ch4:fig:teeth} and stored in vector \verb+data.teeth+. The model selection methods are: strengthened SIC (Section \ref{ch4:fig:teeth}) and SDLL (Section \ref{ch4:sec:sdll}), both applied on a solution path returned by the WBS2 method (Section \ref{ch4:multi:wbs}). The R code (which uses the \verb+breakfast+ package) together with its output is shown below.
    \begin{verbatim}
    > teeth <- rep(rep(c(0, 1), each = 10), 10)
    > set.seed(1)
    > data.teeth <- teeth + rnorm(200) * 1/2
    > model.ic(sol.wbs2(data.teeth))
    Change-point locations estimated by:

    wbs2.ic : none
    > model.sdll(sol.wbs2(data.teeth))
    Change-point locations estimated by:

    wbs2.sdll: 10, 20, 30, 40, 50, 61, 67, 80, 91, 96, 105, 
        120, 130, 140, 150, 160, 170, 180, 190
    \end{verbatim}
    
    The SDLL method estimates the correct number of change-points and their locations with high accuracy, whereas the strengthened SIC estimates zero change-points -- even though if one assumes a change-point model, the change-point locations are relatively obvious visually from Figure \ref{ch4:fig:teeth}. This is related to the fact that SIC (and therefore also the sSIC, which in its default execution is close to SIC) is better at estimating infrequent change-points than frequent ones, whereas SDLL behaves particularly well in frequent change-point scenarios \citep{f20}. 

    \begin{figure}[t]
    \centering
    \includegraphics[width=0.6\linewidth]{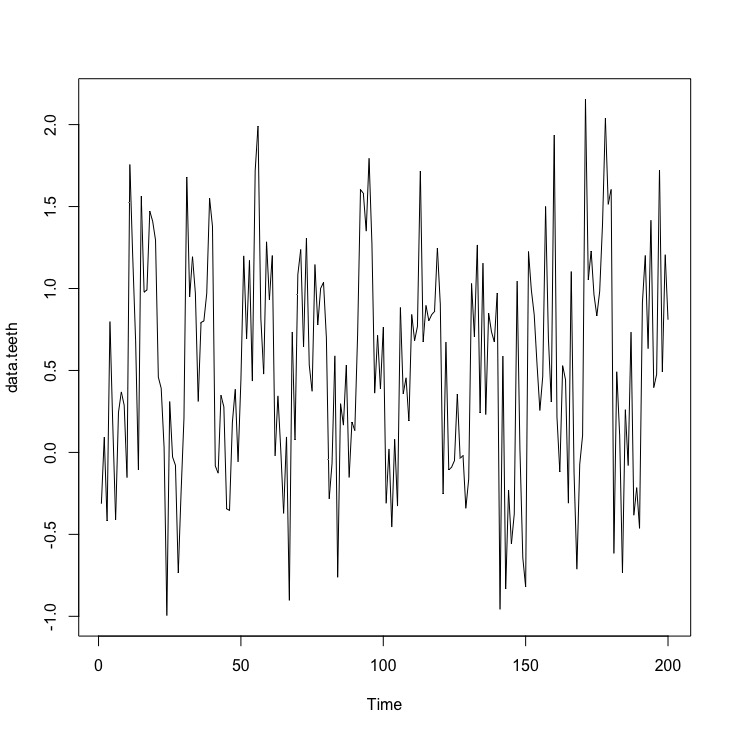}
    \caption{Noisy ``teeth'' signal.\label{ch4:fig:teeth}}
    \end{figure}
    
    \item {\em Data-guided simulations.} To the extent possible, it is a good idea to apply the method being tested to datasets that bear some similarity to the real dataset at hand. This is, by definition, possible only partially (the user does not have a complete knowledge of the true signal $\mathbf{f}$, so it is not possible to simulate from it exactly), but carrying it out even in a superficial manner can sometimes provide useful insight. Amongst other details such as the signal-to-noise ratio and the approximate or suspected number of change-points, attention should be paid to the distribution of the noise. For example, even if the assumption of the lack of serial independence and constant variance appears appropriate, integer-valued data (for which the Gaussian assumption may only be appropriate approximately) may lead to over-detection for some methods (we will provide examples in a future chapter on change-point inference); it is therefore a good idea to attempt to reproduce in the simulations the distribution of the noise from the real data.
    
\end{enumerate}

The more mathematically-minded readers may have a preference for judging a new method by the theory available for it, e.g. by preferring a method with (near-)minimax-optimality to one that does not provably exhibit it. Our view is that while theory can often provide valuable insight, the ultimate proof of a change-point detection method is its practical performance. Many (though not all) of the theoretical results available concern the asymptotic behaviour of methods as $n \to \infty$, and rely on assumptions that are difficult to verify in practice. Further, lack of optimality results does not mean that a method does not exhibit them, at least for certain signal classes -- perhaps those results just have not been shown yet. 

\vspace{10pt}

\noindent {\em If the default parameters fail -- adjusting the tuning parameters.} It may be the case that the application of any method with its chosen (perhaps by default) parameter values produces results that are unsatisfactory -- either because of the difficulty of the problem, or perhaps because the input violates some of the assumptions used by the method (such as the lack of serial correlation in the noise, Gaussianity or constant variance). In such situations, one obvious modification is to apply the same method with a different value of the tuning parameter (such as the penalty of threshold), or perhaps with a whole range of tuning parameters at once.

\section{Example: Well-log Data}
\label{ch4:sec:well-log}

We now turn to a simple application of change-point methods to both demonstrate some of the challenges in applying methods to data and to compare different methods. As we will see, a key feature of this example is that, like most applications, the data does not fit the simplifying modelling assumptions that are used to construct change-point estimators. This means that default implementation of the methods tend to perform poorly, particularly with regard to estimating the number of changes. However, often we can obtain reliable estimates if methods are implemented in an appropriate manner for a given application.

We will analyse the well log data, first presented in \cite{ruanaidh2012numerical} and shown in Figure \ref{Fig:ch4wellEDA}. This data comes from measurements from a probe that has been lowered through a bore hole in the ground. As the probe moves through different rock strata, the signal in the data will change to reflect the different properties of the rock. To detect the different rock strata from the data involves fitting a standard change-in-mean model, where we assume the mean of the data is piecewise constant, with each segment corresponding to a different stratum of rock. This is a relatively simple change-point application, with the changes being easy to detect by eye, but even in this case a routine application of change-point algorithms can lead to poor results.

We will first present some exploratory analysis of the data, and then consider and compare different methods for detecting the changes.

\subsection{Exploratory Data Analysis}

 We can gain some insight into the features of the noise in the data by a simple initial analysis. To do this we first obtain a rough estimate of the signal in the data by running a moving median filter (discussed in Section \ref{ch4:sec:noise_variance}). 

{\footnotesize
\begin{verbatim}
## input well-log data
data <- scan("https://raw.githubusercontent.com/pfryz/cpdds/main/data/oil.dat")
## (alternatively, download or clone the Github repo and scan the 
## oil.dat file from a local location)
### moving median filter
n <- length(data)
h = 50
est <- rep(NA, n)
for (t in 1:n) {
  est[t] <- median(data[max(1, t-h):min(n, t+h)])
}
### residuals of moving median filter
residuals <- data - est
\end{verbatim}
}

The idea of this moving median filter is that we estimate the mean at time $t$ by the median of the data in a window around $t$. In the above code we have, somewhat arbitrarily, set the window width to be 51, i.e. $h=25$ time points before or after $t$. Provided $h$ is less than the shortest segment length, then, for any $t$, more than half the observations in the window centered at $t$ will be from the segment that contains $t$. Thus the median gives a simple but reasonable estimate for the mean of the data at the time $t$.

\begin{figure}[t]
\centering
\includegraphics[scale=.95]{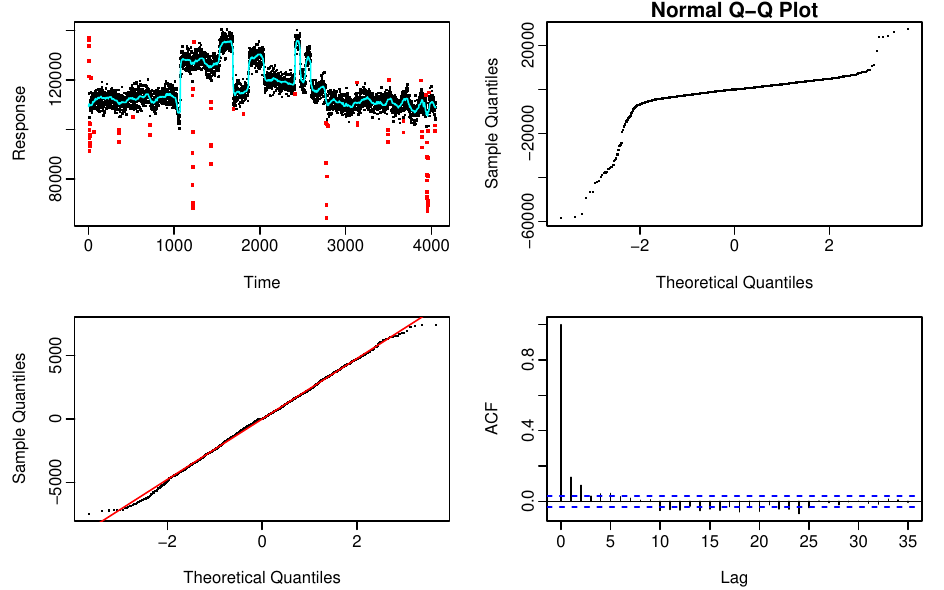}
\caption{ Top left: the well log data (dots) and moving median estimator (light blue line); the outliers that are identified based on the residuals from the moving median are in red. Top right: normal QQ-plot of the residuals from the moving median filter. Bottom left: normal QQ-plot of the residuals once outliers have been removed. Bottom right: auto-correlation plot of residuals, after removal of outliers.
}
\label{Fig:ch4wellEDA}       
\end{figure}

Once we have our moving median estimator we can then construct residuals. Normal QQ-plots of the residuals are shown in Figure \ref{Fig:ch4wellEDA}, and show that the residuals have a much heavier tail than a normal distribution. However if we remove the largest residuals, say all residuals whose absolute value is greater than 7500, we see that the remaining residuals have an excellent fit to a normal distribution. Henceforth we will call these data points, associated with these large residuals, outliers. The final plot in Figure \ref{Fig:ch4wellEDA} shows the auto-correlation of the residuals after we have removed the outliers. There is some significant autocorrelation at small lags, but this is relatively modest, with e.g. the estimated lag-1 autocorrelation being $\approx 0.14$.

From this initial analysis we see that this fits the assumptions of a change in mean in Gaussian data well except for (i) some outliers, and (ii) slight auto-correlation in the noise. We will focus on the challenge of dealing with outliers in a future chapter of this book. Thus, here we will consider only (ii), and to avoid the impact of outliers we will remove them.

\begin{verbatim}
## remove outliers
data.clean <- data[abs(residuals) < 7500]
residuals.clean <- residuals[abs(residuals) < 7500]
\end{verbatim}

\subsection{Penalised Cost Approach}

Our first approach to estimating the change-points will be based on the ideas from Section \ref{ch4:sec:optima}, that is to find the change-points that minimise the residual sum of squares plus a penalty for each change. We estimate the piecewise constant mean $f_{1:n}$ by minimising
\begin{equation} \label{ch5:eq-pencost}
\min_{f_{1:n}} \left( \sum_{t=1}^n (X_t-f_t)^2+\lambda||\Delta f_{1:n}||_0 \right).
\end{equation}
This is equivalent to maximising a penalised Gaussian likelihood.

There are various functions in {\texttt{R}} to do this, and we will use {\texttt{cpt.mean}} from the {\texttt{changepoint}} package, which implements the PELT algorithm of Section \ref{ch4:sec-PELT}. 

Before we can analyse the data we need to choose an appropriate value for the penalty, $\lambda$. As detailed in Section \ref{ch4:sec:optima}, if the noise is independent and identically distributed Gaussian with variance $\sigma^2$ then $\lambda=2\sigma^2\log n$ is the default choice, and leads to estimators of the change points that have excellent theoretical properties. To use this choice of $\lambda$ we need to estimate the variance of the noise. A standard way of doing this is to use the MAD estimator, see Section \ref{ch4:sec:noise_variance}: 
\begin{verbatim}
## MAD estimator of noise variance
sigma.mad <- mad(diff(data.clean))/sqrt(2)
\end{verbatim}
This gives an estimate of $\hat{\sigma}\approx2100$. An alternative, also discussed in Section \ref{ch4:sec:noise_variance} is to use the residuals from a moving median estimator.
\begin{verbatim}
### estimator of noise variance from residuals
sigma <- sqrt(mean(residuals.clean^2))
\end{verbatim}
This gives an estimate of $\hat{\sigma}\approx2400$. 

Why is there so much difference in these two estimates of $\sigma$? The answer is likely to be due to the non-independence of the noise. As if the lag-1 autocorrelation of the noise is $\rho$ then the variance of $X_t-X_{t-1}$ will be $2(1-\rho)\sigma^2$. Thus even relatively small autocorrelation, as we have for this data, can lead to the first estimator noticeably under-estimating the variance of the noise. In the following we use the second estimate for $\sigma$,  with $\hat{\sigma}\approx2400$.

\begin{figure}[t]
\centering
\includegraphics[scale=.95]{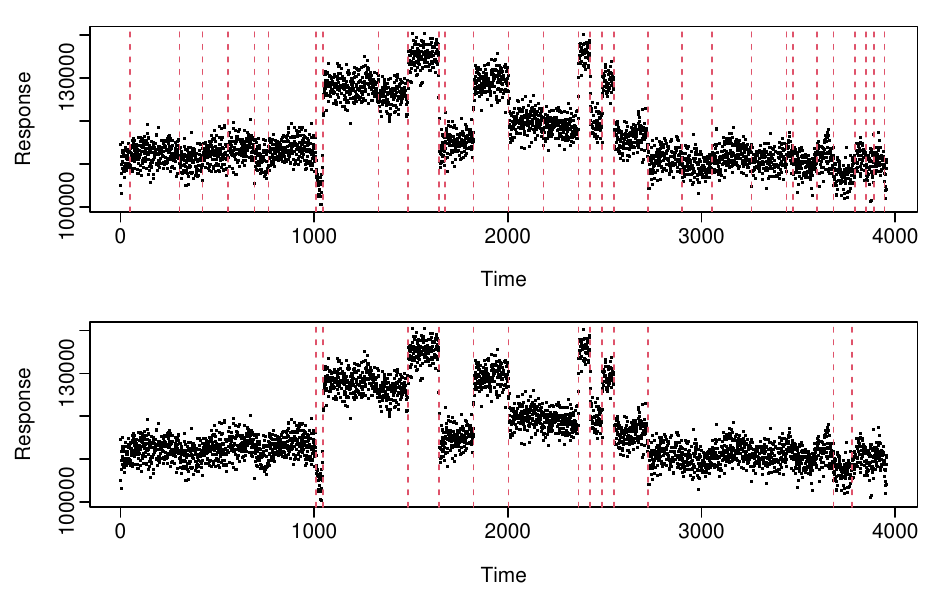}
\caption{ Results for estimating change point for the well log data, after removal of outliers, based on minimising the residual sum of squares: data (black dots) and estimated change points (vertical red dashed lines). Estimates with a penalty of $\lambda=2\hat{\sigma}^2\log n$ for each change-point (top row); and  estimates under the constraint of 13 change-points (bottom row).  
}
\label{Fig:ch4-Well_PELT}       
\end{figure}

To estimate the change points by minimising the penalised cost we use the commands
\begin{verbatim}
library("changepoint")
out.PELT1 <- cpt.mean(data.clean, penalty = "Manual", 
  method = "PELT", 
  pen.val = 2 * sigma^2 * log(length(data.clean)))
\end{verbatim}
The argument {\texttt{penalty = "Manual"}} of {\texttt{cpt.mean}} is used as we are setting our own penalty, and {\texttt{pen.val}} is the numerical value of the penalty. To see the list of estimated change points we use {\texttt{cpts(out.PELT1)}}, and there are 32 estimated change points and these are plotted in Figure \ref{Fig:ch4-Well_PELT}.

Clearly this default implementation has produced a poor segmentation of the data, with too many change points. This is because, even for a simple application like here, the strong assumptions on the log-likelihood we use to construct our cost, and that underpin the default choice of the penalty, do not hold. However, \cite{lm00} shows that minimising the residual sum of squares can give good estimates of the location of changes if we have a good estimate of the number of changes. We can test this by fixing the number of changes, and then minimising the residual sum of squares using the Segment-Neighbourhood algorithm of Section \ref{ch4:sec-SegNeigh}. To implement this with the number of change-points set to 13, which looks a reasonable estimate by eye, we use the commands:
\begin{verbatim}
out.SN13 <- cpt.mean(data.clean, method = "SegNeigh", Q = 14, 
  penalty = "None")
\end{verbatim}
Here we set the argument \texttt{Q} to be the number of segments, that is one more than the number of change-points. 

We plot the resulting change-point estimates in Figure \ref{Fig:ch4-Well_PELT}, and we can see that these estimates do estimate the locations well. We will return how to implement this penalised cost approach in a way that can also give a reliable esimate of the number and the locations of the changes below. First we will compare with other change-point estimation methods.

\subsection{Wild Binary Segmentation}

To estimate the change-points using the Wild Binary Segmentation procedure, we
use the following.
\begin{verbatim}
library("breakfast")
out.wbs <- sol.wbs(data.clean)
\end{verbatim}
The output of this implementation of Wild Binary Segmentation is a solution path -- that is a list of putative change-point locations in order of the evidence for that change. We can see the full solution path with
\begin{verbatim}
out.wbs$solution.path
\end{verbatim}
The solution path immediately gives the set of estimated change-points if we know the number of changes, $N$, as we just take the first $N$ entries in the solution path. 

\begin{figure}[t]
\centering
\includegraphics[scale=.95]{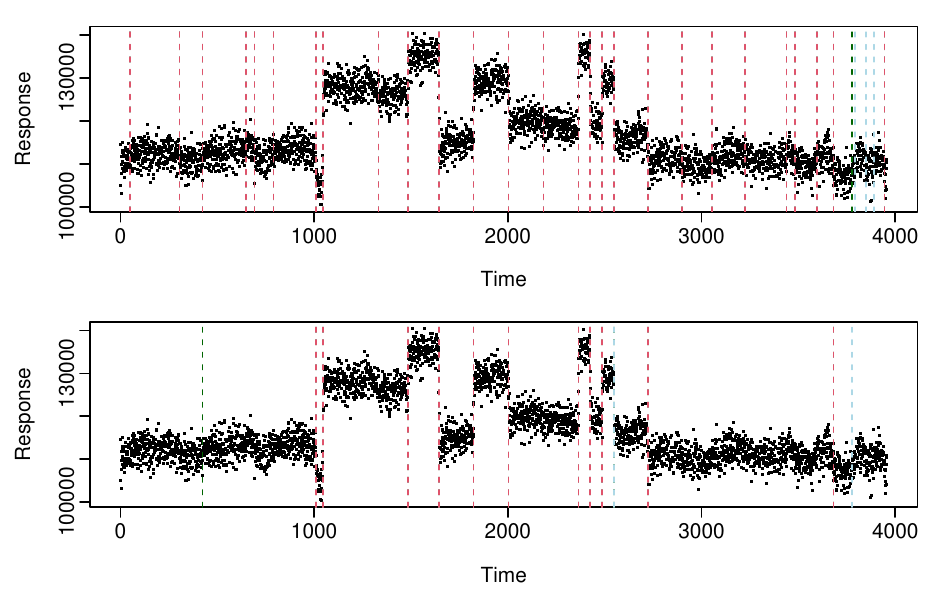}
\caption{Estimates of changes from Wild Binary Segmentation. Estimates under the strengthened Schwarz criterion for Wild Binary Segmentation with 10,000 intervals and with 100,000 intervals (top); estimates of 13 changes for Wild Binary Segmentation with 10,000 intervals and those from minimising the residual sum of squares (bottom). Data is shown by black dots; changes estimated by both methods (red dashed lines), just by Wild Binary Segmentation with 10,000 intervals (dark green dashed lines), and just by the other methods (light blue dashed lines).   
}
\label{Fig:ch4-Well_WBS}       
\end{figure}

The {\texttt{breakfast}} package gives a number of different methods to estimate the number of changes. We will focus using the strengthened Schwarz criterion \citep{f14a}. As above, this also leads to a poor segmentation of the data with 28 changes; see Figure \ref{Fig:ch4-Well_WBS}. Using other estimates for the number changes, such as based on thresholding the CUSUM statistic or using the steepest drop to low levels criterion, also leads to more estimated changes than looks reasonable (respectively 29 and 30).

Two other aspects of Wild Binary Segmentation are of note. The first is that the algorithm depends on the choice of a number of intervals of data. Changing these can give different results. To see this, we can repeat the above the analysis but using more intervals
\begin{verbatim}
out.wbs.slow <- sol.wbs(data.clean, M = 100000)
\end{verbatim}
Here we have increased the number of intervals, indicated by the argument {\texttt{M}}, by a factor of 10 from the default choice. The resulting estimate changes, under the strengthened Schwarz criterion, is slightly different as shown in the top plot of Figure \ref{Fig:ch4-Well_WBS}. This includes two more estimated changes, and one of the changes being in a slightly different location.

It is also interesting to compare the estimated changes by Wild Binary Segmentation with those from the penalised cost approach. To see this we  fix the number of changes to be 13, and show the two sets of estimated changes in the bottom plot of Figure \ref{Fig:ch4-Well_WBS}. Most changes are estimated in the same location, and there are just two differences. One difference is a minor difference in location: a change after observation 2547 versus after observation 2548; the other is that Wild Binary Segmentation places a change near the start of data, and the penalised cost places an additional change near the end. By construction the changes estimated by the penalised cost method will have a lower residual sum of squares -- as that method's estimates will always be the segmentation with smallest residual sum of squares for the given number of changes. In this case the difference in fit is of the order of 40 times the noise variance.


\subsection{MOSUM}

We now turn to estimating the changes using MOSUM. One disadvantage of this approach is that we need to specify a bandwidth. The MOSUM estimates with a bandwidth of 100 can be obtained using:
\begin{verbatim}
library("mosum")
out100 <- mosum(data.clean, G = 100)
\end{verbatim}
where the argument \texttt{G=100} specifies the bandwidth. This implementation of MOSUM uses a data-driven estimate of the noise variance that can vary locally, based on the data in the region around the change. (Though similar results are seen if we estimate the variance globally as for other methods.)

\begin{figure}[t]
\centering
\includegraphics[scale=.95]{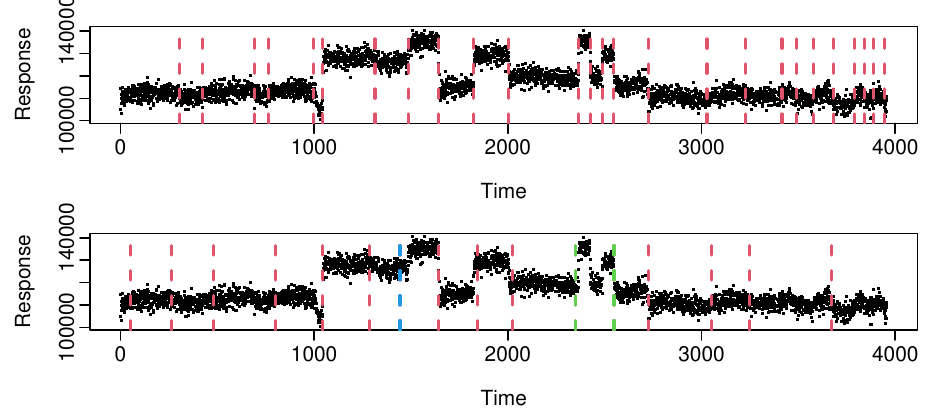}
\caption{Estimates of changes from MOSUM for a bandwidth of $100$ (top) and $200$ (bottom). Estimated change-points shown by vertical dashed lines. For the bottom plot we show an example of a poor estimated location (blue vertical line) and a region of missed changes (between green vertical lines) caused by too large a bandwidth.   
}
\label{ch4:fig:wellMOSUM1}       
\end{figure}

Results for both this bandwidth and a bandwidth of 200 are shown in Figure \ref{ch4:fig:wellMOSUM1}. As with other methods, we see that the MOSUM method overestimates the number of changes due to the autocorrelation in the noise. It looks like this is less of an issue than for other methods, but this is an artefact of the choice of bandwidth, and the fact that the MOSUM methodology avoids estimating changes close to each other. For the bandwidth of $200$ we see that each estimated change is roughly 200 to 300 time-points apart from each other. Further  we can see the problem with estimates if we use too large a bandwidth. It has poor localisation of one change, and misses two changes that are close to others, both due to the method deterring estimates that are much closer than the bandwidth.

\begin{figure}[t]
\centering
\includegraphics[scale=.95]{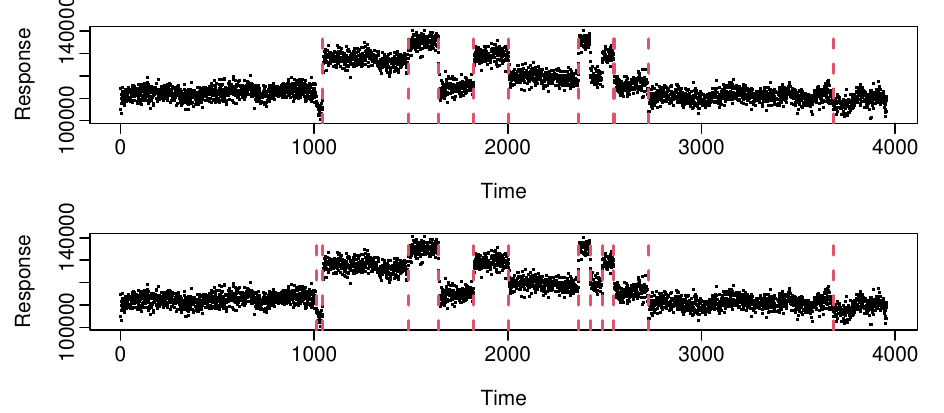}
\caption{Estimates of changes from MOSUM with an inflated threshold. Results for a bandwidth of $100$ (top) and using a grid of bandwidths of size 25, 50, 100 and 200 (bottom). Estimated change-points shown by vertical dashed lines.    
}
\label{ch4:fig:wellMOSUM2}       
\end{figure}

As with other methods, we can avoid the over-estimation of the number of changes by inflating the threshold; the objective here is that is the changes that the MOSUM method finds greatest evidence for should align with the changes one would detect by eye. To see this, we show the estimates from MOSUM with a bandwidth of 100 but with an inflated penalty (chosen by trial and error) in the top plot of Figure \ref{ch4:fig:wellMOSUM2}. In this case, we miss a change due to too large a bandwidth. A more reliable set of estimates can be obtained by using a grid of bandwidths, and using methods for merging results across different bandwidth choices (together with an inflated penalty). See the bottom plot of Figure \ref{ch4:fig:wellMOSUM2}. 

\subsection{Robust Implementation} \label{ch4:sec:robust}

We have seen that for each of the penalised cost approach, Wild Binary Segmentation and MOSUM, default implementations lead to estimates with too many changes. However, if we can estimate the number of changes accurately then they all give good estimates for the location of the changes. Here we consider how we can adapt the implementation of these methods so that they can more reliably estimate the number of the changes. We will show this whilst focussing on the penalised cost approach, but similar ideas will work for the other methods.

First it is helpful to remind ourselves why default implementations lead to over-estimation of the number changes. All methods have a threshold or penalty that, essentially, specifies how much evidence for a change there needs to be before we add a change. The default thresholds or penalties have been chosen under assumptions that the noise is i.i.d. Gaussian. However if these assumptions do not hold then the default choices will be inappropriate. If we have either heavier tailed noise, or positively auto-correlated noise, then it is more likely that, by chance, we will see patterns in the data that look like a change. 

For the well-log data, we saw that the assumptions of Gaussianity were reasonable, but that there was some auto-correlation. By eye this can be particularly seen towards the end of the data. It appears that even with relatively weak auto-correlation the default threshold and penalty choices lead to substantial over-estimation of changes.

Ideally we would address this by developing a method that models the auto-correlation in the data. We will show how this can be done in a future chapter of this book. But here we will investigate whether we can tweak the implementation of a method that assumes i.i.d. Gaussian noise so that it gives reliable esimates. We will consider two approaches, first inflating the penalty or threshold; the other comparing fits from segmentations with different numbers of changes.

\cite{lm00} shows that estimating changes in mean using a penalised cost like (\ref{ch5:eq-pencost}) can work well if we tune the penalty $\lambda$. \cite{bardwell2019most} suggest a simple inflation of the penalty to account for auto-correlation in the noise, and this is to use a penalty $\lambda=2c\sigma^2 \log(n)$, where
\[
c=1+2\sum_{t=1}^\infty \rho_t,
\]
and $\rho_t$ is the auto-correlation of the noise at lag $t$. The intuition behind this inflation factor is that the penalised cost involves terms associated with each segment of the form of 
\[
\sum_{i=s+1}^t \left(X_i-\frac{1}{t-s}\sum_{i=s+1}^t X_i \right)^2.
\]
For i.i.d. noise the variance of such a term is $(t-s-1)\sigma^2$, whereas for auto-correlated noise, if $t-s$ is large (relative to range of auto-correlation in the noise), then the variance of this term is $c(t-s-1)\sigma^2$. We will call $c\sigma^2$ the long-run variance.

\begin{figure}[t]
\centering
\includegraphics[scale=.95]{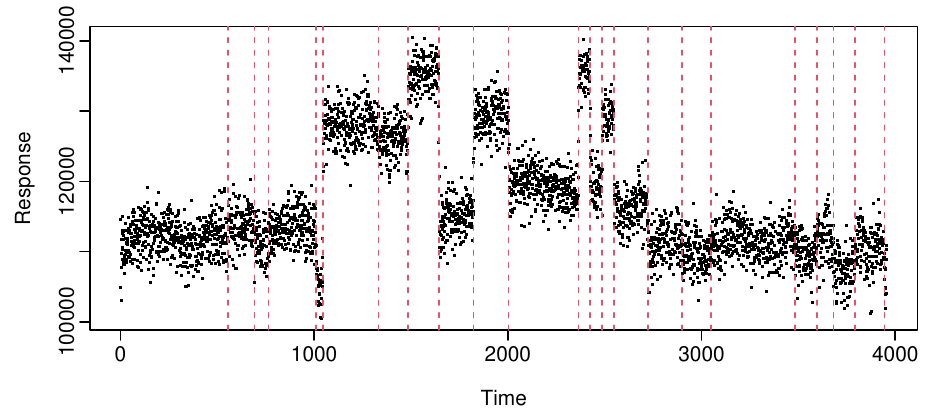}
\caption{Estimated change-points from a penalised cost approach with an inflated penalty $\lambda=2c\hat{\sigma}^2\log(n)$, where $c\hat{\sigma}^2$ is an estimate of the long-run variance that accounts for the auto-correlation in the noise. Data is shown by black dots and estimated changes by red dashed lines.
}
\label{Fig:ch4-Well_Robust1}       
\end{figure}

For our data, and the estimate of the auto-correlation of the residuals from the moving median estimator, we can implement such a penalty as follows.
\begin{verbatim}
auto.cor <- acf(residuals.clean)
c <- 1 + 2 * sum(auto.cor$acf[2:9])
out.PELT2 <- cpt.mean(data.clean, penalty = "Manual", 
  method = "PELT",
  pen.val = 2 * c * sigma^2 * log(length(data.clean)))
\end{verbatim}
This leads to fewer, but still too many, estimated change points; see Figure \ref{Fig:ch4-Well_Robust1}. In part this is because of the challenge in estimating $\sigma$ and $c$. The problem here is that the moving median estimator over-fits to the data and thus leads to an underestimate of $\sigma$ and of $c$. This over-fitting can be reduced slightly by using a larger window in the moving median filter -- an interested reader can repeat the above analysis using $h=50$ in the moving median filter.

An alternative approach, that we recommend when analysing a single data set, is to estimate the change points by considering a range of penalties. For the least-squares fit, we can do this in a computationally efficient manner using the CROPS algorithm \citep{hef17}, and this can be implemented within the {\texttt{cpt.mean}} function.
\begin{verbatim}
out.CROPS <- cpt.mean(data.clean, penalty = "CROPS", 
  method = "PELT",
  pen.val = 2 * sigma^2 * log(length(data.clean)) * c(1,20))
\end{verbatim}
Here the input to {\texttt{pen.val}} is a vector of length 2 with the upper and lower value of penalties to try. The output is then a list of segmentations -- showing how the segmentations change as we vary the penalty. 

\begin{figure}[t]
\centering
\includegraphics[scale=.95]{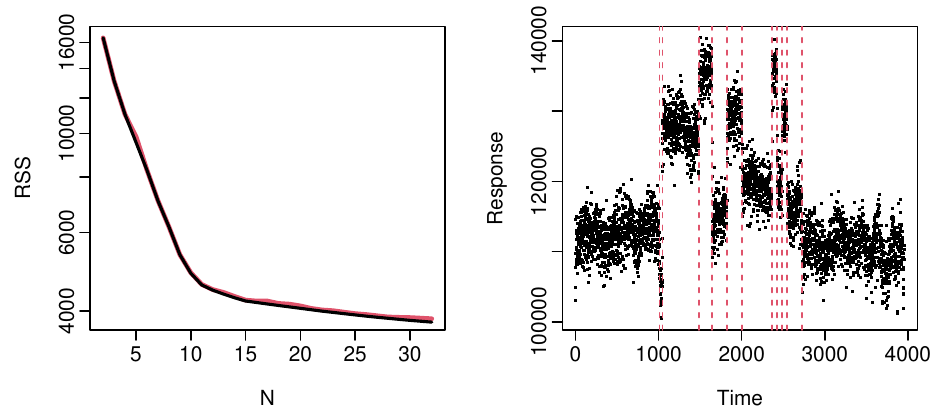}
\caption{ (Left) Residual sum of squares (in units of the estimated marginal variance of the noise) against number of estimated change-points for the penalised cost approach (black) and Wild Binary Segmentation (red). (Right) Estimated change-points for the penalised cost method with the number of changes estimated as $N=11$, corresponding to the elbow of the left-hand plot.
}
\label{Fig:ch4-Well_Robust2}       
\end{figure}

As well as comparing these segmentations on the data we can look at summaries of how the fit to the data changes with the number of change-points. There are two natural ways of doing this, one is to plot the number of change-points, $N$, against the residual sum of squares of a segmentation $l_{1:N}$
\[
\sum_{k=0}^{N} \left\{
\sum_{t=l_k+1}^{l_{k+1}} \left(X_t-\frac{1}{l_{k+1}-l_k}\sum_{t=l_k+1}^{l_{k+1}} X_i \right)^2
\right\},
\]
where $l_0=0$ and $l_{N+1}=n$. The other is to plot $N$ against the minimum penalty that gives that estimate for $N$. In both cases we can then look for ``elbows'' in the plots and a value of $N$ that corresponds to an elbow is a natural estimate of the number of change-points. The challenge with such a heuristic is that often there may be multiple elbows, and that choosing one is somewhat subjective. In our case a plot of $N$ against the residual sum of squares gives a clear elbow at $N=11$, which corresponds to a plausible segmentation; see Figure \ref{Fig:ch4-Well_Robust2}. 

We can produce a similar plot for other methods -- and this plot would be almost identical. To see this we show the plot of residual sum of squares against number of change-points for Wild Binary Segmentation in Figure \ref{Fig:ch4-Well_Robust2}. It also gives an elbow at $N=11$, and the resulting estimated segmentation is almost identical to that shown for the penalised cost approach (the only difference is the location of one change differing by one time-point).

\section{Bibliographical Notes}

An early compilation of references on the change-point problem, including the multiple change-point case, is in \cite{s80}. \cite{cz64} and \cite{y84} adopt the Bayesian viewpoint to estimate the local normal mean under a multiple change-point model. The theory presented in this chapter for estimating the multiple change-point locations via least-squares is based on \cite{ya89} in the known change-point number case and \cite{y88} in the unknown change-point number case; the latter work also shows the consistency of the SIC penalty for the number.

Binary segmentation is usually attributed to \cite{v81_4}; that work and \cite{v93} are the first or among the first to study the theoretical properties of this approach. The MOSUM approach can be traced back to \cite{bh78} and earlier papers by the same authors.

Some relevant recent papers on the multiple change-point problem with i.i.d. Gaussian noise (with possible extensions beyond this class) that do not appear in this chapter or are only mentioned very briefly either for space reasons, and/or because they appeared after the bulk of the chapter had been written, include 
\cite{vflr20}, who propose a multiscale penalty that favours well-separated change-points and discusses its optimality properties, \cite{zwl20}, who propose to estimate the number of change-points via sample-splitting, \cite{ck20}, who propose a localised SIC paired with MOSUM and study its optimality properties, and \cite{wyr18}, who study the optimality of a variant of Wild Binary Segmentation and an $l_0$-penalised procedure.

Papers that promote the use of multiscale residual testing to detect change-points, such as 
\cite{fms14} and \cite{lm16}, are described in a future chapter of this book on statistical inference in change-point problems, as the methods described therein are equipped with devices for uncertainty quantification. We defer the description of the NOT methodology \citep{bcf19} to another future chapter as it is mainly motivated by and applicable to detecting changes in general piecewise-polynomial signal settings in addition to piecewise-constant ones.



Recent years saw a proliferation of informative review papers on the multiple change-point problem. We mention in particular \cite{jfml13}, \cite{nhz16}, \cite{tov18}, \cite{bw20}, \cite{fr20} and \cite{ck20a}.

\section{Summary}

The following are some of the key messages from this chapter.

\begin{itemize}
    \item
    The multiple change-point detection problem can be split into the estimation of the number and the locations of change-points.
    Accurate estimation of the number of change-points is more difficult than estimating their locations.
    \item
    There is a link between multiple change-point detection and variable selection in linear regression; however, many commonly used methods for variable selection such as variants of the lasso cannot be used successfully in the change-point detection context as the design matrix does not satisfy the required decorrelation properties.
    \item
    One large class of multiple change-point estimators is estimators defined as optima. In this class, if the number of change-points is known, locations can be estimated by fitting a piecewise-constant function to data via least-squares. This leads to rate-optimal location estimators under suitable assumptions on the spacings between change-points and jump sizes.
    \item
    In order to estimate, additionally, the number of change-points in the optimisation approach, the optimised criterion needs to include a penalty for over-fitting. Perhaps the most commonly used penalty is SIC (also known as BIC), which can be shown to lead to consistent estimation of the number of change-points under appropriate assumptions. There are many other penalties, some of which penalise for change-points being located too close to each other.
    \item
    Important advances have recently been made in the computation of estimators defined as optima.
    \item
    Another large class of methods for estimating multiple change-points work by estimating them one by one. The main difference between these methods relates to how they search the data to find each next most likely change-point candidate. Amongst others, Binary Segmentation and its descendants and relatives are in this class.
    \item
    Typical estimators of the number of change-points for methods in this class use thresholding or penalisation.
    \item
    In multiple change-point detection, the term ``solution path'' refers to the sequence of models, for a given search algorithm, with 0, 1, \ldots, candidate change-points. Some search algorithms produce nested, while others -- non-nested solution paths.
    \item
    An important class of model selectors (i.e. estimators of the number of change-points) act on the solution path and examine its certain features (such as ``elbows'' in the residual sums of squares from model fits along the solution path) to propose the best model. With a slight abuse of terminology, such methods are referred to as adaptive.
    \item
    There are various important considerations to make when using multiple change-point detection software; these include speed, design and functionality, and accuracy and performance on similar data.
\end{itemize}

\bibliographystyle{plainnat}
\end{document}